\documentclass[useAMS,usenatbib]{mn2e}
\usepackage{graphicx,amssym}

\newcommand{\bc}{\begin{center}}
\newcommand{\ec}{\end{center}}

\title[Metallicity scaling relations]
      {Galaxy metallicity scaling relations in the {\sc EAGLE} simulations}
\author[De Rossi et al.]
       {\parbox{17cm}{Mar\'{\i}a~Emilia De Rossi$^{1,2}$ \thanks{Email:mariaemilia.dr@gmail.com}, Richard~G. Bower$^{3}$, 
        Andreea~S. Font$^{4}$, Joop~Schaye$^{5}$ and Tom Theuns$^{3}$}
       \\     
       \\
       $^{1}$ Universidad de Buenos Aires, Facultad de Ciencias Exactas y Naturales y Ciclo B\'asico Com\'un. Buenos Aires, Argentina\\  
       $^{2}$ CONICET-Universidad de Buenos Aires, Instituto de Astronom\'{\i}a y F\'{\i}sica del Espacio (IAFE). Buenos Aires, Argentina\\ 
       $^{3}$ Institute for Computational Cosmology, Physics Department, University of Durham, South Road, Durham DH1 3LE, UK \\
       $^{4}$ Astrophysics Research Institute, Liverpool John Moores University, 146 Brownlow Hill, Liverpool, L3 5RF, UK\\
       $^{5}$ Leiden Observatory, Leiden University, PO Box 9513, 2300 RA, Leiden, the Netherlands
}
\begin{document}

\date{Accepted  ???? ??. 2017 ???? ??}

\pagerange{\pageref{firstpage}--\pageref{lastpage}} 
\pubyear{2017}

\maketitle

\label{firstpage}

\begin{abstract}
We quantify the correlations between gas-phase and stellar metallicities and 
global properties of galaxies, such as stellar mass, halo mass, age and gas fraction,
in the Evolution and Assembly of GaLaxies and their Environments ({\sc EAGLE})
suite of cosmological hydrodynamical simulations.  
The slope of the correlation
between stellar mass and metallicity of star-forming (SF) gas ($M_*-Z_{\rm SF,gas}$ relation)
depends somewhat on resolution, with the higher-resolution run reproducing a steeper slope.
This simulation predicts a non-zero metallicity evolution,
increasing by $\approx 0.5$ dex at $\sim 10^9 {\rm M}_{\sun}$ since $z = 3$.
The simulated relation between stellar mass, metallicity and star formation rate 
at $z \la 5$ agrees remarkably well with the observed fundamental metallicity relation.
At $M_* \la 10^{10.3} {\rm M}_{\sun}$ and fixed stellar mass, 
higher metallicities are associated with lower specific star formation rates, lower gas fractions and older stellar populations. 
On the other hand, at higher $M_*$, there is a hint of an inversion of the dependence of metallicity on these parameters.
The fundamental parameter that best correlates with the metal content, in the simulations,
is the gas fraction.  The simulated gas fraction-metallicity relation exhibits small scatter and
does not evolve significantly since $z = 3$. 
In order to better understand the origin of these correlations, we analyse a set of lower resolution simulations
in which feedback parameters are varied. We find 
that the slope of the simulated $M_*-Z_{\rm SF,gas}$ relation is mostly determined by stellar feedback at 
low stellar masses ($M_* \la 10^{10} {\rm M}_{\sun}$), and at high masses ($M_* \ga 10^{10} {\rm M}_{\sun}$) by the 
feedback from active galactic nuclei.  
\end{abstract}
                                                                                                                 

\section{Introduction}
\label{sec:introduction}

The metallicity properties of galaxies encode crucial information about the different physical
processes (e.g. star formation, infall and outflows of gas, etc.) 
that drive their evolution \citep[e.g.][]{tinsley1980, dave2010, finlator2016}. 
In this context, the determination of scaling relations between the metallicity of
galaxies and other key properties such as their stellar masses ($M_*$) or gas fractions is a matter of great
interest and the subject of an on-going debate in the community as it can help to constrain models 
of structure formation.    

In the local Universe, there is a well-known correlation between stellar mass and metallicity, 
such that more massive galaxies are more metal-enriched
\citep[e.g.][]{lequeux1979, tremonti2004}.  This mass-metallicity relation (MZR) has also been investigated at higher
$z$ but with an apparent offset towards lower metallicities with respect to the local MZR 
\citep[e.g.][]{savaglio2005, erb2006, maiolino2008}.  However, a comparison between observational
works at different redshifts is not straightforward because of 
selection biases, aperture effects and the use of different metallicity indicators \citep[e.g.][]{steidel2014}.
Thus, the detailed dependencies of the MZR and their level of evolution are still debated
\citep[e.g.][]{kewley2008,telford2016}.

Recently, several authors have reported secondary dependences of metallicity at a given stellar mass, arguing
that the MZR is just the projection onto two-dimensions (the mass-metallicity plane) 
of more fundamental relations contained in higher-dimensional parameter space 
(e.g. the space defined by mass, metallicity and star formation rate).
In particular, \citet{ellison2008} reported that galaxies with
higher specific star formation rates (SFRs) or larger half-light radii show
systematically lower gas-phase metallicities than systems with similar $M_*$ but 
lower specific SFRs or smaller sizes.  
More recently, \citet{mannucci2010} have reported the existence of a 
`fundamental metallicity relation' (FMR) between $M_*$, metallicity and star formation rate (SFR)
that exhibits little scatter and does not show significant evolution, at least below $z \approx 2.5$
\citep[see also][]{laralopez2010a}. 
According to the FMR, systems with higher SFRs tend to have lower metallicities at a given stellar 
mass, consistently with the behaviour reported by \citet{ellison2008}.
Thus, part of the observed evolution
of the MZR might be due to the fact that surveys at high $z$ tend to select systems with elevated
SFRs.  

More recently, \citet{bothwell2013} suggested that the FMR might be a consequence of a 
more fundamental correlation between $M_*$, metallicity and gas fraction:
systems with higher gas fractions tend to exhibit lower metallicities at a fixed
mass.  These studies have been extended at different $z$ by many authors in recent years 
\citep[e.g.][]{cresci2012,hunt2012,henry2013,kacprzak2016,laralopez2013a,
stott2013,cullen2014,maier2014, nakajima2014, zahid2014a, bothwell2016} but,
as in the case of the MZR, the large uncertainties and diverse observational techniques
involved in observational works
prevent convergence towards a clear determination of these fundamental metallicity relations of galaxies,
as discussed by \citet[][]{telford2016}.

The tightness of the observed FMR seems to depend strongly on the adopted abundance determination method 
and might be affected by systematic errors such as those associated with stellar mass estimates, aperture effects, etc.
\citet{telford2016}, for example, applied new abundance diagnostics to data from the Sloan Digital Sky Survey
(SDSS), obtaining an anti-correlation between metallicity and SFR, at a given mass, which is weaker than that found
by \citet{mannucci2010}, by 30-55\%. 
\citet{andrews2013}, on the other hand, reported a stronger anti-correlation between
metallicity and SFR, at a fixed mass, than \citet{mannucci2010} for nearby galaxies. According to \citet{yates2011}, the
trend of increasing metallicity with decreasing SFR inverts at high stellar masses while
\citet{salim2014} suggested that this apparent turnover might be an artefact due to the
signal-to-noise cuts imposed on the observational sample. 
Finally, other authors claimed that there is no significant dependence of the observed MZR on
the SFR \citep[e.g.][]{hughes2013, sanchez2013, sanchez2017}.

From the theoretical point of view, different works have tried to address the origin
and evolution of metallicity scaling relations \citep[e.g.][]{tissera2005, 
dave2012, dayal2012, yates2011, lilly2013, romeo2013, yates2013, vogelsberger2014, ma2016, genel2016, weinberg2016}.
These models and simulations reproduce qualitatively the observed trends but
they show discrepancies regarding the exact value of the slope and level of evolution of the
predicted relations.
Supernova (SN)-driven outflows have often been invoked as a key ingredient for establishing a
MZR \citep{larson1974, tremonti2004, dalcanton2007, kobayashi2007}.  Given the shallower
potential wells of low-mass galaxies, metal-enriched material can be more efficiently ejected from
these systems, keeping their metallicities low.
Note, however, that the mass-loading of a wind, $\dot M_w/\dot M_\star$, 
is not necessarily the same as the metal mass loading of a wind, $\dot M_Z/\dot M_\star$
\citep[e.g.][]{maclow1999, creasey2015}, which makes the problem more complex.
Besides, less efficient star formation in low-mass galaxies
could also cause the lower chemical enrichment of smaller galaxies
and explain the origin of the MZR \citep[e.g.][]{brooks2007, mouchine2008, calura2009}.
The evolution of the MZR could also be affected by the different SF histories associated with galaxies 
with different morphologies \citep[see e.g.][]{calura2009}.
In addition, the infall of metal-poor gas onto the outer parts of galaxies or inflows triggered 
by mergers with other
systems could play an important role
\citep[e.g.][]{koppen1999, dalcanton2004, finlator2008, dave2010}.
\citet{derossi2007}, for example, have shown that a correlation between mass and metallicity can arise naturally
in a hierarchical scenario solely as a consequence of the regulation of SF by merger events. 
However, the lack of SN-feedback model prevented \citet{derossi2007} from reproducing the
observed slope of the MZR due to overcooling.
According to \citet{koppen2007}, the MZR might simply reflect 
variations in the stellar initial mass function (IMF).

\citet{derossi2015b,derossi2015} and  \citet{derossi2016} have studied the origin of the MZR and FMR by using   
the Galaxies-Intergalactic Medium Interaction Calculation ({\sc GIMIC}, \citealt{crain2009})
suite of cosmological hydrodynamical simulations.  These authors found that the star-forming (SF) gas
and stellar components of simulated galaxies follow local 
mass-metallicity relations very similar to those observed but with
less scatter.  
The simulated relations seem to be driven mainly
by infall of metal-poor gas as well as by the efficient action of SN feedback.
However, the {\sc GIMIC} simulations do not predict the observed level of
evolution of the MZR, because of the old average stellar ages ($\sim 10$ Gyr) of simulated galaxies.  
In addition, the {\sc GIMIC} simulations do not reproduce the observed flattening of the MZR at
the high-mass end.  \citet{derossi2015} claimed that the latter issue is probably related to
the lack of AGN feedback in {\sc GIMIC}.

More recently, \citet{schaye2015} showed that the Evolution and Assembly of GaLaxies 
and their Environments ({\sc EAGLE}, \citealt{schaye2015}) simulations are able to reproduce
a correlation between stellar mass and gas-phase metallicity at $z=0$, which agrees well with observed data
\citep{tremonti2004, zahid2014a}
for the high-resolution version of those simulations. 
{\sc EAGLE} high-resolution simulations have also been found to predict an evolution of the 
MZR consistent with the observational trend reported by \citet{zahid2013b} \citep[e.g.][]{guo2016}.
In addition, by analysing intermediate-resolution {\sc EAGLE} runs, 
\citet{lagos2015} found that metallicity
can be robustly determined from neutral gas fractions, or from $M_*$ and SFR.
In particular, the strength of the stellar feedback implemented in {\sc EAGLE} has an important impact on the $z=0.1$ simulated MZR \citep{crain2015}.
Furthermore, according to the results of \citet{segers2016a}, the AGN model included in {\sc EAGLE} yields 
a relation between $M_*$ and stellar-$\alpha$-element-to-iron ratio ($[{\rm \alpha / Fe}]_*$) consistent
with observations of massive ($M_* > 10^{10.5} {\rm M}_{\sun}$) early type galaxies.   
Simulations can therefore play an important role in examining the relative importance of different
physical processes that drive the evolution of correlations.  In addition, they are valuable tools 
in examining observational biases by allowing the comparison of intrinsic correlations to those inferred
from simulations after applying observed selections to mock observables.

In this article, we extend previous works by analysing in detail the evolution of 
metallicity of galaxies as a function of mass and redshift using the {\sc EAGLE} 
suite of cosmological simulations. 
We focus on the analysis of the {\em high-resolution} version of the simulations which implement
the so-called {\em recalibrated model} (see below). We show that the improved {\sc EAGLE} subgrid prescriptions lead to a better description
of the evolution of the MZR than the {\sc GIMIC} simulations, preserving many key features of the observed relation that were also reproduced by {\sc GIMIC}.
In particular, the AGN feedback model in {\sc EAGLE} yields a better description
of the global metal enrichment of massive galaxies, generating the observed turn-over of the MZR at the high-mass
end, as was also shown by \citet{segers2016b}.

The plan of the paper is as follows. The simulation and the sample selection
are described in Section \ref{sec:simulation}.
The simulated mass-metallicity relation obtained from SF gas abundances is presented in 
Section \ref{sec:gas_mzr} while that derived from stellar abundances are discussed in 
Section \ref{sec:mzr_stars}.  In Section \ref{sec:yields}, we compare observed and simulated
effective yields.  The scatter of the simulated MZR and 
its dependence on secondary parameters, such as gas fraction,
SFR and mass-weighted stellar age, is discussed in Section \ref{sec:mzr_scatter}.
In Section \ref{sec:models}, we explore different sets of simulations corresponding to different resolutions
and models in order to reveal the processes that
determine the main features of the simulated metallicity scaling relations.
Finally, our conclusions are summarised in Section \ref{sec:conclusions}.

\section{The {\sc EAGLE} simulations}
\label{sec:simulation}

The {\sc EAGLE} suite\footnote{See 
http://eagle.strw.leidenuniv.nl and http://www.eaglesim.org/ for different data
products, images and movies.  In addition, {
a database with many integrated properties of {\sc EAGLE} galaxies  \citep{mcalpine2015}
and particle data \citep{eagle2017} is publicly available.}}
of cosmological simulations \citep{schaye2015, crain2015} constitutes
a set of different hydrodynamical simulations run with different 
resolutions, box sizes and subgrid physics models.  
The simulations were performed by using a modified 
version of the {\sc GADGET-3} Smoothed Particle Hydrodynamics (SPH) code 
(last described by \citealt{springel2005a}). The modifications to the SPH implementation
are collectively referred to as ANARCHY (Dalla Vecchia, in prep.; see also \citealt{schaller2015a}). 

The {\sc EAGLE} simulations track the joint evolution of dark matter and baryons  
within periodic comoving volumes of side-length up to 100 comoving megaparsec (cMpc) from $z=127$ to $z=0$.  
A reference model has been implemented, for which the subgrid parameters associated
with energy feedback were calibrated to obtain good agreement with the $z=0.1$ galaxy
stellar mass function (GSMF), whilst also reproducing the observed sizes of present-day disk galaxies.
\citet{furlong2014} show that the simulated GSMF 
also broadly reproduces the data up to $z \approx 7$. In addition to the reference model,
other variations of subgrid parameters have been explored as discussed in detail
in \citet{crain2015}.

To distinguish more easily the different runs within the {\sc EAGLE} suite of simulations, 
the name of a given simulation includes
a suffix that indicates the box length in comoving megaparsec (e.g. L100)
and the cube root of the initial number of particles per species (e.g. N1504).
Simulations with the same subgrid model as the primary one (the reference model) are
denoted with the prefix "Ref-" (e.g. Ref-L100N1504).  As discussed in \citet{schaye2015}, for higher resolution
simulations, a "Recal-" model has been implemented in addition to the reference one.
The former model uses subgrid parameters that have been recalibrated
following similar procedures to those applied to the reference run to improve the fit to
the $z\sim0$ GSMF when working with the high-resolution simulations
(Recal-L025N0752).

Within the {\sc EAGLE} suite, the Recal-L025N0752 simulations have been found to
reproduce the observed trends for the MZR (\citealt{schaye2015}, see also below), because
of its higher feedback efficiency.  Thus, our analysis will be focused on this run. 
We will also perform comparisons with other simulations carried out within a fixed simulated volume
("L025" runs) but using different resolutions and model parameters to assess their effects 
on the obtained trends.  
Variations of AGN feedback
parameters have only been tested within comoving volumes of side-lengths of 50 cMpc
("L050" runs); we will employ this set of simulations to analyse the impact of AGN feedback. 
In Table \ref{tab:simus}, we present the set of simulations analysed in this work, indicating
their main subgrid parameters, which are described below.

\subsection{Cosmological parameters and subgrid implementation}
\label{sec:simulation_details}

A flat $\Lambda$CDM cosmology is assumed with cosmological parameters consistent
with those inferred by the Planck Collaboration (2014):
${\Omega}_{\Lambda} = 0.693$, ${\Omega}_{\rm m} = 0.307$, ${\Omega}_{\rm b} = 0.04825$,
${\sigma}_8 = 0.8288$, $h = 0.6777$, $n_{s} = 0.9611$ and $Y=0.248$ where 
${\Omega}_{\rm m}$, ${\Omega}_{\Lambda}$ and ${\Omega}_{\rm b}$ are the average densities of matter, dark
energy and baryonic matter in units of the critical density at $z=0$, 
${\sigma}_8$ is the square root of the
linear variance of the matter distribution when smoothed with a top-hat filter of radius 8 $h^{-1}$ cMpc,
$H_0 \equiv h \ 100 \ {\rm km \ s^{-1} \ Mpc^{-1}}$ is the Hubble parameter, 
$n_{s}$ is the scalar power-law index of the power spectrum of primordial adiabatic perturbations,
and $Y$ is the primordial abundance of helium.
Below, we briefly describe the main characteristics of {\sc EAGLE} subgrid physics; more details can
be found in \citet{schaye2015}.

\begin{table*} 
\begin{center}
\caption{Parameters that are varied in the simulations. Columns list: simulations identifiers, the side length of the volume ($L$) and the particle number per species (i.e. gas, DM) per dimension ($N$), 
the power-law slope of the polytropic equation of state of SF gas ($\gamma_{\rm eos}$), 
the power-law index of the star formation law ($n$), 
the asymptotic maximum ($f_{\rm th,max}$) and minimum ($f_{\rm th,min}$) values of $f_{\rm th}$ (Equation \ref{eq:fth}), 
the parameters that control the characteristic density and the power-law slope of the density dependence 
of the energy feedback from star formation ($n_{\rm H,0}$ and $n_{\rm n}$, respectively), the subgrid accretion disc viscosity parameter ($C_{\rm visc}$), and the temperature increment of stochastic AGN heating ($\Delta T_{\rm AGN}$). The upper section comprises models that have been calibrated to reproduce the $z = 0.1$ GSMF and the lower section comprises models featuring single-parameter variations of Ref. Numbers in bold indicate variations with respect to the reference model (Ref). 
This table has been adapted from Table $1$ in \citet{crain2015} for the simulations used in
this work.}
\begin{tabular}{r c c c c c c c c c c c}
\hline
\hline
Identifier & Side length L & ${\rm N}$ & $\gamma_{\rm eos}$ & $n$               & $f_{\rm th,max}$ & $f_{\rm th,min}$ & $n_{\rm H,0}$     & $n_{\rm n}$ & $C_{\rm visc}/2\pi$ & $\Delta T_{\rm AGN}$  \\
           & [cMpc]      &     &                    & [${\rm cm}^{-3}$] &                  &                  & [${\rm cm}^{-3}$] &             &                     & $\log_{10}$ [K]       \\
\hline
\textit{Calibrated models} \\
Recal-L025N0752     & 25  & 752   & $4/3$      & $1.4$     & $3.0$      & $0.3$       & $\bf 0.25$  & $\bf 1/\ln{10}$ & $\bf 10^3$ & $\bf 9.0$ \\
Ref-L025N0752       & 25  & 752   & $4/3$      & $1.4$     & $3.0$      & $0.3$       & $0.67$      & $2/\ln{10}$     & $10^0$     & $8.5$ \\
Ref-L025N0376       & 25  & 376   & $4/3$      & $1.4$     & $3.0$      & $0.3$       & $0.67$      & $2/\ln{10}$     & $10^0$     & $8.5$ \\
Ref-L050N0752       & 50  & 752   & $4/3$      & $1.4$     & $3.0$      & $0.3$       & $0.67$      & $2/\ln{10}$     & $10^0$     & $8.5$ \\
Ref-L100N1504       & 100 & 1504  & $4/3$      & $1.4$     & $3.0$      & $0.3$       & $0.67$      & $2/\ln{10}$     & $10^0$     & $8.5$ \\
FBconst-L050N0752   & 50  & 752   & $4/3$      & $1.4$     & $\bf 1.0$  & $\bf 1.0$   & $\bf -$     & $\bf -$         & $\bf 10^3$ & $8.5$ \\
\textit{Ref. variations} \\
eos1-L025N0376      & 25  & 376   & $\bf 1$    & $1.4$     & $3.0$      & $0.3$       & $0.67$      & $2/\ln{10}$     & $10^0$     & $8.5$ \\
eos5/3-L025N0376    & 25  & 376   & $\bf 5/3$  & $1.4$     & $3.0$      & $0.3$       & $0.67$      & $2/\ln{10}$     & $10^0$     & $8.5$ \\
KSLow-L025N0376 & 25  & 376   & $4/3$      & $\bf 1.0$ & $3.0$      & $0.3$       & $0.67$      & $2/\ln{10}$     & $10^0$     & $8.5$ \\
KSHi-L025N0376  & 25  & 376   & $4/3$      & $\bf 1.7$ & $3.0$      & $0.3$       & $0.67$      & $2/\ln{10}$     & $10^0$     & $8.5$ \\
WeakFB-L025N0376    & 25  & 376   & $4/3$      & $1.4$     & $\bf 1.5$  & $\bf 0.15$  & $0.67$      & $2/\ln{10}$     & $10^0$     & $8.5$ \\
StrongFB-L025N0376  & 25  & 376   & $4/3$      & $1.4$     & $\bf 6.0$  & $\bf 0.6$   & $0.67$      & $2/\ln{10}$     & $10^0$     & $8.5$ \\
NOAGN-L050N0752     & 50  & 752   & $4/3$      & $1.4$     & $3.0$      & $0.3$       & $0.67$      & $2/\ln{10}$     & $\bf -$     & $\bf -$ \\
AGNdT8-L050N0752    & 50  & 752   & $4/3$      & $1.4$     & $3.0$      & $0.3$       & $0.67$      & $2/\ln{10}$     & $10^0$     & $\bf 8.0$ \\
AGNdT9-L050N0752    & 50  & 752   & $4/3$      & $1.4$     & $3.0$      & $0.3$       & $0.67$      & $2/\ln{10}$     & $10^0$     & $\bf 9.0$ \\
\hline
\end{tabular}
\label{tab:simus}
\end{center}
\end{table*}

Radiative cooling and heating rates are computed on an element-by-element basis
for gas in ionization equilibrium in the presence of a \citet{haardt2001} ionizing
UV/X-Ray background and the Cosmic Microwave Background.  The {\em total} metallicity variable, $Z$,
and the 11 elements (H, He, C, N, O, Ne, Mg, Si, S, Ca, and Fe) that are important for
the radiative cooling at $T > 10^4$ K \citep{wiersma2009a} are
tracked individually.

Star formation is implemented stochastically following 
\citet{schaye2008},
but including a metallicity-dependent density threshold $n_{\rm H}^*$, as:

\begin{equation}
n_{\rm H}^* = 10^{-1} {\rm cm}^{-3} \big(\frac{Z}{0.002}\big)^{-0.64}, 
\end{equation}

\noindent
that yields the Kennicutt-Schmidt relation.
It exhibits a dependence on gas metallicity $Z$ that captures the transition
from the warm, atomic to the cold, molecular gas phase \citep{schaye2004}. 
A temperature floor 
$T_{\rm eos} ( {\rho}_{\rm g} )$ is applied, which is associated with the equation of state
$P_{\rm eos} \propto {\rho}_{\rm g}^{{\gamma}_{\rm eos}}$ 
(${\gamma}_{\rm eos} = 4/3$, for standard runs), normalised to $T_{\rm eos} = 8 \times 10^3$ K
at $n_{\rm H} = 10^{-1} {\rm cm}^{-3}$, a typical temperature for the warm interstellar medium (ISM) 
\citep[e.g.][]{richings2014}. In this way, cold dense gas is prevented from artificial
fragmentation due to a lack of resolution.
When gas particles reach densities $n_{\rm H} > n_{\rm H}^*$ and 
$\log_{10} (T/{\rm K}) < \log_{10} T_{\rm eos}/{\rm K} + 0.5$ , they are eligible for
star formation and are assigned an SFR, ${\dot m}_*$ \citep{schaye2008}:

\begin{equation}
\label{eq:sfr}
{\dot m}_* = m_{\rm g} \ A \ (1 {\rm M}_{\sun} {\rm pc}^{-2})^{-n} \ (\frac{\gamma}{G} 
             f_{\rm g} P)^{(n-1)/2} ,
\end{equation}

\noindent
where $m_{\rm g}$ is the gas particle mass, $\gamma = 5/3$ is the ratio of specific
heats, $G$ is the gravitational constant, $f_{\rm g}$ is the mass fraction in gas 
and $P$ is the total pressure.  For the {\sc EAGLE} simulations used in this work, 
$f_{\rm g} = 1$. The parameters $A= 1.515 \times 10^{-4} \ {\rm M}_{\sun} \ {\rm yr}^{-1} \ {\rm kpc}^{-2}$ 
and $n=1.4$ are obtained directly from 
the observed Kennicutt-Schmidt relation \citep{kennicutt1998}, when scaled
to the \citet{chabrier2003} IMF.
Throughout this paper, when performing comparisons with observational results 
that assume another IMF, 
data are converted to a Chabrier IMF for consistency.

The chemical enrichment model follows the prescriptions of \citet{wiersma2009b}.
As mentioned, the simulations track the stellar mass losses of 11 elements associated with
three stellar evolutionary channels:
($i)$ stellar winds and core-collapse (type~II) supernovae resulting from massive stars ($M>6{\rm M}_\odot$), 
($ii)$ type Ia supernovae assumed to result from catastrophic mass transfer in close binary stars, 
and ($iii$) winds from asymptotic giant branch (AGB) stars. Stellar evolutionary tracks and 
yields that depend on the initial metal abundance
are implemented.
For type~II supernovae, yields from \citet{portinari1998} were used because 
they consider mass loss from massive stars.  In the case of AGB stars, 
yields of \citet{marigo2001} were implemented as they constitute a self consistent set 
with yields from \citet{portinari1998}.
For type Ia supernovae, \citet{wiersma2009b}
used the last version of the standard "W7" model \citep{thielemann2003} (the reader is referred
to \citealt{wiersma2009b} for more details about yield choices and implementation).
As discussed in detail in \citet{wiersma2009b}, nucleosynthetic yields are uncertain by factors
of a few and the abundance evolution is sensitive to the particular
choice of yield tables.

Stochastic thermal feedback from star formation is applied in the {\sc EAGLE} simulations. 
The feedback model is described by
\citet{dallavecchia2012} and is based on a stochastic selection of neighbouring gas 
particles that are heated by a temperature increment of $10^{7.5}$K.  Taking into account the local
metallicity and gas density, a fraction $f_{\rm th}$ of energy from core-collapse supernovae
is injected into the ISM 30 Myr after the birth of a stellar population \citep{schaye2015, crain2015}.
In the model, $f_{\rm th}$ is given by:

\begin{equation}
\label{eq:fth}
f_{\rm th} = f_{\rm th,min} + 
\frac{f_{\rm th,max}-f_{\rm th,min}}{1 + \big( \frac{Z}{0.1 Z_{\sun}} \big) ^{n_Z} 
\big( \frac{n_{\rm H, birth}}{n_{\rm H,0}} \big) ^{-n_n}}  , 
\end{equation}

\noindent
where $n_{\rm H, birth}$ is the density inherited 
by the star particle from its parent gas particle, $Z$ is the metallicity, 
$f_{\rm th,min}$ and $f_{\rm th,max}$ are the asymptotic values of $f_{\rm th}$ 
while $n_Z$, $n_n$ and $n_{\rm H,0}$ are free parameters.
It is assumed that $n_Z = n_n$.
The parameters $n_{\rm H,0}$ and $n_n$ were chosen to reproduce the present-day GSMF and galaxy sizes.

When the halo mass of a system increases above $10^{10} h^{-1} {\rm M}_{\sun}$, 
seed black holes (BHs) of mass $10^5 h^{-1} {\rm M}_{\sun}$ are placed inside them 
following \citet{springel2005b}.  
BHs grow by subsequent gas accretion events\footnote{ 
When the subgrid BH mass becomes higher that its host
particle mass, the BH can stochastically accrete neighbouring
gas particles (see \citealt{schaye2015} for more details).}
and mergers
at a rate computed according to the modified Bondi-Hoyle accretion rate of 
\citet{rosasguevara2015} and \citet{schaye2015}.  
To regulate the Bondi rate in high-circulation flows,
a viscosity parameter $C_{\rm visc}$ is introduced.  
AGN feedback is implemented thermally and stochastically similarly to energy feedback from star formation.
Particles surrounding the BH are chosen randomly and heated by a temperature 
$\Delta T_{\rm AGN}$.
Increasing $\Delta T_{\rm AGN}$ leads to more energetic individual feedback
events, generally resulting in smaller radiative
losses in the ISM. In addition, larger values of $\Delta T_{\rm AGN}$ generate a 
more intermittent feedback process.
We note that a single feedback mode is included in {\sc EAGLE} as 
the current implementation naturally yields an AGN feedback that mimics 'quasar-mode' and 'radio-mode'
feedbacks at high and low accretion rates, respectively \citep{rosasguevara2016}.

\subsection{Set of studied simulations}
\label{sec:simulation_set}

Table \ref{tab:simus} summarises the different {\sc EAGLE} simulations analysed in this work and their main
parameters.

As noted by \citet{schaye2015}, the correlation between $M_*$ and star forming (SF)
gas metallicity ($Z_{\rm SF,gas}$) depends on resolution and the choice of feedback parameters.  
In particular, Recal-L025N0752 reproduces the 
slope and normalization of certain observational data sets 
better because of the increase of the feedback efficiency with respect to 
the reference model.
The increase of the energy feedback from star formation required to match the observed 
GSMF, simultaneously decreases the metal content of the ISM of low-mass galaxies
leading to a better agreement with observations.
As discussed in Section \ref{sec:models} (see also Section \ref{sec:gas_mzr}), a decrease in resolution (which in our case also implies the use of
the reference model and so, different feedback parameters) does not alter
the main trends of the fundamental metallicity scaling relations but can moderately
affect their detailed features.

\begin{figure*}
\begin{center}
\resizebox{8.85cm}{!}{\includegraphics{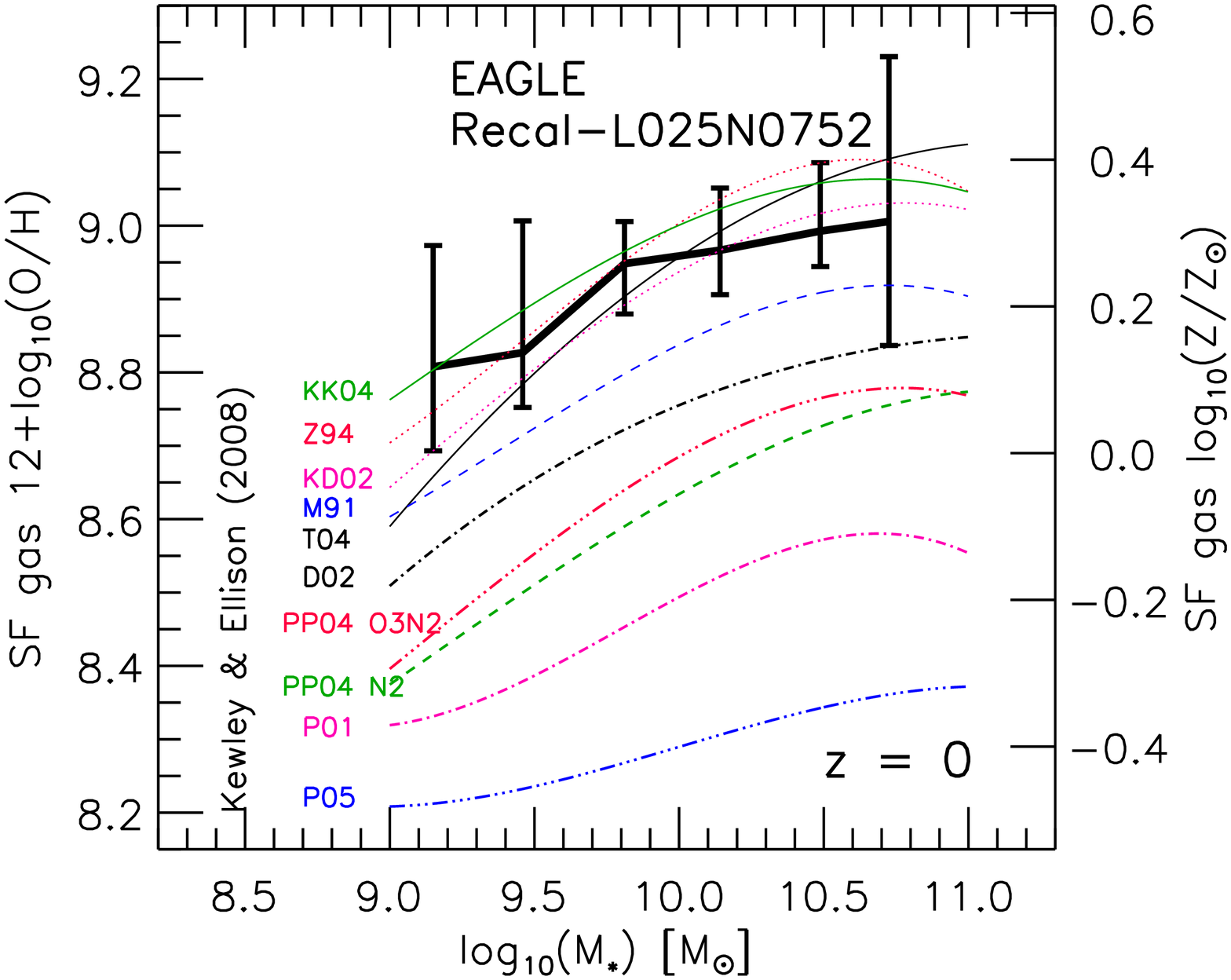}}
\hspace{-0.25cm}
\resizebox{8.85cm}{!}{\includegraphics{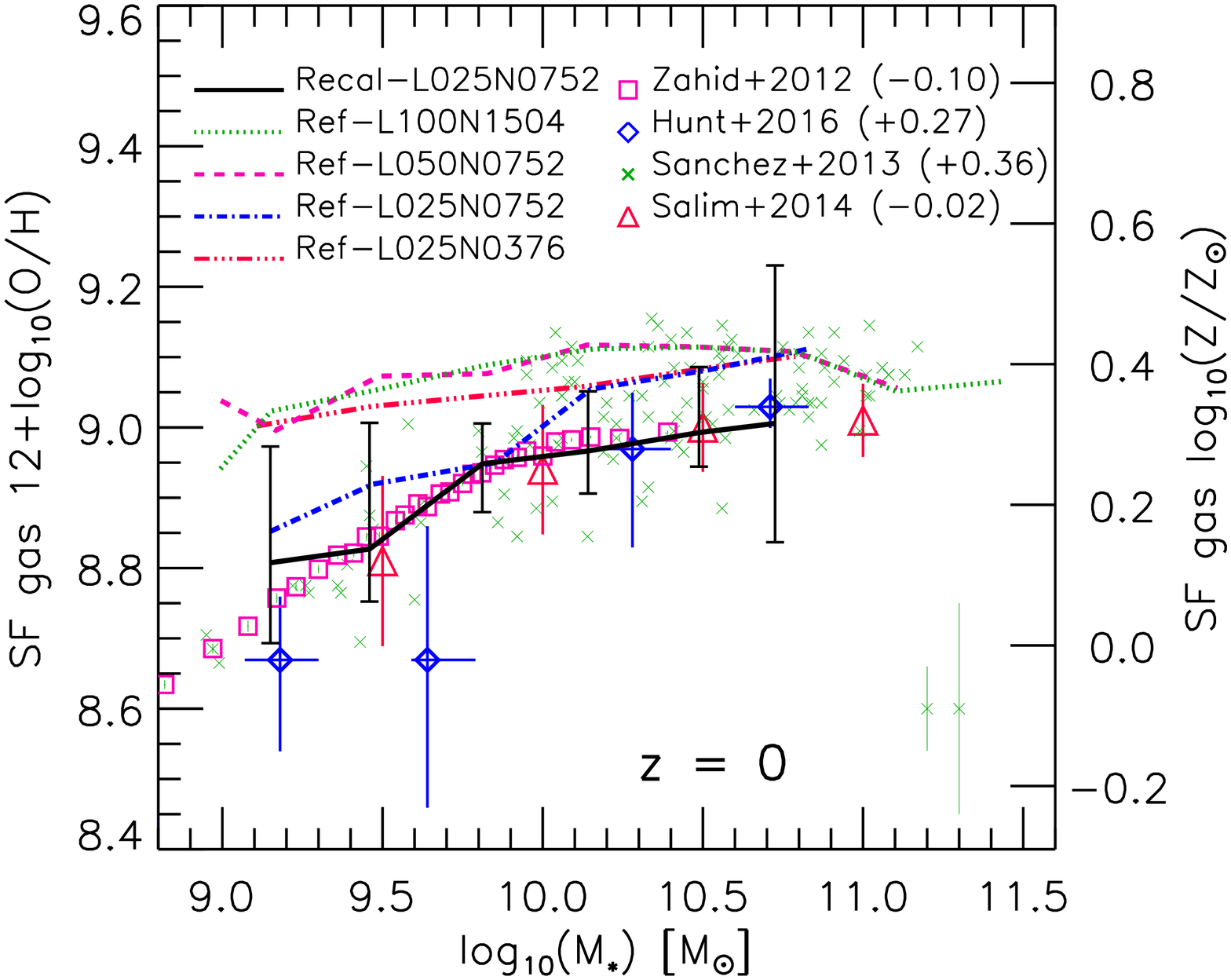}}
\end{center}
\caption[]
{
Left panel: Median $M_* - {\rm O/H}|_{\rm SF,gas}$ relation at $z=0$ obtained from Recal-L025N0752 {\sc EAGLE} simulation
(thick black line). Error bars depict the 25th and 75th percentiles.
Thin lines depict the SDSS best-fit
relations derived by \citet{kewley2008} by using a diverse set of metallicity calibrators,
as indicated in the figure (see text for details).
Right panel: Median $M_* - {\rm O/H}|_{\rm SF,gas}$ relations at $z=0$ obtained from different {\sc EAGLE} simulations
(curves with different colours).  In the case of the Recal-L025N752
simulations, error bars depict the 25th and 75th percentiles.
Observational results reported by different authors are represented with symbols and error bars.
Error bars at the bottom right corner represent the maximum and median errors reported by \citet{sanchez2013}.
As specified in the figure,
observational data were renormalized by adding -0.1, 0.27, 0.36 and -0.02 dex 
in order to match the Recal-L025N752 relation at $M_* \approx 10^{10.5} \ {\rm M_{\sun}}$ (see the text for details).
The conversion between oxygen abundances along the left $y$ axis and total metallicities
shown along the right $y$ axis has been carried out taking 
$12+ \log_{10} ({\rm O/H})\sun = 8.69$ \citep{allende2001}.
}
\label{fig:gas_mzr_z0}
\end{figure*}

Our analysis will be focused on the high-resolution simulation Recal-L025N0752
because it agrees better with the 
slope and normalization of certain
observed metallicity scaling relations.
For intermediate-resolution simulations, the predicted MZR is too flat at
low stellar masses (see Fig. \ref{fig:gas_mzr_z0} and corresponding discussion 
in Sec. \ref{sec:gas_mzr}). 
Thus, {\em unless otherwise specified, 
we will show results from the Recal-L025N0752 simulation in this work.}
In addition, for assessing resolution effects and comparison with the
reference model at a fixed volume, we will use the
simulations Ref-L025N0376 and Ref-L025N0752.  To test the impact on our results of
the slope of the $P - \rho$ relation imposed at high $\rho$, 
we will analyse simulations eos1-L025N0376 and eos5\/3-L025N0376.
The effects of changing the power-law index in the star formation law
will be addressed by studying simulations KSLow-L025N0376 and KSHi-L025N0376.
Simulations WeakFB-L025N0376 and StrongFB-L025N0376 
will allow us to compare a weak and strong stellar feedback model, respectively.
Simulation FBconst-L050N0752 will allow us to evaluate the effects 
of injecting into the ISM a fixed quantity of energy
per unit stellar mass formed, independent of local conditions.
We will compare the effects of varying the AGN feedback temperature, 
at a fixed volume and resolution, by contrasting simulations Ref-L050N0752, 
AGNdT8-L050N0752 and AGNdT9-L050N0752. We will also analyse the simulation
NOAGN-L050N0752, for which the BH implementation is turned off.  Thus,
in the later simulations,
BH gas accretion and AGN feedback are disabled entirely.

\subsection{Identifying galaxies in {\sc EAGLE}}
\label{sec:sample}

Dark matter overdensities were identified by applying the 
``Friends-of-Friends'' (FoF) method, assuming
a linking length of 0.2 times the average inter-particle spacing \citep{davis1985}.
In the case of baryonic particles, they are assigned to the same FoF-group
as their nearest dark matter neighbour.  Self-bound substructures, which
can contain both dark matter and baryons, are then identified by applying
the {\sc SUBFIND} algorithm \citep{springel2001, dolag2009}.
A FoF halo can contain several {\sc SUBFIND} sub-groups, or subhalos.
We define the central galaxy as the subhalo with the lowest value of the gravitational potential
while any remaining subhalos are classified as satellite galaxies.
Unless otherwise specified, we include both central and satellite galaxies
in our sample.

In order to mimic the aperture of instruments used for observations, we computed
integrated quantities inside a given radius.  Unless otherwise indicated, we follow
\citet{schaye2015} and
report global properties of galaxies inside a sphere with radius 30 proper kpc (pkpc).  In general, for the analysis carried
out here, aperture effects do not change the general trends but can generate moderate
variations in the slope and normalization of metallicity scaling relations (see the Appendix A)
particularly for massive galaxies.

\section{The $M_{*} - Z_{\rm SF,gas}$ relation}
\label{sec:gas_mzr}

In this section, we study the correlation between stellar mass and gas-phase metallicity
of simulated galaxies and compare our results with different observational findings.
The mass-metallicity relation is typically inferred by measuring oxygen abundances in star-forming gas
HII regions. 
Thus, we calculated global oxygen abundances (${\rm O/H}|_{\rm SF,gas}$) considering gas particles 
with ${\dot m}_* > 0$ (see Eq. \ref{eq:sfr}) at radius $r \le 30$ kpc in a given subhalo.
It is worth noting that, had we considered {\it all} gas particles (i.e. SF gas and non SF -NSF- gas) 
for these estimates, the gas metallicities would have decreased by $\approx 0.1-0.3$
dex (the exact values depend on mass and redshift).  This is because the NSF gas is less metal-enriched 
than the SF gas (see below).

\subsection{The local $M_{*} - Z_{\rm SF,gas}$ relation}
\label{sec:gas_mzr_z0}

\citet{schaye2015} have compared the local $M_* - {\rm O/H}|_{\rm SF,gas}$ relation 
measured by \citet{tremonti2004} and \citet{zahid2014a} with results from the
{\sc EAGLE} intermediate- and high-resolution simulations,
Ref-L100N1504 and Recal-L025N0752, respectively.  The two sets of observed data
are both based on SDSS data, but metallicities were obtained using different techniques.
Both observed MZRs agree at $M_* \sim 10^{11} \ {\rm M}_{\sun}$ but the relation
reported by \citet{tremonti2004} is steeper at low masses.
The intermediate-resolution simulation agrees with the flatter relation given by \citet{zahid2014a} 
to better than 0.1 dex for $M_* > 10^{9.5} \ {\rm M}_{\sun}$ but, at lower masses, this
simulation does not reproduce the steep
observed slopes obtained by \citet{tremonti2004} or \citet{zahid2014a}.  On the other hand, the high-resolution simulation
predicts a steeper slope for the $M_* - {\rm O/H}|_{\rm SF,gas}$ relation, 
in better agreement with \citet{tremonti2004} and \citet{zahid2014a} results.
Thus, the observed MZR seems to be better reproduced
by the simulation Recal-L025N0752 than by Ref-L100N1504.

The shape and normalization of the observed MZR are still a matter of extensive
debate. 
The use of different metallicity indicators, different methods
for estimating stellar masses, selection biases, aperture effects and dust can affect the slope and
zero-point of the MZR.
In Fig. \ref{fig:gas_mzr_z0}, we extend the analysis by \citet{schaye2015}  
studying the observed MZR and comparing it with the
$M_* - {\rm O/H}|_{\rm SF,gas}$ relation obtained 
from simulations with different resolutions and different simulation volumes. 
In the left panel, we compare the 
$M_* - {\rm O/H}|_{\rm SF,gas}$ relation obtained from the EAGLE Recal-L025N0752 simulation with the SDSS best-fit 
relations derived by \citet{kewley2008} by using a set of diverse metallicity calibrators:
KK04 \citep{kobulnicky2004}, Z94 \citep{zaritsky1994}, KD02 \citep{kewley2002}, 
M91 \citep{mcgaugh1991}, T04 \citep{tremonti2004}, D02 \citep{denicolo2002}, 
PP04 O3N2, PP04 N2 \citep{pettini2004}, P01 \citep{pilyugin2001} and P05 \citep{pilyugin2005}.
We see that the differences between observational results can reach 
$\approx 0.7$ dex in some cases. The slope of the observed MZR is also affected by the choice of the metallicity
calibrator, with the steeper relation obtained by the T04 method and the shallower relation
inferred when using the P05 technique.

In the right panel of Fig. \ref{fig:gas_mzr_z0}, we compare results from simulations with observational data
reported by different authors: 
\citet{zahid2012} (median metallicities in bins of $M_*$ 
and standard errors on the mean), \citet{sanchez2013} (metallicities for individual galaxies), 
\citet{salim2014} (median metallicities and standard deviations
in $M_*$ bins) and \citet{hunt2016} (medians of metallicities with the 75\% and 25\% quantile levels).
For the sake of clarity, only the maximum and median errors reported by \citet{sanchez2013} are represented
at the bottom right corner.
\citet{zahid2012} determined metallicities of galaxies in the SDSS by using the strong-line 
calibration KK04.
\citet{sanchez2013} reported data from the CALIFA survey, 
with metallicities derived from the strong-line PP04 O3N2-calibrator.
In the case of \citet{salim2014}, they calculated metallicities from  SDSS data by applying the
strong-line technique of \citet{mannucci2010}. 
Data from \citet{hunt2016} correspond to a compilation of different samples of galaxies
at $z=0$, with metallicities calculated using the PP04 N2 strong-line method.
As the normalization of the observed MZRs obtained by these authors is affected by the different 
metallicity calibrators used, we renormalized these
observational results in order
to match the median Recal-L025N0752 relation at $M_* \approx 10^{10.5} \ {\rm M}_{\sun}$. 
We added -0.10, 0.36, -0.02 and 0.27 dex to the data reported by \citet{zahid2012},
\citet{sanchez2013}, \citet{salim2014} and \citet{hunt2016}, respectively.
In this way, we avoid normalization issues and can 
focus on the comparison of the shapes of the relations.

In the case of simulations, we estimated median relations from mass bins containing more
than 10 galaxies. The number of galaxies per bin is $N_{\rm bin} \ga 30$, for simulation Ref-L100N1504 and
$N_{\rm bin} \ga 20$, for simulation Ref-L050N0752. 
For simulations Ref-L025N0376, Ref-L025N0752 and Recal-L025N0752, $N_{\rm bin} \ga 10$, considering
the whole analysed mass range and $N_{\rm bin} \ga 20$, if we remove the highest mass bin.

Independent of the volume, intermediate-resolution simulations (Ref-L025N0376, Ref-L050N0752, Ref-L100N1504)
yield similarly flat shapes for the $M_* - {\rm O/H}|_{\rm SF,gas}$ relation, departing from observations.  
On the other hand, high-resolution simulations yield steeper MZR slopes more consistent with the observational trend.  
We can see a general good agreement between results from high-resolution simulations and observations
by \citet{zahid2012} and \citet{salim2014}. 
In particular, the recalibrated model predicts a MZR that reproduces encouragingly well the
relation found by \citet{salim2014}, even without including the normalization adjustment (-0.02 dex).
Generally, the steeper (shallower) slope obtained at low (high) masses for Recal-L025N0752 simulation
improves the agreement with the observed trend.

We note that, because of their smaller volume ($25^3 \ {\rm cMpc}^3$), higher resolution simulations
cannot sample high-density environments.  This issue leads to a dearth
of galaxies, specially towards higher masses, and might affect the comparison with observations. 
However, intermediate resolution simulations run in different volumes 
show similar metallicity relations, suggesting a weak influence of environment on the obtained trends.
By analysing {\sc GIMIC} simulations, \citet{derossi2015} also found that metallicity scaling 
relations are not significantly affected by the large-scale environment.
Recent observations suggest a small dependence of the MZR on the environment \citep[e.g.][]{wu2017}, too.

Finally, it is worth recalling that the determination of abundances in the simulations are 
affected by the choice of the nucleosynthetic yields that are uncertain by a factor of a few
\citep{wiersma2009b}. Thus, given the issues affecting observational results and the uncertainties
in simulated yields, the comparison between models and observations should be taken with care.

\begin{figure*}
\begin{center}
\vspace{-0.25cm}
\resizebox{18cm}{!}{\includegraphics{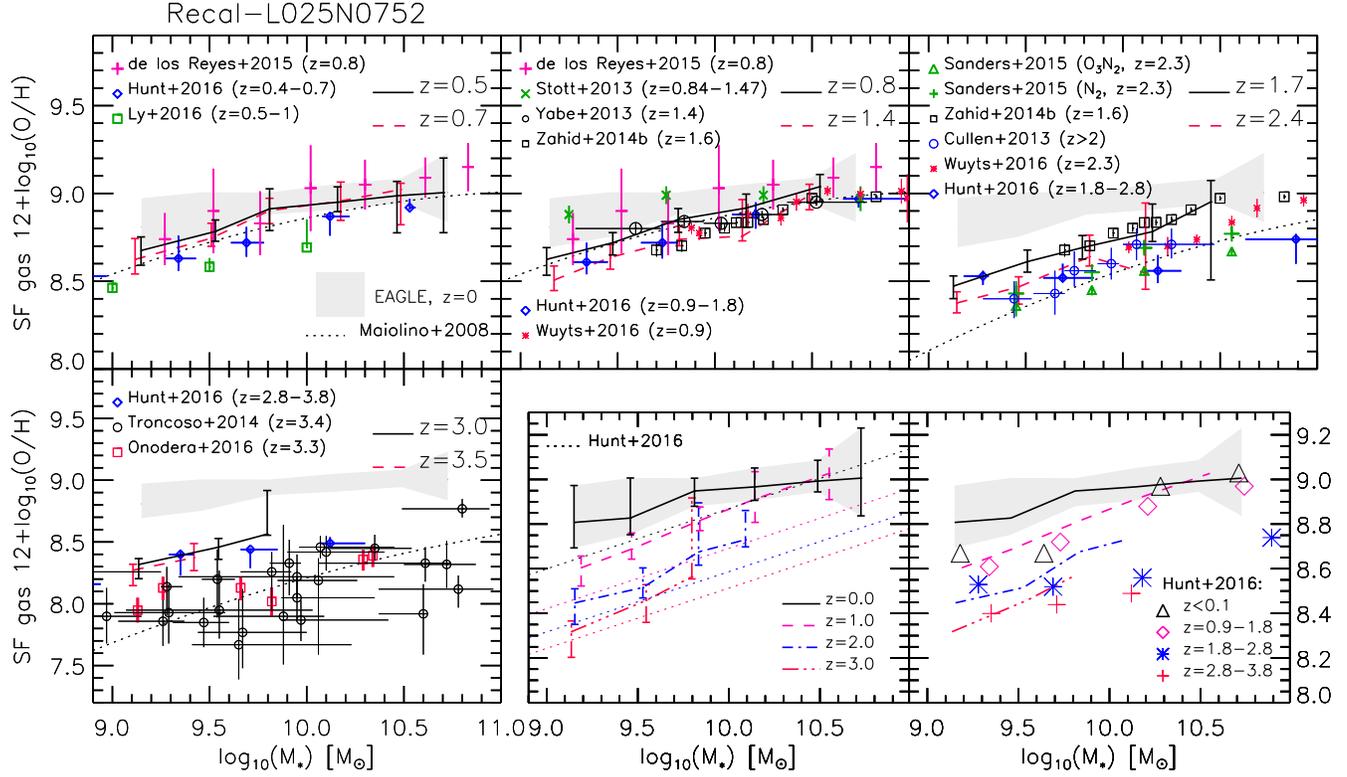}}
\vspace{-1cm}
\end{center}
\caption[]
{
Large panels: Simulated $M_* - {\rm O/H}|_{\rm SF,gas}$ relation at different $z$. 
Curves depict the median relation and error bars the 25th and 75th percentiles
corresponding to simulation Recal-L025N752.
Different observational results at $z$ close to simulations 
are shown as symbols with error bars.  
Fitted relations to observations given in \citet{maiolino2008} 
are represented with dotted lines.
Small left panel: Comparison between simulation results (curves with error bars) and fitted relations
given in \citet{hunt2016} at $z= 3,2,1$ and 0 (from bottom to top, dashed lines).
Small right panel: Comparison between simulation results (curves) and 
observed median relations at different $z$ reported by \citet{hunt2016} (symbols).
In all cases, observed data
were renormalized so that the observed local MZR considered by each author agrees with
simulation results at 
$M_* \approx 10^{10.5} \ {\rm M}_{\sun}$ at $z=0$ (see the text for details).  
As a reference, the area enclosed by the 25th and 75th percentiles corresponding to the
simulation at $z=0$ is reproduced
in all panels as a grey shadow.
}
\label{fig:gas_mzr_z}
\end{figure*}

\subsection{Evolution of the  $M_{*} - Z_{\rm SF,gas}$ relation}
\label{sec:gas_mzr_z}

In this section, we analyse the $M_* - {\rm O/H}|_{\rm SF,gas}$ relation as a function of $z$ and 
compare it to observations.  For this study, the high-resolution 
simulation Recal-L025N0752 is used, which is the 
simulation run that best reproduces the slope and normalization of certain observed
stellar mass-gas-phase metallicity relations at $z\approx 0$ \citep{schaye2015}.

In Fig. \ref{fig:gas_mzr_z}, we show the evolution of the simulated $M_* - {\rm O/H}|_{\rm SF,gas}$ 
relation.  In each of the four large panels, the simulated relation is shown
at two redshifts close to the observed data, with lines denoting the median relations and error bars 
the 25th and 75th percentiles.
As a reference, the area enclosed by the 25th and 75th percentiles corresponding to $z=0$ is reproduced
in all panels as a grey shadow.
We constructed the median relations considering
only mass bins containing more than 10 galaxies ($N_{\rm bin} \ga 10$).
We also show observational findings reported at similar redshifts to those associated with the simulations.
In the following, we summarize briefly 
the different observational data used here (for more details,
the reader is referred to the corresponding papers):
\citet{delosreyes2015} (medians and scatter at $z\approx0.8$), 
\citet{hunt2016} (medians with 75\% and 25\% quantile levels at $0.4 <z \le 0.7$, $0.9 < z \le 1.8$,
$1.8 <z \le 2.8$ and $2.8 < z \le 3.8$),
\citet{ly2016} (medians with the 16th and 84th percentiles at $z=0.5-1.0$), 
\citet{stott2013} (median values and standard errors at $z\approx 0.84-1.47$),
\citet{yabe2013} (medians and bootstrap errors at $z\approx 1.4$),
\citet{zahid2014} (fitted metallicities and observational uncertainties at $z \approx 1.6$),
\citet{wuyts2016} (mean values and standard errors at $z \approx 0.9$ and $z\approx 2.3$),
\citet{sanders2015} (mean values and uncertainties at $z \approx 2.3$),
\citet{cullen2014} (fitted metallicities with uncertainties at $z \ga 2$),
\citet{troncoso2014} (individual measurements and their uncertainties at $z\approx 3.4$) and,
\citet{onodera2016} (results from stacking analysis at $z\approx 3.3$).
Dotted lines in the large four panels indicate the fits to observations at different $z$ given by \citet{maiolino2008}.
In the case of \citet{maiolino2008} and \citet{troncoso2014}, we show the relations
reported for masses estimated from \citet{bruzual2003} templates.

The use of different samples, selection criteria and, mainly,
the implementation of different metallicity calibrators yield discrepancies between 
different observations at similar $z$.
In particular, as already noted for the local relation, 
the normalization of the MZR is still a matter of extensive
debate in the literature. This problem is even worse at higher $z$ due to different selection
biases and the difficulties in measuring metallicities of very distant galaxies. 
In particular, an unbiased comparison of the MZR at
various redshifts is not guaranteed by simply using one diagnostic/calibrator.
This would be only true if the implemented metallicity calibrator/diagnostic is equally valid for
the HII regions of galaxies at
different redshifts, and does not have any mass/SFR/redshift dependent biases.
Locally calibrated metallicity diagnostics may not be appropriate for measuring
metallicities at higher redshift \citep{steidel2014, strom2017}.
Also, the N2 and $R_{23}$ diagnostics are known to become insensitive to metallicity
at 12+log(O/H) $\sim$ 9.0 \citep[e.g.][]{kewley2002, liang2006}.
Nevertheless, in order to mitigate differences in normalizations generated by the use of different
metallicity calibrators (see Section \ref{sec:gas_mzr_z0}) and ease the comparison of the MZR
evolution, we renormalized observed relations taking into account
the local MZR used by each author to compare their observed relations at $z>0$.
Those local MZRs were obtained by some authors from previous works while
other authors re-estimated the local MZRs using self-consistent methods to those applied
at higher $z$
(for further details, see the corresponding papers). 
As each author compared the corresponding high-$z$ MZR with a local MZR
derived from the same metallicity calibrator, 
the effects of the choice of the metallicity indicator
on the level of evolution of the MZR should diminish.
Thus, in Fig. \ref{fig:gas_mzr_z}, an offset
is applied to observed relations at $z>0$ so that their associated
local observed MZRs yields $12 + {\log}_{10}({\rm O/H}) = 9.0$ at 
$M_* = 10^{10.5} \ {\rm M_{\sun}}$ (consistent with $z=0$ results from
{\sc EAGLE} Recal-L025N0752).\footnote{
Offsets applied to the data: 
-0.02 dex \citep{maiolino2008, troncoso2014, onodera2016},
+0.10 dex \citep{delosreyes2015}, 
+0.20 dex \citep{ly2016}, 
+0.25 dex \citep{sanders2015},
+0.27 dex \citep{hunt2016},
+0.30 dex \citep{cullen2014},
+0.32 dex \citep{zahid2014, wuyts2016} and 
+0.35 dex \citep{stott2013, yabe2013}.}
In this way, the comparison between findings from the simulation and the level of evolution reported
in observational works 
is more straightforward: if the shape of the MZR does not change significantly with $z$,
different normalizations in Fig. \ref{fig:gas_mzr_z} can be related to different levels 
of evolution in the simulation with respect to observations and the association is exact at  
$M_* \approx 10^{10.5} \ {\rm M}_{\sun}$.

\begin{figure}
\begin{center}
\resizebox{8.0cm}{!}{\includegraphics{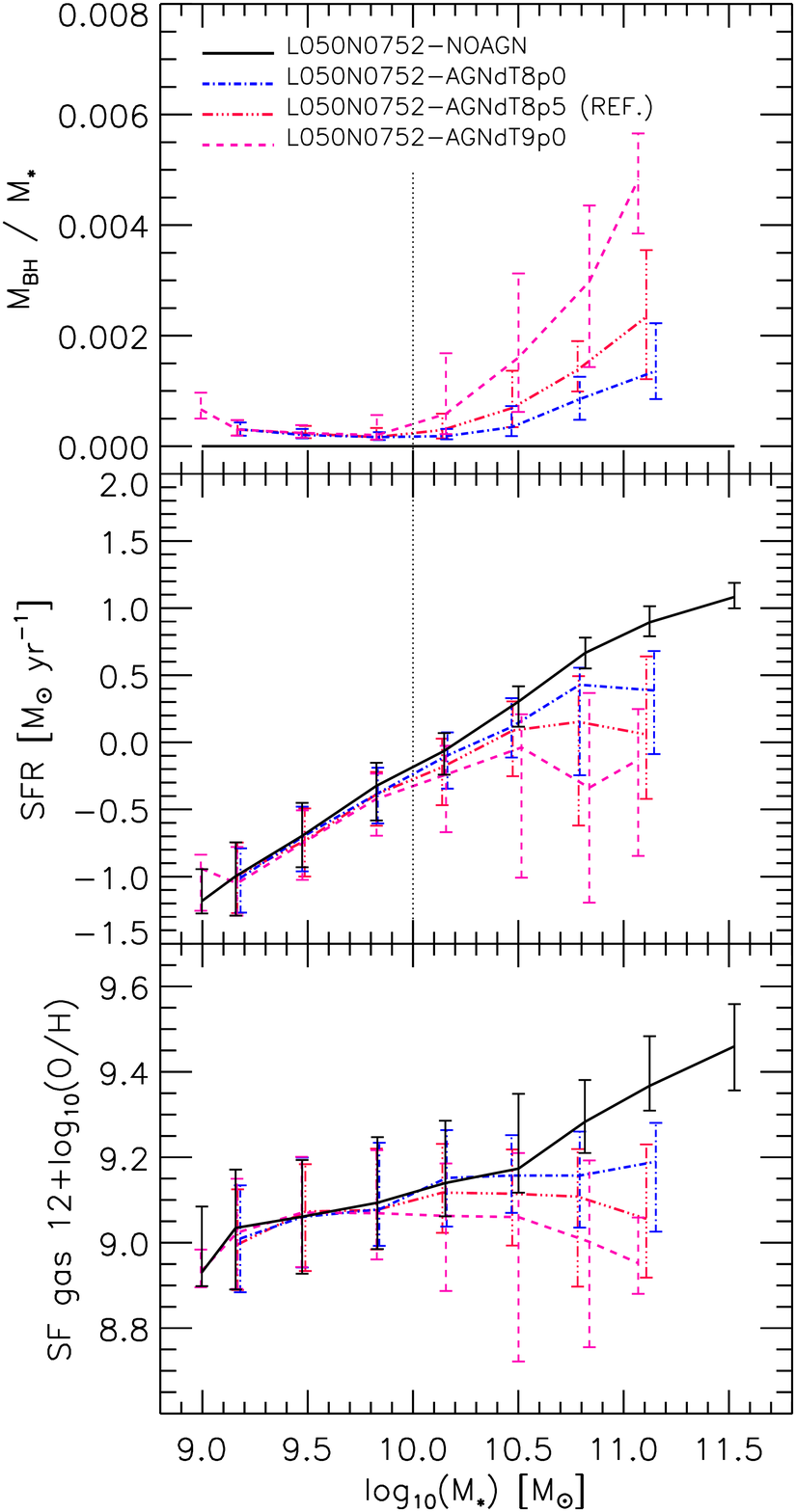}}
\end{center}
\caption[]
{
Simulated 
$M_* - M_{\rm BH}/M_*$ (upper panel),
$M_* - {\rm SFR}$ (middle panel) and
$M_* - {\rm O/H}|_{\rm SF,gas}$ (lower panel)
relations at $z=0$ for different models. 
Results obtained from simulations 
``L050N0752'' with different AGN feedback parameters are presented:
NOAGN (AGN feedback suppressed entirely),
AGNdT8 ($\Delta T_{\rm AGN} = 10^{8}$K),
reference model ($\Delta T_{\rm AGN} = 10^{8.5}$K) and
AGNdT9 ($\Delta T_{\rm AGN} = 10^{9}$K).
Note that AGN effects set in above $M_* \sim 10^{10}\ {\rm M}_{\sun}$ (dashed vertical line).
}
\label{fig:mzr_agns1}
\end{figure}

\begin{figure}
\begin{center}
\resizebox{8.0cm}{!}{\includegraphics{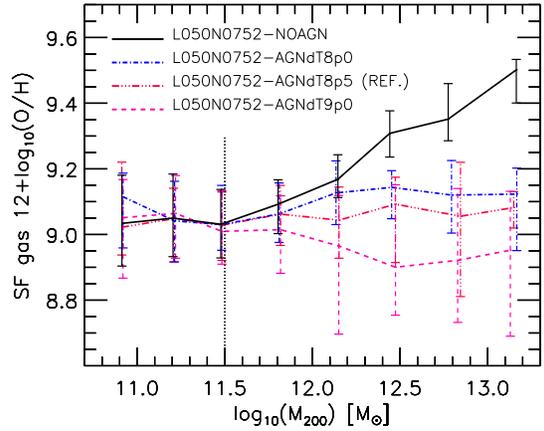}}
\end{center}
\caption[]
{
Relation between ${\rm O/H}|_{\rm SF,gas}$ and halo mass for {\em central}
simulated galaxies at $z=0$ for the same models shown in Fig. \ref{fig:mzr_agns1}.
Note that AGN effects set in above $M_{200} \sim 10^{11.5}\ {\rm M}_{\sun}$, corresponding
to $M_* \sim 10^{10}\ {\rm M}_{\sun}$. 
}
\label{fig:mzr_agns2}
\end{figure}

We see that, at all $z$, 
the {\sc EAGLE} simulation Recal-L025N0752 reproduces the observed trend of increasing 
${\rm O/H}|_{\rm SF,gas}$ with increasing $M_*$ well, 
exhibiting a slope very closed to the observed one. 
Also, at a given $M_*$, metallicity tends to decrease towards higher redshifts,
in agreement with the observed trend.  
The predicted level of evolution 
at $z \la 1.5$ agrees remarkably well with the observed one.
At $z \approx 2$, the simulation predicts a MZR in good agreement with
\citet{zahid2014} but other observations suggest a higher level of
evolution, with the larger discrepancies towards higher $M_*$
\citep[e.g.][]{hunt2016}.
At $z \approx 3$, simulations tend to produce more metal-rich galaxies than inferred by
\citet{maiolino2008} and \cite{onodera2016}. On the other hand, 
observational findings by \citet{hunt2016} at $z\approx 3$ seem to yield values closer to
those from the simulations, in particular, towards lower $M_*$.

In the small bottom panels of Fig. \ref{fig:gas_mzr_z}, we compare the evolution of the simulated 
$M_* - {\rm O/H}|_{\rm SF,gas}$ relation to
that derived by \citet{hunt2016}.
Their ``MEGA'' sample comprises 1000 galaxies with a common O/H calibration (PP04 N2) and
spans almost two orders of magnitude in metallicity, a factor of $10^6$ in SFR, and a
factor of $10^5$ in stellar mass. As explained above, 
we renormalized MEGA's relations to $12+ {\log}_{\rm 10} ({\rm O/H}) = 9.0$
at $M_* \approx 10^{10.5} \ {\rm M}_{\sun}$ at $z=0$.
In the small left panel, we compare simulation findings with  
multi-variable linear regressions on the MEGA dataset for 12+log(O/H) as a function of $M_*$ and 
$z$ \citep[see][for details]{hunt2016}.
We see that the simulation predicts an evolution of the MZR
in good agreement with the fitted relations obtained for the MEGA sample at $z=0-3$.
On the other hand, the small right panel shows that the MEGA median relations at different $z$ bins
exhibit some discrepancies with simulations. At the low-mass end 
($M_* \approx 10^{9.5} \ {\rm M}_{\sun}$), simulations predict the
observed level of evolution at $z \approx 1-3$ but overpredict, by $\la 0.2$ dex, the evolution
at $z \approx 0-1$ (observations predict a negligible evolution in this case).
At the high-mass end ($M_* \approx 10^{10.5} \ {\rm M}_{\sun}$), simulations and observations
predict a negligible evolution at $z=0-1$.  Although simulations lack massive systems at
high $z$, the extrapolation of simulation relations towards higher masses suggests a lower
level of evolution compared to that obtained for the MEGA sample.

\citet{ly2016} carried out the first systematic study of the evolution 
of the MZR to $z \approx 1$ using only the electron temperature ($T_{e}$) method.
These authors reported good agreement between their findings and
the evolution of the MZR obtained from the {\sc EAGLE} high-resolution simulation Recal-L025N0752.
\citet{guo2016} also found a good agreement between the $M_* - {\rm O/H}|_{\rm SF,gas}$ relation
derived from this simulation
and observations by \citet{zahid2013b} at $z\la 1$. At $1 \la z \la 2$, the latter authors
obtained a slightly lower level of
evolution (by $\approx 0.2$ dex) in simulations with respect to observations\footnote{ 
Note that \citet{guo2016} computed the MZR considering all particles inside 
{\sc EAGLE} galaxies while, in the present work, we only include particles within an aperture of 30 pkpc.}
(the reader is referred to \citealt{guo2016} for a comparison between {\sc EAGLE} results and predictions
from some semi-analytical models).
On the other hand, as mentioned, the predicted metallicity evolution below $z\approx3$ seems to be 
lower than what the data by \citet{maiolino2008} suggest
(e.g. $\approx 1$ dex at $M_* \approx 10^9 \ {\rm M}_{\sun}$ and $\approx 0.8$ dex at $M_* \approx 10^{10} \ {\rm M}_{\sun}$).
More recently, \citet{onodera2016} have also reported a stronger evolution ($\approx 0.7$ dex) for the observed MZR below $z \approx 3.0-3.7$. 
As discussed in Section \ref{sec:introduction}, because of the observed FMR, the lower abundances of
observed galaxies at $z \ga 1$ might  be partly explained by selection effects present 
in observational surveys that tend to be biased towards systems with higher SFRs at a given $M_*$ and, 
hence, lower metallicities. In Section \ref{sec:mzr_scatter}, we show that {\sc EAGLE}
galaxies are also consistent with the existence of a FMR since $z \approx 5$; 
thus, at a given $z$ and $M_*$, there is an anti-correlation between SFR and metallicity 
in the simulations.

Our results regarding the increasing metallicity of the SF gas with time 
are consistent with \citet{segers2016b}, who show that recycling of stellar mass loss 
in {\sc EAGLE} becomes increasingly important
for fuelling star formation towards lower redshift.
Also, \citet{segers2016b} determined a characteristic mass, $M_* \approx 10^{10.5} {\rm M_{\sun}}$, 
below which,
the contribution of recycled mass increases with mass and above which,
it decreases with mass.
The negligible evolution that we obtained for the $M_* = 10^{10.5} - 10^{11.0} {\rm M}_{\sun}$ bin in 
Fig. \ref{fig:gas_mzr_z} and the flattening of the slope of the $z=0$ MZR at high masses (Fig. \ref{fig:gas_mzr_z0}) 
might reflect the transition determined by \citet{segers2016b}.
These authors claimed that 
this transition reflects the transition from stellar to AGN feedback.

We recall that otherwise we use 30 kpc spherical apertures.
We have verified that the main trends of the scaling relations presented so far 
are preserved if the
global properties are measured in different apertures (see Appendix A).  

Finally, it is worth noting that \citet{derossi2015b} found a negligible evolution of the
MZR in the case of the {\sc GIMIC} simulations.  
The improved subgrid physics implemented in {\sc EAGLE} leads to a better agreement with the data. 
A comparison of {\sc GIMIC} and {\sc EAGLE} MZRs is presented in Appendix B.  
There, we show that the implementation of energy feedback from star formation
that depends on the local density and metallicity leads to an evolution of the MZR consistent
with the observed trend (see, e.g. \citealt{fu2012} and \citealt{dave2017}, for similar findings
derived from other models and simulations).  On the other hand, a constant energy feedback, as that implemented in {\sc GIMIC},
leads to a negligible MZR evolution.

\subsection{Impact of AGN feedback}
\label{sec:mzr_agns}

In the previous section, we analysed the evolution of the 
$M_* - {\rm O/H}|_{\rm SF,gas}$ relation using the high-resolution simulation
Recal-L025N0752.  However, to investigate the role of AGN feedback on the MZR,
we need to focus on the trends at high masses and, thus, we will need
to use intermediate resolution simulations.  
In the overlap region at high masses, the trends followed by the simulated $M_* - {\rm O/H}|_{\rm SF,gas}$ relation seem to be more 
robust against resolution (Fig. \ref{fig:gas_mzr_z0}) and different resolution runs agree.  
At $z = 0$, the intermediate- and high-resolution simulations
exhibit similarly flat slopes at $M_* \ga 10^{10} \ {\rm M}_{\sun}$, though with a decrease by $\approx 0.1$ dex in the normalization when using the recalibrated model compared to the reference one.  
The analysis of different simulation outputs 
($z\approx$ 0.0, 0.1, 0.18, 0.27, 0.37, 0.50, 0.62 and 0.74)
shows that the flattening of the
predicted MZR at the high-mass end extends to $z \approx 0.7$ (see Fig. \ref{fig:gas_mzr_z}),
in agreement with some observations \citep[e.g.][]{delosreyes2015}. In this section, we will
show that the flattening of the simulated MZR is mainly regulated 
by AGN feedback.\footnote{We note that, for MZR studies, galaxies with evidence of current AGN activity 
are removed from observed samples.  However, BH feedback may have occurred in cyclic episodes 
that affected also the selected observed sub-samples.
In all {\sc EAGLE} massive galaxies, the activity of SMBHs also varies with time.
}

Within the {\sc EAGLE} suite of cosmological simulations, the impact of the AGN parameters was explored
using the intermediate-resolution ``L050N0752'' version of the simulations. In addition to the Ref-L050N0752 run, three other
simulations are analysed here: NOAGN-L050N0752, AGNdT8-L050N0752 and AGNdT9-L050N0752 (see Table \ref{tab:simus}).
The simulation NOAGN-L050N0752 does not include AGN feedback. In the case of simulations AGNdT8-L050N0752 and AGNdT9-L050N0752, 
the temperature increase of the
gas caused by AGN feedback has been set to $\Delta T_{\rm AGN} = 10^8$K and $\Delta T_{\rm AGN} = 10^9$K, 
respectively.\footnote{Note that the simulation AGNdT9-L050N0752 analysed by \citet[][]{schaye2015} assumes $C_{\rm visc}/2 \pi = 10^2$, while 
here  $C_{\rm visc}/2 \pi = 10^0$ (Table \ref{tab:simus}, \citealt{crain2015}), as we are interested in studying single-parameter variations.}
In Fig. \ref{fig:mzr_agns1}, we compare the $z=0$  $M_* - M_{\rm BH}/M_*$ (where $M_{\rm BH}$ depicts the black hole mass), 
$M_* - {\rm SFR}$ and $M_* - {\rm O/H}|_{\rm SF,gas}$ relations 
predicted by the four aforementioned simulations.
All considered mass bins contain more than 10 galaxies inside them.
We can see that BH growth and the corresponding AGN feedback sets in at $M_* \sim 10^{10} \ {\rm M}_{\sun}$, with $M_{\rm BH}/M_*$ increasing with $M_*$, as expected.
As discussed in \citet{crain2015}, since $M_{\rm BH}$ is determined to first order by halo mass 
(\citealt{booth2010}, see also \citealt{bower2017}), 
the offset of the $M_* - M_{\rm BH}/M_*$ relation at $M_* \ga 10^{10} \ {\rm M}_{\sun}$ is related to the different halo mass 
associated with galaxies in this mass range.
At a given $M_*$, SFR tends to decrease as $\Delta T_{\rm AGN}$ increases (middle panel of Fig. \ref{fig:mzr_agns1}) because the less frequent
but more energetic feedback episodes associated with a higher $\Delta T_{\rm AGN}$ are more
efficient at regulating SFR in massive galaxies.
In particular, if AGN feedback is suppressed, higher values of $M_*$ are obtained.
Differences between the MZRs obtained from the NOAGN, AGNdT8, AGNdT9 and reference simulations (bottom panel of Fig. \ref{fig:mzr_agns1})
are also appreciable 
at $M_* \ga 10^{10} {\rm M}_{\sun}$, reaching a metallicity offset of $\approx 0.3$ dex at $M_* \sim 10^{11} {\rm M}_{\sun}$.
It is clear that, when AGN feedback is turned off, the MZR does not exhibit a flattening at high masses.
On the contrary, the MZR slope increases towards higher masses in the case of the NOAGN model.
As $\Delta T_{\rm AGN}$ increases, the MZR slope decreases and, in particular, 
above $\Delta T_{\rm AGN} \approx 5 \times 10^8$K,
AGN feedback can generate an inversion of the relation between stellar mass and metallicity 
turning it from a correlation into an anti-correlation. These trends reflect the behaviour obtained for the
$M_* - {\rm SFR}$ relation: higher $\Delta T_{\rm AGN}$ leads to less on-going star formation.
Additionally, we see that a 
higher $\Delta T_{\rm AGN}$ leads to larger scatter in $M_{\rm BH}/M_*$, SFR and ${\rm O/H}|_{\rm SF,gas}$ at a given $M_*$ 
at $10^{10} \la M_* / {\rm M}_{\sun} \la 10^{11}$.
Taking into account the different AGN feedback histories that can take place in real galaxies, we can conclude
that part of the dispersion of the observed MZR at the high-mass end 
may be associated with AGN feedback,
in which case its study could provide important constraints on galaxy formation models (see also Section \ref{sec:models}).

In Fig. \ref{fig:mzr_agns2}, we plot ${\rm OH}|_{\rm SF,gas}$ as a function of the
halo mass ($M_{200}$) for {\em central} simulated galaxies.\footnote{The halo mass, $M_{200}$ is defined as
the total mass contained within the virial radius $R_{200}$, which is
the radius within which the mean internal density is 200 times
the critical density, $3H^{2}/8 \pi G$, centred on the dark matter particle
of the corresponding FoF halo with the minimum gravitational
potential.}
More than 15 galaxies are included in each mass bin.
We obtained a similar trend to that derived previously for $M_*$.
At a given halo mass, higher $\Delta T_{\rm AGN}$ yields lower metallicities towards higher masses.
When increasing $\Delta T_{\rm AGN}$ from $10^8$K to $10^9$K, the metallicity associated with central
galaxies in massive halos decreases by around $0.2$ dex.
If AGN feedback is suppressed, the median metallicity of massive galaxies increases significantly,
reaching an offset $\ga 0.3$ dex at $M_{200} \approx 10^{13} \ {\rm M}_{\sun}$
relative to the case of $\Delta T_{\rm AGN}=10^8 {\rm K}$.
In section \ref{sec:agn_effects}, we try to disentangle the diverse effects caused by AGN
feedback in galaxies that lead to the decrease of the global metallicity of these systems.
A further investigation is left for a future work.

\begin{figure}
\begin{center}
\resizebox{8.5cm}{!}{\includegraphics{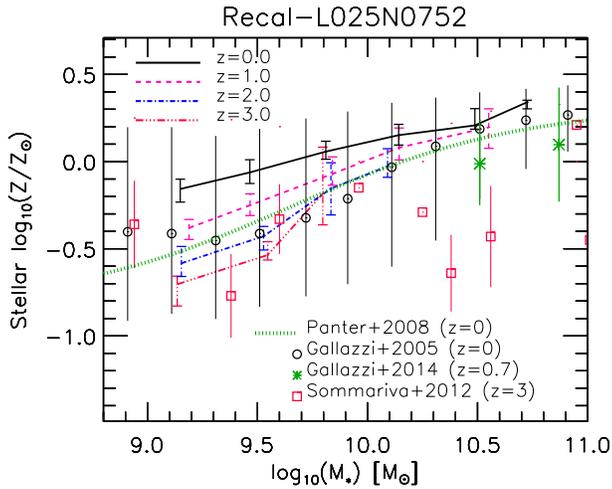}}
\end{center}
\caption[]
{
$M_* - Z_*$ relation at different $z$ for simulation Recal-L025N0752.  The simulation results are
shown as curves depicting the median relations and error bars indicating the 25th and 75th
percentiles. Observational data at $z\approx 0.0, 0.7$ and 3.0 are shown with symbols, as indicated
in the figure (see the text for details).   
The fitted relation to SDSS data given in \citet{panter2008} is shown with a dotted green line.
For simulations, $Z_{\sun} = 0.0127$ \citep{wiersma2009b}. Observed relations are re-scaled to this
value, with the exception of data from \citet{sommariva2011}, who reported $12+ \log_{10} ({\rm O/H})$ in units of
$12+ \log_{10} ({\rm O/H})\sun = 8.69$ \citep{allende2001}.
}
\label{fig:mzr_stars}
\end{figure}

\begin{figure}
\begin{center}
\resizebox{8.5cm}{!}{\includegraphics{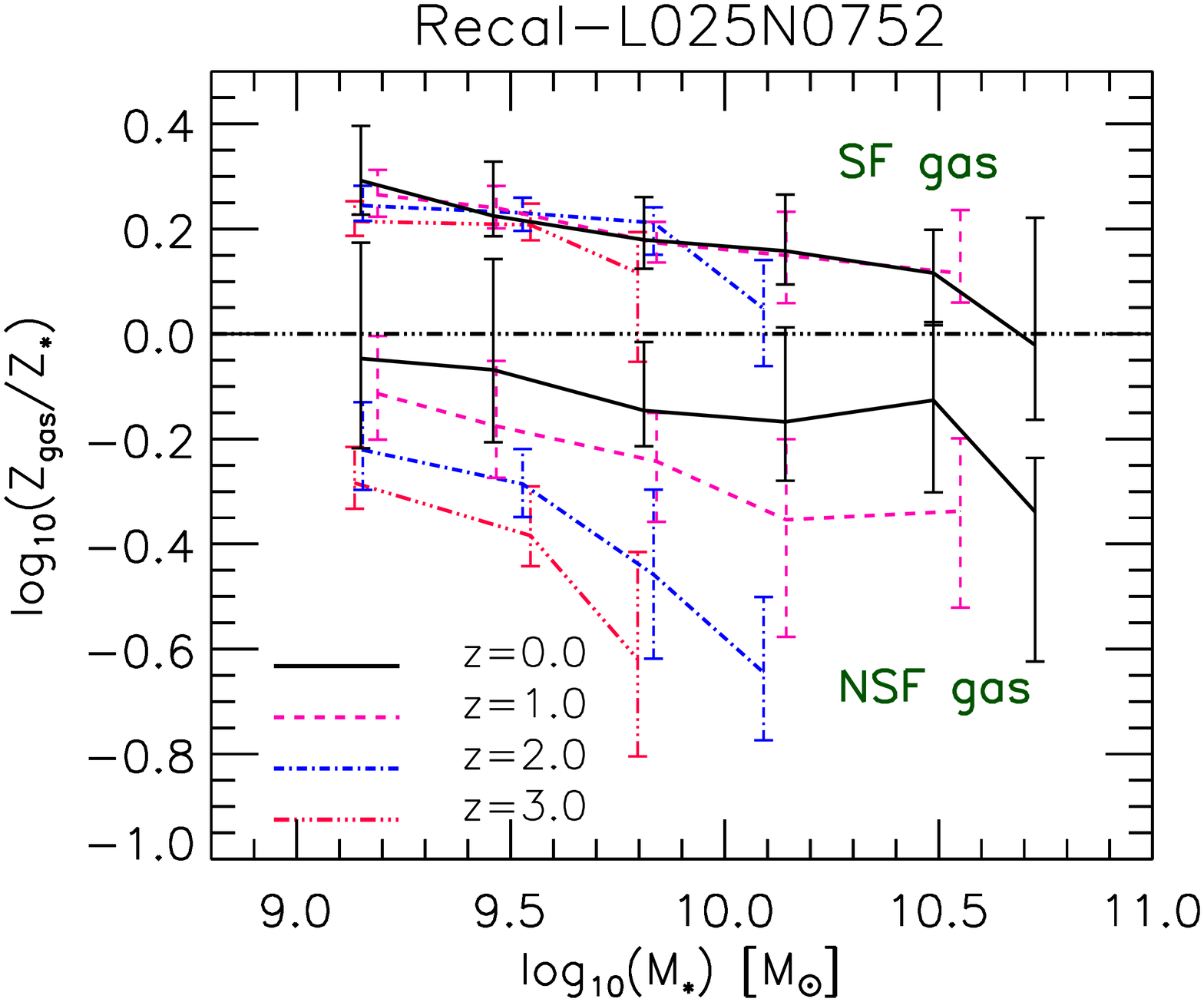}}
\end{center}
\caption[]
{
Ratio between the gas-phase and stellar metallicities of {\sc EAGLE} galaxies at
different $z$.  The results for the SF gas and NSF gas components are presented
separately as indicated in the figure. The curves with error bars indicate the median relations together
with the 25th and 75th percentiles. The horizontal line corresponds to the case in which the gas and stellar
phases have identical metallicities ($Z_{\rm gas} = Z_*$).
}
\label{fig:ZgZs_vs_M}
\end{figure}

\section{The $M_{*} - Z_{\rm *}$ relation}
\label{sec:mzr_stars}

The predicted relation between $M_*$ and stellar metallicity ($Z_*$) is shown 
in Fig. \ref{fig:mzr_stars} (curve with error bars) for Recal-L025N0752. 
The number of galaxies per mass bin is $N_{\rm bin} \ga 10$ at all considered $z$.
Observational data at $z\sim 0$ from \citet{gallazzi2005} are represented as
black circles with error bars, which indicate the median relation together with the 16th and 84th percentiles.
The fitted relation to SDSS data given in \citet{panter2008} is shown with a dotted green line.
We also show findings by \citet{gallazzi2014} at $z=0.7$ as green asterisks with error bars,
depicting the median relation together with the 16th and 84th percentiles.
Results at $z \approx 3$ reported by \citet{sommariva2011} are represented with 
red diamonds and error bars, which
denote the measured metallicities and corresponding uncertainties.
For some galaxies, \citet{sommariva2011} were not able to determine metallicity uncertainties so that
no error bars are shown in those cases (the reader is referred to that paper for further details).

We see that, at a given $M_*$, $Z_*$ tends to decrease with $z$, as previously reported
by \citet{guo2016} for this simulation\footnote{For this analysis, 
\citet{guo2016} considered all particles inside {\sc EAGLE} 
galaxies while, in this work, we take into account particles within an aperture of 30 pkpc.}
(we referred the reader to that work for a comparison
between {\sc EAGLE} results and semi-analytical models).
We can see that, at $z=0$, the median simulated and observed relations tend to agree at high masses.  
At the low-mass end, simulation predicts higher metallicities, on average, than observations.
It is clear that the scatter in the data at $z\approx0$
is larger than predicted by simulations (see also \citealt{schaye2015}).
This was also noted by \citet{derossi2015} when using the {\sc GIMIC} simulations 
and might be related to uncertainties in observed stellar metallicity determinations.
\citet{gallazzi2005} adopted a Bayesian statistical approach to obtain $Z_*$ for individual galaxies, estimating 
the 68 per cent confidence interval within which this parameter
is constrained.   
Observational errors on $Z_*$ have an average value of 0.12 dex but can extend to $\approx 0.30$ dex
\citep[see][for further details]{gallazzi2005}.

At $z \approx 0.7$, observed $Z_*$ is only reported at $M_* \ga 10^{10.5} \ {\rm M}_{\sun}$ 
and exhibits lower median values than simulated ones by $\approx 0.2$ dex.
\citet{gallazzi2014} estimated $Z_*$ for individual galaxies similarly to \citet{gallazzi2005}, obtaining
an average uncertainty on stellar metallicity of 0.3 dex, slightly worse than for the local observed sample.
Thus, simulations and observations results agree within the error. 
In the mass range where simulations and observations overlap at $z \approx 3$, the simulated
$M_* - Z_*$ relation is consistent with the stellar metallicities reported by \citet{sommariva2011}.

It is worth noting that the simulated $Z_*$ is calculated weighting by mass while, in 
the case of \citet{gallazzi2005, gallazzi2014} and \citet{sommariva2011},
the observed $Z_*$ is a luminosity-weighted quantity. 
On the other hand, the approach used by \citet{panter2008} yields a mass fraction 
weighted metallicity. From Fig. \ref{fig:mzr_stars}, we see that the slope of the simulation
$M_* - Z_*$ relation shows a better agreement with the latter authors.
To analyse this issue in more detail, we re-calculated 
the simulated $Z_*$ 
weighting by light in different photometric bands (u, g, r, i, z, Y, J, H, K).
\footnote{For these calculations, we used photometric data of stellar particles
available in the public EAGLE database \citep{mcalpine2015}.}
We obtained that light-weighted metallicities tend to be higher than mass-weighted metallicities 
by $\la 0.2$ dex, with the larger differences towards lower masses. 
In addition, the scatter of the 
simulated $M_* - Z_*$ relation slightly increases, by $\la 0.1$ dex, when using light-weighted metallicities
instead of mass-weighted metallicities.  
Thus, the use of light- instead of mass-weighted metallicities 
does not generate significant changes in the trends predicted for the simulated
$M_* - Z_*$ relation.

The $M_* - {\rm [Fe/H]}_*$ relation 
has been widely studied in the literature for early-type galaxies \citep[e.g.][]{conroy2014}.
Observed ${\rm [Fe/H]}_*$ increases with $M_*$ for more massive galaxies, as expected.  
We have verified that the simulation reproduces this trend, 
however, a more detailed analysis is beyond the scope of this paper.
For a study of the $M_* - {\rm [ \alpha / Fe]}_*$ and ${\rm age} - {\rm [ \alpha / Fe]}_*$
relations in {\sc EAGLE} simulations, the reader is referred to \citet{segers2016a}.

\begin{figure}
\begin{center}
\resizebox{9.cm}{!}{\includegraphics{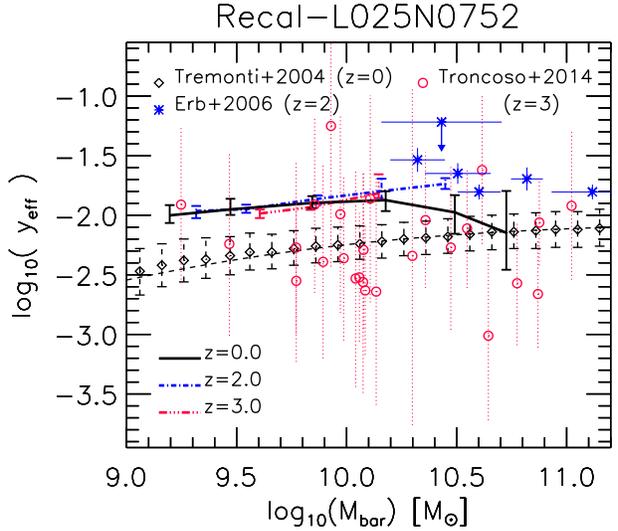}}
\end{center}
\caption[]
{
Effective yields as a function of baryonic mass.
Curves with error bars depict the simulated median
relations with the associated 25th and 75th percentiles.
The diamonds with error bars indicate the median relation together with the 16th and 84th percentiles
corresponding to observational results by \citet[][]{tremonti2004} at $z\approx0$.  
Observational results at $z\approx2$ \citep{erb2006} and $z\approx3$ \citep{troncoso2014} 
are shown with asterisks and circles, respectively. 
The dashed black curve without error bars depict a fit to the observed relation 
at $z\approx0$.  
Observed yields were renormalized so that the observed $12 + {\log}_{10}({\rm O/H}) = 9.0$ at $M_* = 10^{10.5} \ {\rm M}_{\sun}$ 
at $z\approx0$ (see the text for details).
}
\label{fig:yeff_vs_M}
\end{figure}

A comparison between the metallicity of the different baryonic components of {\sc EAGLE}
galaxies is presented in Fig. \ref{fig:ZgZs_vs_M}.  We show how the ratio between 
gas ($Z_{\rm gas}$) and stellar ($Z_*$) metallicity varies as a function of $M_*$ considering both gas components:
the star forming gas and the non star forming gas (NSF gas) phases.  
The SF gas component of a given galaxy was calculated by summing the mass of all gas particles
with ${\dot {\rho}}_* > 0$ (see Section \ref{sec:simulation_details}), while the NSF gas component
has ${\dot {\rho}}_* = 0$.
Each mass bin associated to the median relations shown in Fig. \ref{fig:ZgZs_vs_M} 
contains more than 10 galaxies at all considered $z$.
The quantity $Z_{\rm gas}/Z_*$ can be considered to be a measure of the current compared
to past average chemical enrichment of the ISM of galaxies.
The SF and NSF gas-components
present clearly distinct levels of enrichment with respect to stars at all $z$ considered.

The SF gas component is more metal-enriched than the stellar component at all $z$ considered and the opposite
is true for the NSF gas.
For the SF gas, the ratio $Z_{\rm SF,gas} / Z_*$ tends to decrease from $\approx 1.5-2.0$ at $M_* \sim 10^9 {\rm M}_{\sun}$
to $\approx 1.0$ at $M_* \ga 10^{10.5} {\rm M}_{\sun}$, with no evidence for a dependence on $z$.
Fig. \ref{fig:mzr_stars} also shows that, at a given $M_*$, $Z_*$ tends to be lower than the metallicity of the SF gas, 
especially in low-mass systems (compare with Figs. \ref{fig:gas_mzr_z0} and  \ref{fig:gas_mzr_z}).
In addition, the $M_* - Z_*$ relation is steeper than the $M_* - Z|_{\rm SF,gas}$ relation, 
with a slope increasing towards high $z$. 
$Z|_{\rm SF,gas}$ can be directly associated with the
recent metal enrichment of the galaxy while $Z_*$ reflects the average chemical evolution
of the system during its periods of star formation.\footnote{Note, however, that, in the case of observations, 
$Z_*$ uses to be luminosity-weighted and tends to be biased to the most luminous stars.}  
Hence, star formation in smaller galaxies tends to be driven by more metal-enriched gas than
that associated with their past star formation, represented by $Z_*$.  
As smaller galaxies have larger gas fractions (see Section \ref{sec:mzr_scatter}), the chemical evolution at the low-mass end develops faster as can be 
seen from Figs. \ref{fig:gas_mzr_z} and \ref{fig:mzr_stars}.
On the other hand, more massive galaxies form stars from gas with a similar level
of enrichment as their average stellar component.  This is 
consistent with the flattening of the $M_* - Z_{\rm SF,gas}$ and $M_* - Z_{\rm *}$ relations at high stellar masses
(see \citealt{zahid2014a} for similar findings derived from analytical methods).
The metallicity of high-mass
galaxies saturates at early times at $12+ \log_{10} ({\rm O/H}) \approx 8.9-9.0$, exhibiting less significant differences between SF gas and stars.

In the case of the NSF gas component, $Z_{\rm NSF,gas} / Z_*$ decreases towards higher stellar masses and increases with
decreasing $z$.  Thus, at lower $z$,  the metallicities of the different
baryonic phases of more massive galaxies are more homogeneous, as a result of accumulated mixing of metals over time. 
These results also suggest that star formation in massive galaxies might have been regulated by 
supply of metal-poor gas in the NSF phase that was transferred to the more enriched SF one.  
As most massive galaxies exhibit low-gas fractions and nearly homogeneous metallicity distributions between
their different baryonic components, their subsequent chemical evolution may be driven by mergers
(see \citealt{yates2014} for a similar analysis in the context of semi-analytical models).

\section{Effective yields}
\label{sec:yields}
In a closed-box model \citep[e.g.][]{schmidt1963,tinsley1980}, where there are neither
inflows nor outflows, the metallicity is a simple function of the stellar yield ($y$)
and gas fraction ($f_{\rm gas} = M_{\rm gas} / (M_{\rm gas} + M_{*})$):

\begin{equation}
\label{eq:yields}
Z = y \ln ( f_{\rm gas}^{-1} ),
\end{equation}

\noindent
where $y$ represents the ratio of the mass in metals synthesised and released to the net
mass that remains locked-up in stars.
If we were to unrealistically assume a constant stellar yield \citep[e.g.][]{garnett2002, tremonti2004}, 
then the level of metal enrichment only depends on the gas fraction.
By inverting Eq. (\ref{eq:yields}), we can obtain the effective yields, 
$y_{\rm eff} = Z_{\rm SF,gas} / \ln ( f_{\rm SF,gas}^{-1}$). Note that we 
calculate $y_{\rm eff}$ from the SF gas phase as this is the component
usually observed.
Clearly, if a galaxy evolves as a closed box, then $y_{\rm eff} = y$ 
while, in both cases of metal-enriched outflows
and infall of primordial gas, $y_{\rm eff} \le y$ \citep{edmunds1990}. Thus, $y_{\rm eff}$ can be used to
diagnose the role of inflows and/or winds on the evolution of galaxies.

In Fig. \ref{fig:yeff_vs_M}, we plot $y_{\rm eff}$ as a function of the mass of baryons
($M_{\rm bar} = M_* + M_{\rm SF, gas}$) for
simulations and observations at $z\approx0$, 2 and 3.
Black, blue and red lines with error bars denote simulated median relations
at $z=0$, 2 and 3, respectively.
Error bars depict the 25th and 75th percentiles.
Observational results by \citet{tremonti2004} at $z\approx0$ (black diamonds),
\citet{erb2006} at $z\approx2$ (blue asterisks)
and \citet{troncoso2014} at $z\approx3$ (red circles) are shown for comparison.
In order to estimate $y_{\rm eff}$, \citet{tremonti2004}, \citet{erb2006} and \citet{troncoso2014} 
derived gas masses from the observed star formation rates assuming a 
Schmidt law \citep{kennicutt1998}.  The reader is referred to those papers
for more details.
Note that the determination of observed yields is also affected by the metallicity calibration 
(Eq. \ref{eq:yields}) and, as we have seen, there are significant discrepancies
between observed metallicities obtained from different calibrators (Fig. \ref{fig:gas_mzr_z0}). 
\citet{erb2006} used the PP04 N2 method to determine metallicities, which yields lower metallicities
that those derived by the T04 method (see Fig. \ref{fig:gas_mzr_z0}).  
\citet{troncoso2014} adopted the same method as in \citet{maiolino2008}, which gives higher
metallicities than the PP04 N2 method, with the differences ranging between $\approx 0.3$ and
0.1 dex depending on mass.
Thus, as explained in Section \ref{sec:gas_mzr_z}, an offset
was applied to observed yields so that the corresponding $12 + {\log}_{10}({\rm O/H}) = 9.0$ at
$M_* = 10^{10.5} \ {\rm M_{\sun}}$ at $z\approx0$.
To renormalize metallicities in the case of \citet{erb2006} and \citet{troncoso2014}, 
we used the local MZR used by each author to contrast their higher-$z$ results.
Each author used a local MZR derived from the same metallicity calibrator employed for
obtaining the corresponding high-$z$ MZR.\footnote{Offsets applied to 
observed effective yields: -0.07 dex \citep{tremonti2004},
+0.35 dex \citep{erb2006} and
-0.02 dex \citep{troncoso2014}.}

We see that there is negligible evolution of the simulated $y_{\rm eff} - M_{\rm bar}$ relation 
at $M_* \la 10^{10.2} \ {\rm M}_{\sun}$ below $z=3$, while more massive systems tend to present higher $y_{\rm eff}$
at higher $z$. Considering the differences between the metallicity calibrations applied by \citet{tremonti2004}
and \citet{erb2006}, observational results also suggest a decrease of $y_{\rm eff}$ for massive galaxies
towards $z\approx0$.
Regarding the scatter of the $y_{\rm eff} - M_{\rm bar}$ relation, in the case of the observations, 
the scatter is large partly because of the significant uncertainties in the determination of gas masses.
On the other hand, we have verified that in {\sc EAGLE} the larger scatter of the simulated $y_{\rm eff}$ at higher
masses is related to AGN feedback. By comparing the AGNdT8-L050N0752 and AGNdT9-L050N0752 models, 
we found that an increase of $\Delta T_{\rm AGN}$ leads to a larger scatter 
at $M_{\rm bar} \ga 10^{10} {\rm M}_{\sun}$. The simulation NOAGN-L050N0752 predicts a negligible scatter 
for the $y_{\rm eff} - M_{\rm bar}$ relation.\footnote{
We note that, in the case of Recal-L025N0752 at $z=0$, the number of galaxies in the highest mass
bin, where the scatter is larger, is $N_{\rm bin} \approx 15$.
For "L050N0752" simulations, at the same bin ($M_{\rm bar} \approx 10^{10.8} \ {\rm M_{\sun}}$) and redshift, 
$N_{\rm bin} \approx 100$ (NOAGN), $N_{\rm bin} \approx 100$ (AGNdT8) and $N_{\rm bin} \approx 50$ (AGNdT9). 
}

\citet{tremonti2004} fitted the observed data using the function:

\begin{equation}
\label{eq:yields_fit}
y_{\rm eff} = \frac{ y_0 }{1 + (M_0 / M_{\rm bar})^{0.57} },
\end{equation}

\noindent
where the asymptotic value $y_0$ represents the true yield (if no metals are lost).
These authors obtained $y_0 = 0.0104$ and $M_0 = 3.67 \times 10^9 \ {\rm M}_{\sun}$ (dashed curve in 
Fig. \ref{fig:yeff_vs_M}) for the observed sample.
The simulated $z=0$ $y_{\rm eff} - M_{\rm bar}$ relation follows roughly the observed trends
at $M_* \la 10^{10} \ {\rm M}_{\sun}$ but it is inverted at higher masses.
Thus, we fitted the simulated relation at $z=0$ using the same function as \citet{tremonti2004}
but including only systems with $M_{\rm bar} \la 10^{10} \ {\rm M}_{\sun}$.
We obtained $y_0 \approx 0.013$ and $M_0 \approx 3.4 \times 10^8 {\rm M}_{\sun}$.
As the fitted curve for simulations follows tightly the median relation, 
for the sake of clarity, we do not include it in 
Fig. \ref{fig:yeff_vs_M}.

In the simulations, metal enrichment is implemented following the assumed stellar yields tables
\citep{portinari1998, marigo2001, thielemann2003} but, subsequently, metal abundances
are modified by different processes such as inflows and outflows of gas.
The trends predicted for $y_{\rm eff}$ provide clues about 
the role of gas infall and/or winds on the evolution of galaxies \citep[e.g.][]{dalcanton2007}.
The decrease of the simulated $y_{\rm eff}$ towards lower masses suggests
that these systems have been affected by efficient outflows of metal-enriched SN ejecta.
Because small galaxies tend to have high mean gas fractions (see Sections  
\ref{sec:mzr_scatter}), the infall of metal-poor gas could
not drive significant changes in their effective yields \citep{dalcanton2007}.

At high masses, $y_{\rm eff}$ tends to decrease towards $z=0$.
In particular, some systems depart from the mean relation towards very low $y_{\rm eff}$ 
due to their very low gas fractions.
A possibility is that AGN feedback drives metal-enriched outflows in more massive galaxies. 
In fact, we have verified that an increase of $\Delta T_{\rm AGN}$ leads to 
lower average $y_{\rm eff}$ at $M_{\rm bar} \ga 10^{10} {\rm M}_{\sun}$.
However, infall of unenriched material probably generates the largest reductions in 
$y_{\rm eff}$ in gas-poor massive galaxies \citep{dalcanton2007} (see 
\ref{sec:agn_effects}, for a more detailed analysis of the effects of AGN feedback 
on massive galaxies).
Merger events can also play a significant role in determining the effective yields 
in massive galaxies.

\begin{figure*}
\begin{center}
\resizebox{15.5cm}{!}{\includegraphics{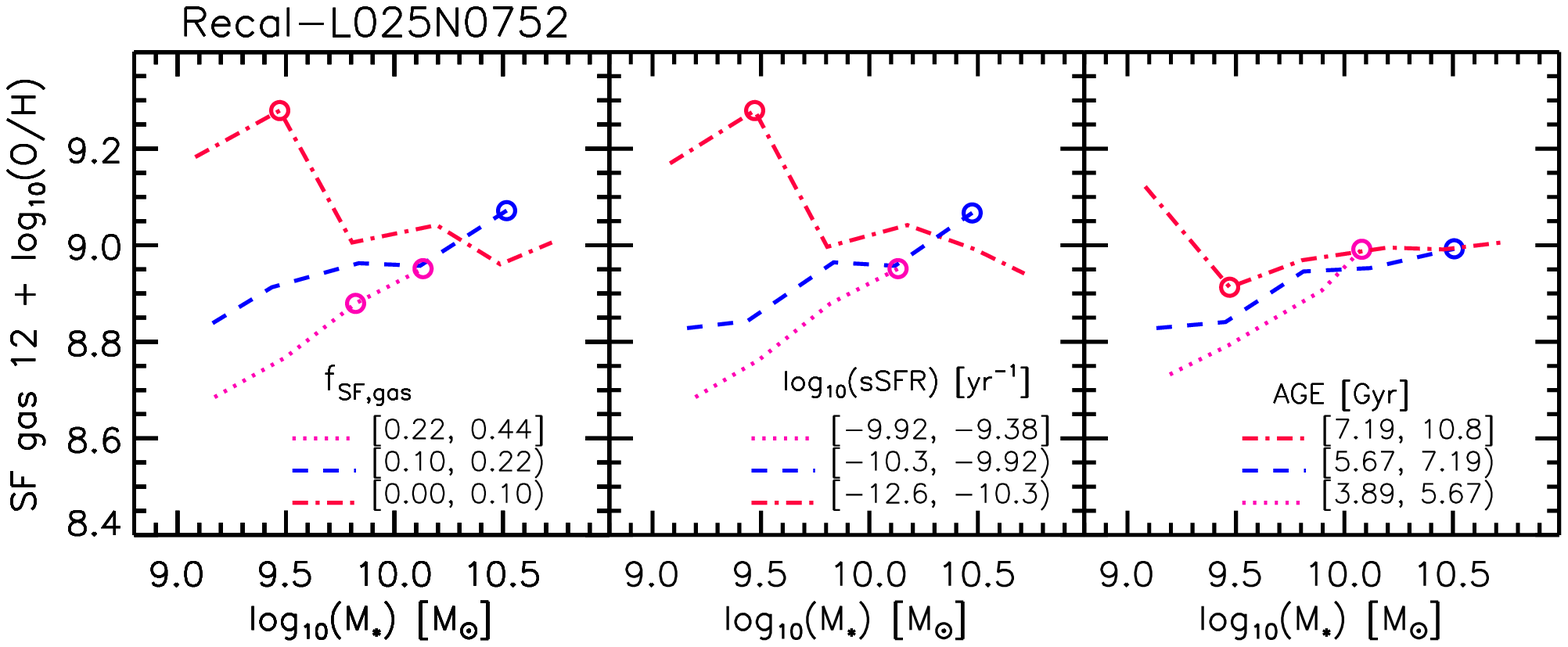}}\\
\resizebox{15.5cm}{!}{\includegraphics{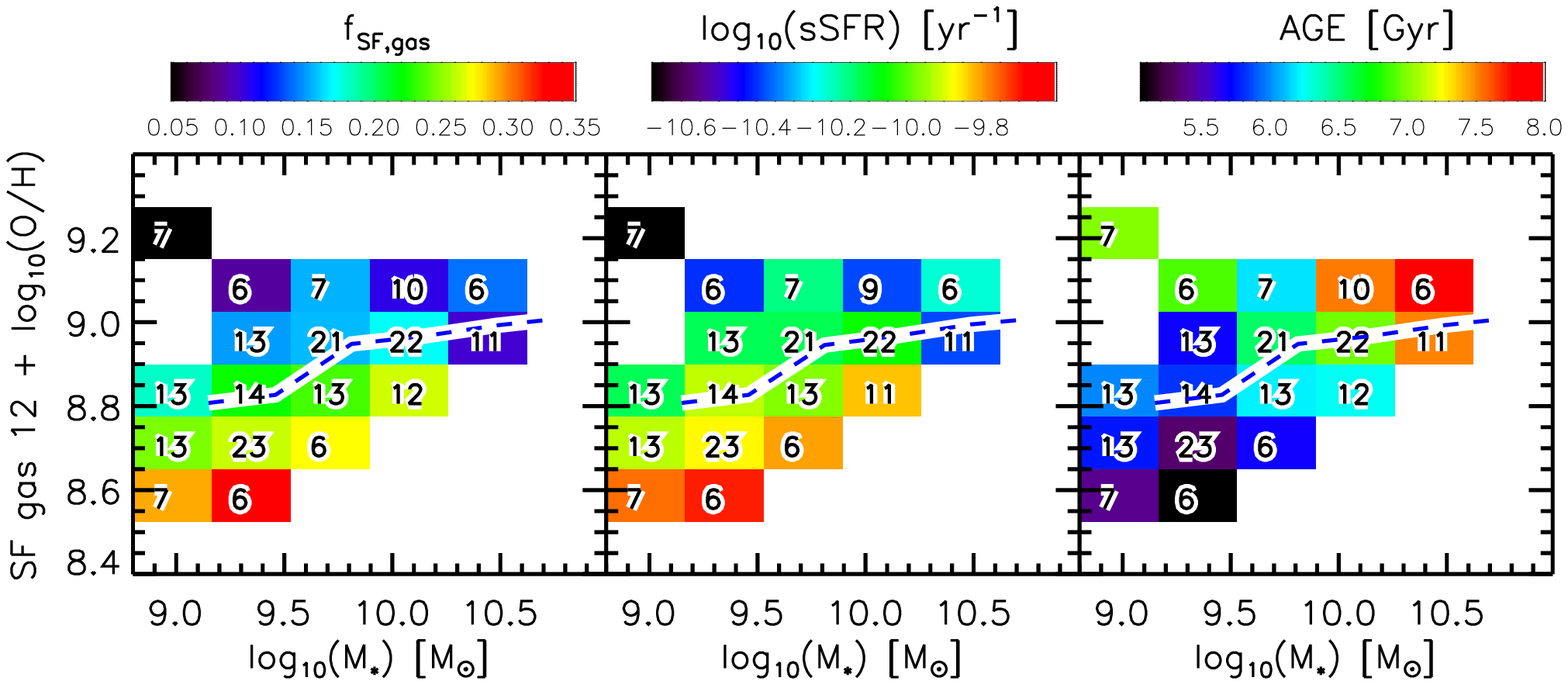}}
\end{center}
\caption[]
{
Top panels: Median $M_* - {\rm O/H}|_{\rm SF,gas}$ relation at $z=0$ binned in $f_{\rm SF,gas}=M_{\rm SF,gas}/(M_{\rm SF,gas}+M_\star)$ (left panel), 
sSFR (middle panel) and stellar mass-weighted mean age (right panel), as indicated in the figure.
All considered mass bins contain $N_{\rm bin} \ge 5$ galaxies; less populated bins ($ 5 \le N_{\rm bin} < 10$)
are marked with a circle.
Bottom panels: 2D colour histograms, where
filled squares are coloured according to the median $f_{\rm SF,gas}$ (left panel), 
sSFR (middle panel) and stellar mass-weighted mean age (right panel) of galaxies.
The number of galaxies included in each bin is indicated inside it.  Only bins with $N_{\rm bin} \ge 5$ are
considered for the analysis.
As a reference, the median $M_* - {\rm O/H}|_{\rm SF,gas}$ relation obtained from the whole galaxy sample is shown 
in the three bottom panels as a blue dashed line.
}
\label{fig:MZR_bin}
\end{figure*}

\section{Fundamental metallicity relation and scatter of the MZR}
\label{sec:mzr_scatter}

The scatter in the observed MZR is partially due to observational uncertainties.
However, as discussed in Section \ref{sec:introduction}, there is evidence for secondary dependences
of the metallicity on properties such as gas fractions or SFR.
In this Section, we explore secondary metallicity dependences of {\sc EAGLE} galaxies 
that can explain part of the scatter of the MZR at fixed mass.  

In Fig. \ref{fig:MZR_bin} (top panels), we analyse the $M_* - {\rm O/H}|_{\rm SF,gas}$ relation obtained for 
Recal-L025N0752 at $z=0$ by binning
the galaxy sample according to star-forming gas fraction (left panels), specific SFR 
(sSFR$= {\dot M_*} / M_*$, middle panels)
and stellar mass-weighted mean age (right panels). 
The gas fractions shown in left panels were calculated using the star-forming gas component 
($f_{\rm SF,gas} = M_{\rm SF,gas} / (M_* + M_{\rm SF,gas})$) in the galaxies
but similar trends are obtained if considering the total gas ($f_{\rm gas} = M_{\rm gas} / (M_* + M_{\rm gas})$) instead.
In each of the upper panels, we show results for three bins corresponding to the aforementioned
properties, as indicated in the figure.  
In the bottom panels, 2D colour histograms are shown, where
filled squares are coloured according to the median $f_{\rm SF,gas}$ (left panel),
sSFR (middle panel) and stellar mass-weighted mean age (right panel).
As a reference, the median $M_* - {\rm O/H}|_{\rm SF,gas}$ relation obtained from the whole galaxy sample is shown
in the three bottom panels as a blue dashed line.

We can see that, at a given stellar mass, 
there is a clear dependence of metallicity on these three parameters.
At fixed $M_*$, systems with higher $f_{\rm SF,gas}$ or higher sSFRs 
have lower values of ${\rm O/H}|_{\rm SF, gas}$, consistently with the trends found
in certain observational studies \citep[e.g.][]{kewley2008, salim2014, bothwell2013}.
We also obtained lower values of ${\rm O/H}|_{\rm SF, gas}$ for younger galaxies,
and all these trends are stronger at lower masses.
Our findings are consistent with a scenario in which higher fractions of metal-poor gas drive 
higher sSFRs in smaller galaxies and, consequently, these subsamples of galaxies tend to be younger 
than more metal-enriched galaxies of similar stellar masses.

In the case of higher $M_*$ galaxies, as we have previously mentioned, the metallicity saturates
towards an asymptotic value at early times.  
However, it is worth noting that there is a hint of an inversion of the dependence of metallicity on the 
$f_{\rm SF,gas}$ and sSFR at the massive end of the MZR.  Massive galaxies with lower $f_{\rm SF,gas}$ and sSFR 
tend to have lower metallicities, contrary to the trend shown by smaller galaxies.
This issue might be a consequence of the influence of AGN feedback which can produce a decrease
in the SF activity in these galaxies and, consequently, the quenching of their chemical evolution. 

In the top panels of Fig. \ref{fig:MZR_bin}, some bins used to construct the median relations 
are poorly sampled ($5 \le N_{\rm bin} < 10$, circles). Also, in the bottom
panels, the 2D histograms include bins with less that 10 galaxies in some cases.  
In order to check if the obtained trends are affected by
sampling effects due to the limited volume of the simulation ($25^3 \ {\rm cMpc^3}$), we repeated the previous
analysis for the simulations Ref-L050N0752 and Ref-L100N1504, for which the number of galaxies per bin in
the 2D histograms increases by up to $\approx 100$ and $\approx 700$, respectively.  Although  
the shape and normalization of the $M_* - {\rm O/H}|_{\rm SF,gas}$  relation 
vary when using the simulations Ref-L050N0752 and Ref-L100N1504 (Fig. \ref{fig:gas_mzr_z0}), the 
secondary dependences of metallicity (at a fixed $M_*$) on $f_{\rm SF,gas}$, sSFR and age are
also present in these larger-volume simulations.  Thus, these dependences seem not to be an artefact caused
by sampling issues.

To corroborate if AGN feedback might generate a decrease in the SF activity and quench the metallicity
evolution of galaxies, we compared 
NOAGN-L050N0752, AGNdT8-L050N0752 and AGNdT9-L050N0752 models.  
We obtained that for higher $\Delta T_{\rm AGN}$,
the inversion of the dependence of metallicity with the $f_{\rm SF,gas}$ and sSFR is more significant (see Section \ref{sec:models}).
When AGN feedback is suppressed, we did not find a clear trend for such inversion. 
Moreover, we found that massive galaxies ($M_* \ga 10^{10.5} {\rm M}_{\sun}$) tend to exhibit lower 
$f_{\rm SF,gas}$ and metallicities as the ratio between the black hole mass and $M_*$ ($M_{\rm BH}/M_*$) increases.

\begin{figure}
\begin{center}
\vspace{-1cm}\resizebox{7.0cm}{!}{\includegraphics{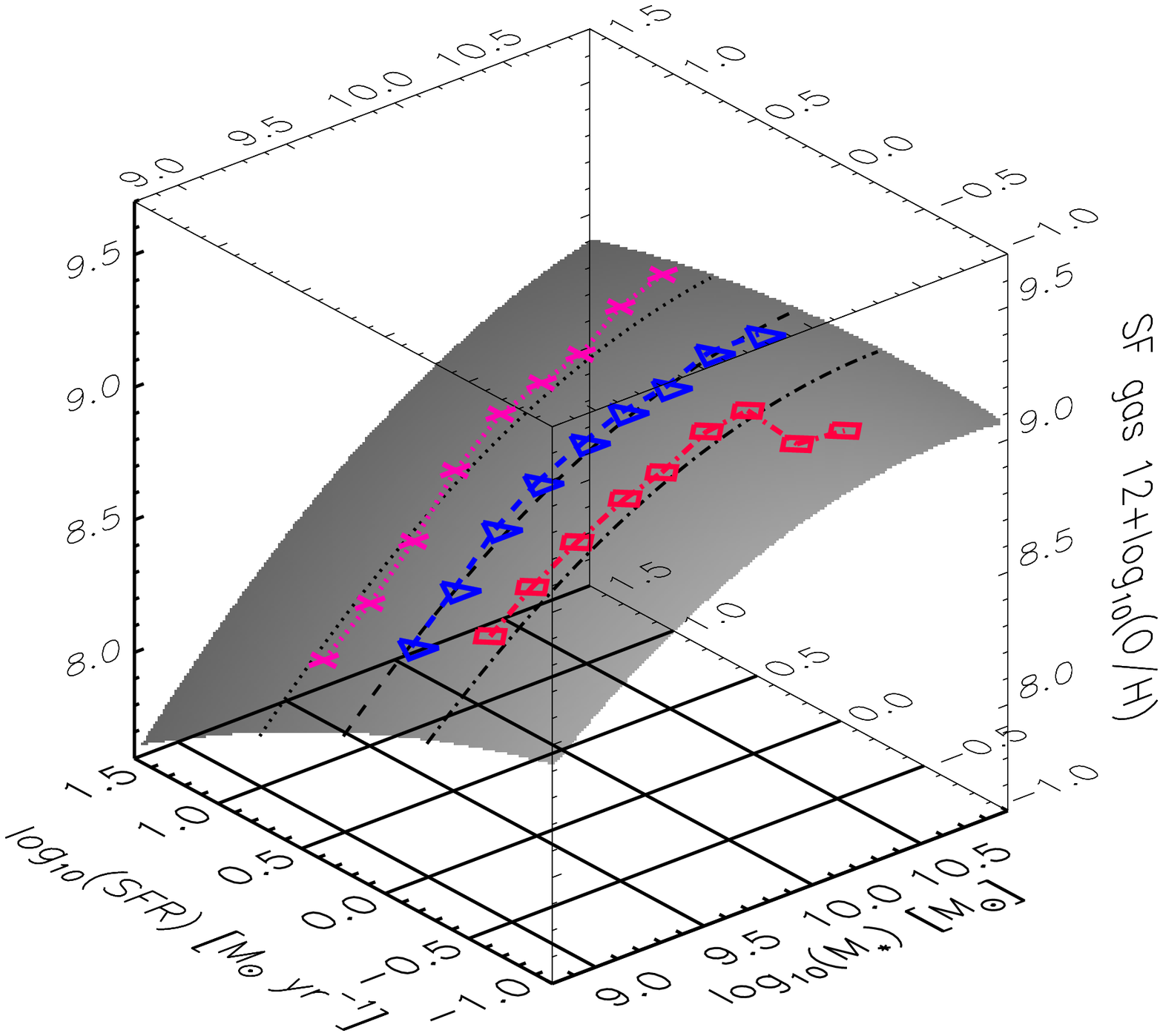}}\\
\vspace{-1cm}\resizebox{7.0cm}{!}{\includegraphics{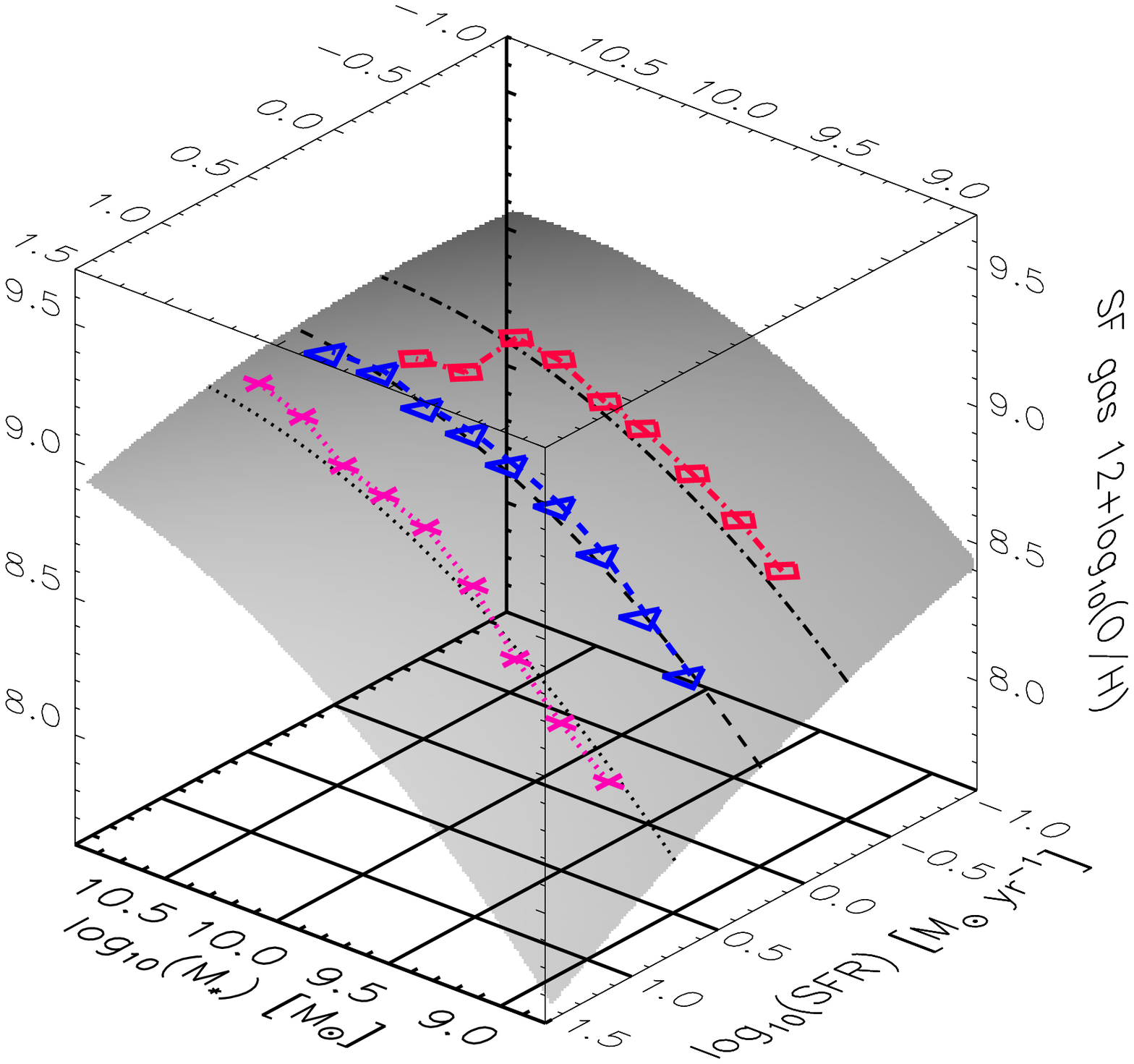}}\\
\vspace{-0.3cm}\resizebox{7.0cm}{!}{\includegraphics{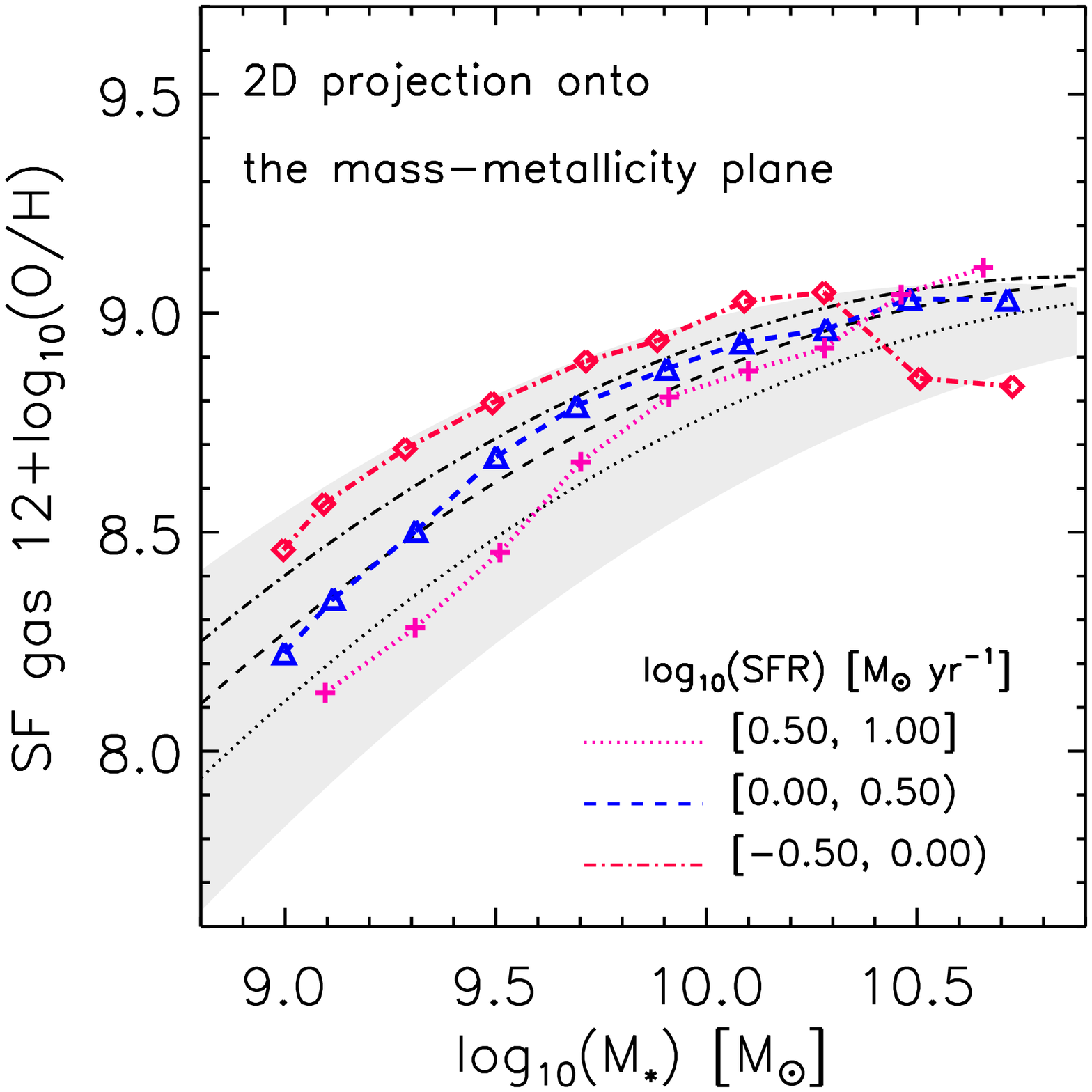}}
\vspace{-0.25cm}
\end{center}
\caption[]
{
Upper panels: Median simulated $M_* - {\rm O/H}_{\rm SF,gas}$ relations binned according 
to SFR in the $M_*$ - SFR - O/H space for Recal-L025N0752 simulations.
Different SFR bins are represented with different lines and symbols: 
$\log_{10} ({\rm SFR} / {\rm M}_{\sun} {\rm yr}^{-1})= -0.5-0.0$ 
(red dash-dotted line with diamonds), 
$\log_{10} ({\rm SFR} / {\rm M}_{\sun} {\rm yr}^{-1}) = 0.0-0.5$ (blue dashed line with triangles) and
$\log_{10} ({\rm SFR} / {\rm M}_{\sun} {\rm yr}^{-1}) = 0.5-1.0$ (pink dotted line with crosses). 
The shaded area represents the best-fit 2D surface taken from \citet{mannucci2010}.
Black lines trace the shaded surface at the center of each SFR bin. 
For the sake of clarity and to illustrate the level of agreement with the data, two projections
of the $M_*$ - SFR - O/H space are shown.
Bottom panel: 2D projection of the aforementioned relations onto the mass-metallicity plane.
For the simulation, all galaxies identified below $z\approx5$ have been considered for the calculations.
Note that in Fig. \ref{fig:MZR_bin} only $z=0$ galaxies were taken into account.
}
\label{fig:FMR}
\end{figure}

The existence of a relation between ${\rm O/H}|_{\rm SF,gas}$, $M_*$ and SFR (FMR) 
has been investigated by many authors in recent years.
However, as discussed in Section \ref{sec:introduction},
the features of this relation and its strength remain poorly constrained because of the large
uncertainties and systematic errors that affect observational studies.
Fig. \ref{fig:FMR} clearly indicates that {\sc EAGLE} galaxies follow a well-defined relation
between $M_*$, SFR and ${\rm O/H}|_{\rm SF,gas}$ at least below $z \approx 5$.  
Only mass bins with $N_{\rm bin} > 10$ were considered for this analysis: 
$N_{\rm bin} = 13-296, \ 25-142$ and $23-82$ at 
$\log_{10} ({\rm SFR} / {\rm M}_{\sun}yr^{-1}) \approx -0.25, \ 0.25$ and 0.75, respectively. 
The predicted 3D relation agrees remarkably well
with those given in \citet{mannucci2010} for observed galaxies in the local Universe, especially at $M_* < 10^{10.5} {\rm M}_{\sun}$
(see also Fig. \ref{fig:simus}, top left panel, for an analysis of residuals).
At a given SFR, metallicity shows a strong correlation with $M_*$
with a flattening of the slope towards higher masses.  At a given $M_*$, the metallicity of
low-mass systems decreases with SFR while, for massive galaxies, the metallicity 
increases with SFR. Similar trends were reported by \citet{yates2011} and \citet{derossi2015} in the context of
semi-analytical models and hydrodynamical simulations, respectively. 
\citet{lagos2015} explored the relation between SF gas metallicity, SFR and $M_*$ 
in the {\sc EAGLE} Ref-L100N1504 simulation. 
They found good agreement with the data at high masses ($M_* \ga 10^{10} \ {\rm M}_{\sun}$) while
low-mass galaxies depart from the measured relation by up to $\approx 0.4$ dex towards higher metallicities.
For the higher-resolution Recal-L025N0752 simulations used here, the agreement with the data is better because
of the use of the recalibrated model (see Section \ref{sec:simulation_set}), which yields
more efficient stellar feedback (see Section \ref{sec:models}).

\begin{figure}
\begin{center}
\resizebox{8.5cm}{!}{\includegraphics{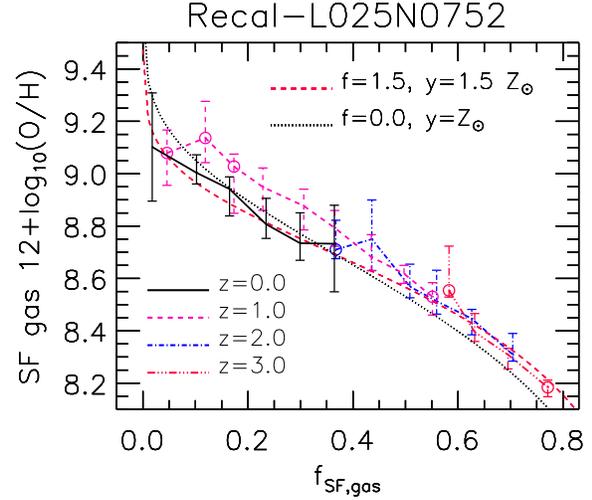}}
\end{center}
\vspace{-0.5cm}
\caption[]
{
SF gas-phase metallicity as a function of star-forming gas fraction 
for {\sc EAGLE} galaxies at different $z$, as indicated in the figure. 
Curves depict the median relation and error bars, the 25th and 75th percentiles.
The number of galaxies per bin is $N_{\rm bin} \ge 7$, with circles indicating less populated
bins ($N_{\rm bin} = 7-9$).
The gas fraction is calculated as $M_{\rm SF,gas} / (M_{\rm SF,gas} + M_{*})$, where
$M_{\rm SF,gas}$ denotes the star-forming gas component. 
The curves without error bars correspond to a closed-box
model with solar yield (dotted black line) and a simple model with
supra-solar yield and an outflow rate $\dot{M} \sim f \times {\rm SFR}$,
with $f=1.5$
(dashed red line, see text for details).
}
\label{fig:OH_vs_fg}
\end{figure}

From Fig. \ref{fig:FMR} we can see that Recal-L025N752 reproduces 
the observed flattening of the slope of the FMR towards high $M_*$.  Nevertheless, 
in the case of massive galaxies with low SFRs, the impact of AGN feedback seems to have been very strong,
generating a departure towards lower metallicity from observations by \citet{mannucci2009}.  
We note, however, that other observational studies 
\citep[e.g.][]{yates2011, laralopez2013b, andrews2013, zahid2013} found lower metallicities in low-SFR massive
systems, which is more consistent with {\sc EAGLE} trends.

\citet{mannucci2010} claimed that the FMR does not evolve significantly
below $z\approx 2.5$.
On the other hand, 
\citet{troncoso2014} reported that observed galaxies at $z \approx 3.4$ lie below the FMR by a factor of $\sim 10$, suggesting that the observed FMR is established at $z < 3$. 
More recent works suggested that the trend of the SFR dependence is preserved at $z < 2.3$ but
its strength evolves \citep{salim2015, brown2016, grasshorn2016}.
{\sc EAGLE} predicts a FMR that is already in place at 
$ z \approx 5$ and does not evolve significantly with time.  

The FMR might be caused by the existence of
a more fundamental relation between $M_*$, metallicity and gas fraction \citep[e.g.][]{bothwell2013, hughes2013}.
\citet{derossi2015} obtained that dependence in the case of GIMIC
galaxies while \citet{lagos2015} found similar results by performing
a principal component analysis over different galaxy properties in the {\sc EAGLE} Ref-L100N1504 simulation.
At a given $M_*$, galaxies with higher gas fractions tend to exhibit higher SFRs
and, as we have discussed, these galaxies have lower gas-phase metallicities.  
Fig. \ref{fig:OH_vs_fg} shows the 
${\rm O/H}|_{\rm SF,gas} - f_{\rm SF,gas}$ relation in Recal-L025N0752 simulations at
$z = 0,1,2$ and 3.
The $M_*$ range associated to each $z$ 
can be seen in previous figures (e.g. Fig. \ref{fig:gas_mzr_z}).
Again, we have calculated gas fraction using the star-forming gas component in the galaxies.
We can see that systems with larger gas fractions exhibit lower metallicities.
As $f_{\rm SF,gas}$ increases from 0 to 0.8, the metallicity decreases by $\approx 1$ dex. 
The scatter around the mean relation is relatively low ($\la 0.3$ dex), with the only exception of 
very gas-poor systems at $z=0$. We have verified that the scatter of the 
${\rm O/H}|_{\rm SF,gas} - f_{\rm SF,gas}$ relation is generated by the different stellar masses
of the systems that are included in each $f_{\rm SF,gas}$ bin.
We found that, at $z \ga 0.6$ and for a fixed $f_{\rm SF,gas}$,
metallicity correlates with $M_*$, as expected.  On the other hand, at $z \la 0.6$, 
metallicity correlates with $M_*$ in the case of systems with $f_{\rm SF,gas} \ga 0.1$, 
while the metallicities of gas-poor systems ($f_{\rm SF,gas} \la 0.1$) tend to decrease with $M_*$.
This behaviour is consistent with the trends obtained in 
Fig. \ref{fig:MZR_bin}, left panels.

Fig. \ref{fig:OH_vs_fg} shows that $f_{\rm SF,gas}$ decreases
and ${\rm O/H}|_{\rm SF,gas}$ increases systematically for decreasing $z$, just as in a closed-box model
and most open-box models.
Thus, for the sake of comparison,
we estimated the correlation between ${\rm O/H}|_{\rm SF,gas}$ and
$f_{\rm SF,gas}$ obtained from a closed-box model and a simple model with outflows.
Following \citet{erb2006}, we implemented a simple model derived from
the closed-box model to include gas outflow at a rate 
$\dot{M}$ (${\rm M}_{\sun} {\rm yr^{-1}}$), which is a fraction $f$ of the SFR.
The metallicity of the ejected gas is assumed to be the same as the metallicity of the gas 
that remains in the galaxy.  Under these considerations, the metallicity is

\begin{equation}
\label{eq:outflow_model}
Z = y (1+f)^{-1} \ln[1+(1+f)(f_{\rm SF,gas}^{-1}-1)].
\end{equation}

Note that the closed-box model (Eq. \ref{eq:yields}) is obtained from this general
expression taking $f=0$, while observations suggest that outflows 
eject gas from galaxies at a rate that is comparable to the SFR 
and may be higher ($f\ga 1$, \citealt{veilleux2005}).
The curves without error bars in Fig. \ref{fig:OH_vs_fg} show the results corresponding to 
a closed-box model with $y= Z_{\sun}$ (dotted black line) and a simple model with $y= 1.5\times Z_{\sun}$
and an outflow rate $\dot{M} = 1.5 \times {\rm SFR}$ (dashed red line).  
Interestingly, the ${\rm O/H}|_{\rm SF,gas}$-$f_{\rm SF,gas}$ relation
obtained from the simulations follows roughly the trend associated with the closed-box model.
This can be explained considering the flat shape and negligible evolution obtained 
for the $y_{\rm eff} - M_{\rm bar}$ relation (Fig. \ref{fig:yeff_vs_M}).  
However, the simple outflow model reproduces better the simulated trends associated with
gas-rich systems at all $z$ and also improves the description of the behaviour of gas-poor systems towards $z=0$.
As discussed before, simulated systems drive metal-enriched
outflows and accrete metal-poor gas as well as undergoing merger events, whose relative efficiencies might vary with time.  
These processes cannot be captured by means of simple analytical models but it is interesting that such
models are able to roughly described the simulation results.

It is worth noting that we have analysed the metallicity evolution for {\em all} (i.e. central and satellite) 
galaxies. In {\sc EAGLE}, satellites tend to have higher metallicities than central galaxies
of similar stellar masses \citep{bahe2016} and this is also observed \citep{pasquali2010,pasquali2012}.  
At a given stellar mass and at $z=0$, for example, the median metallicity of satellites is higher
than for central galaxies by $\approx 0.2$ dex at $M_* \sim 10^9 {\rm M}_{\sun}$
and by $\approx 0.1$ dex at $M_* \sim 10^{10.5} {\rm M}_{\sun}$.  Also, the MZR of satellites presents a scatter that 
is a factor $\ga 2$ larger than for central galaxies. 
Therefore, the inclusion of satellites in the main sample increases the scatter
in the MZR. As discussed by \citet{derossi2015}, 
ram-pressure stripping could lead to higher global metallicities of satellites by removing
the reservoir of metal-poor gas usually located in the outer regions of galaxies.
The origin of the satellite metallicity enhancement in Ref-L100N1504 was studied by \citet{bahe2016}.
These authors found an excess gas and stellar metallicities
in satellites in agreement with observations, except for stellar metallicities 
at $M_* \la 10^{10.2} {\rm M}_{\sun}$ where the predicted excess is smaller than observed.
Stripping of low-metallicity gas and 
suppression of metal-poor inflows seem to play an important
role on driving the enhancement of gas metallicity.

Finally, as the predicted relation between star-forming gas fraction and metallicity shows no significant evolution 
and only involves two key global galaxy properties, 
we conclude that it is more fundamental than the FMR, at least in these simulations.
It is worth noting that similar conclusions were obtained by \citet{zahid2014a} by using 
empirical-constrained analytical models of chemical evolution. According to these authors, the
redshift independence of the relation between metallicity and stellar-to-gas ratio suggests
that it is a universal relation. The MZR would originate from this universal relation due to the
evolution of the gas content of galaxies. Our results seem to support these previous findings.

\section{Discussion}
\label{sec:models}

\subsection{Subgrid parameters}
\label{sec:eagle_models}

We have seen that the {\sc EAGLE} Recal-L025N0752 simulation reproduces 
the observed trends for the MZR and FMR remarkably well.  In this Section, we 
explore the effects of varying some model parameters to assess
their influence on the observed features of the fundamental metallicity relations.
The different simulations used for this study and the corresponding parameter variations can be found in Table \ref{tab:simus}.
In addition to checking resolution effects on the FMR ("N0376" vs "N0752" runs) and comparing the reference and recalibrated models, we 
study the impact of AGN feedback (NOAGN vs AGNdT9 models) and the stellar 
feedback efficiency (WeakFB vs StrongFB models).  
We also analyse the effects of changing the slope of the equation of state
imposed on the ISM (eos1 vs eos5/3 models)
and the star formation law (KSLow vs KSHigh models).

\begin{figure*}
\begin{center}
\resizebox{6.0cm}{!}{\includegraphics{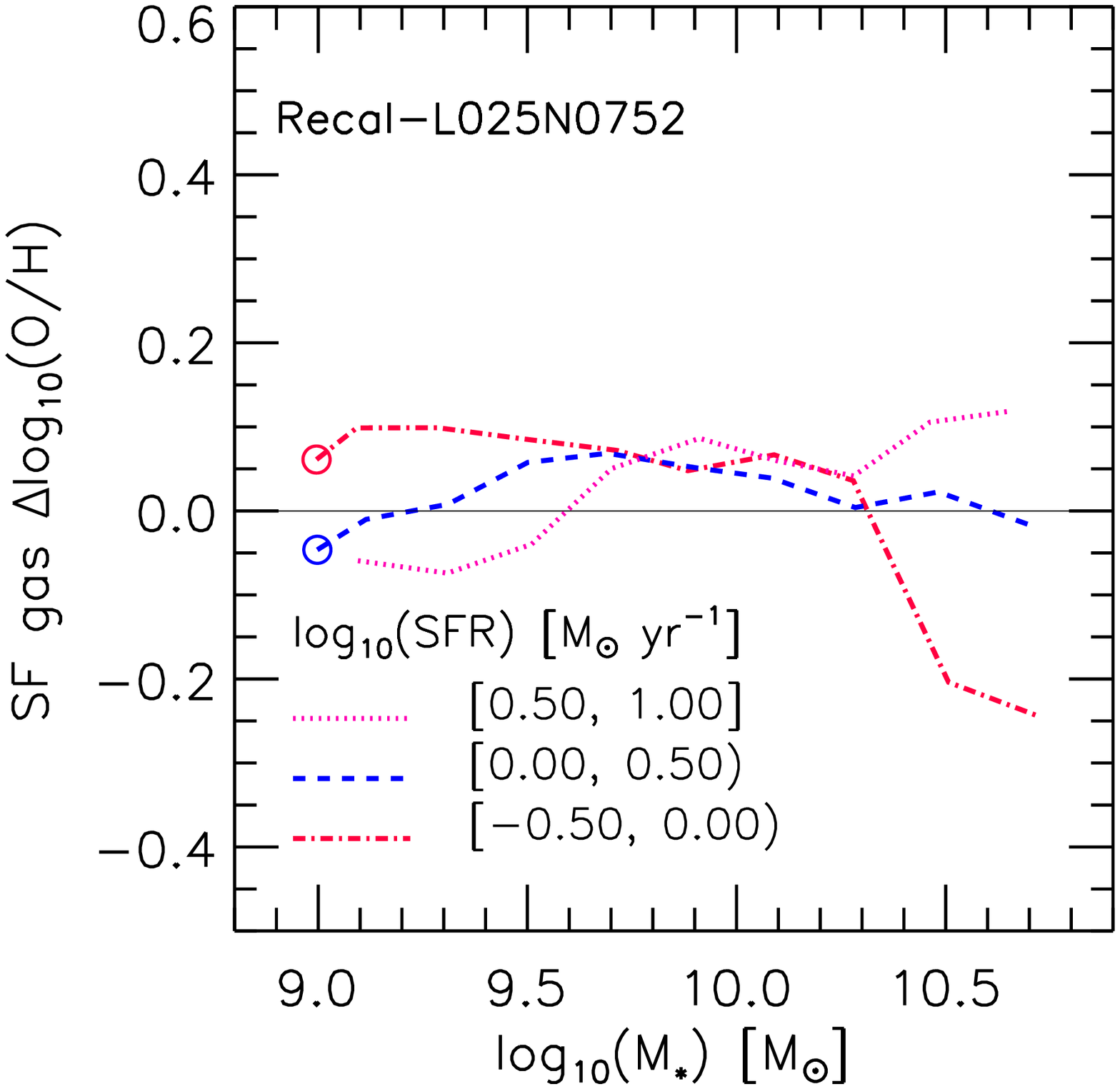}}
\hspace{-0.5cm}\resizebox{6.0cm}{!}{\includegraphics{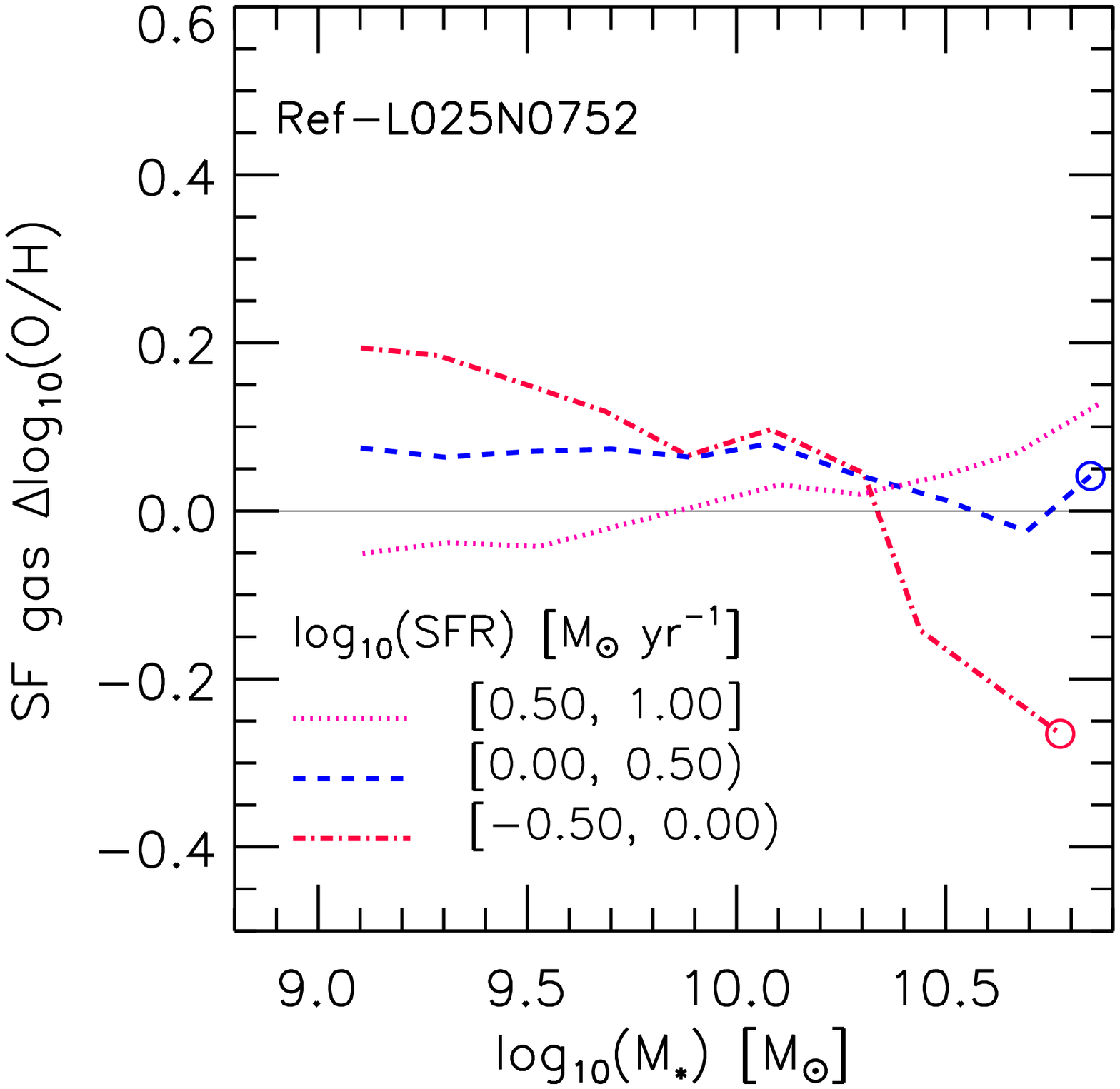}}
\hspace{-0.5cm}\resizebox{6.0cm}{!}{\includegraphics{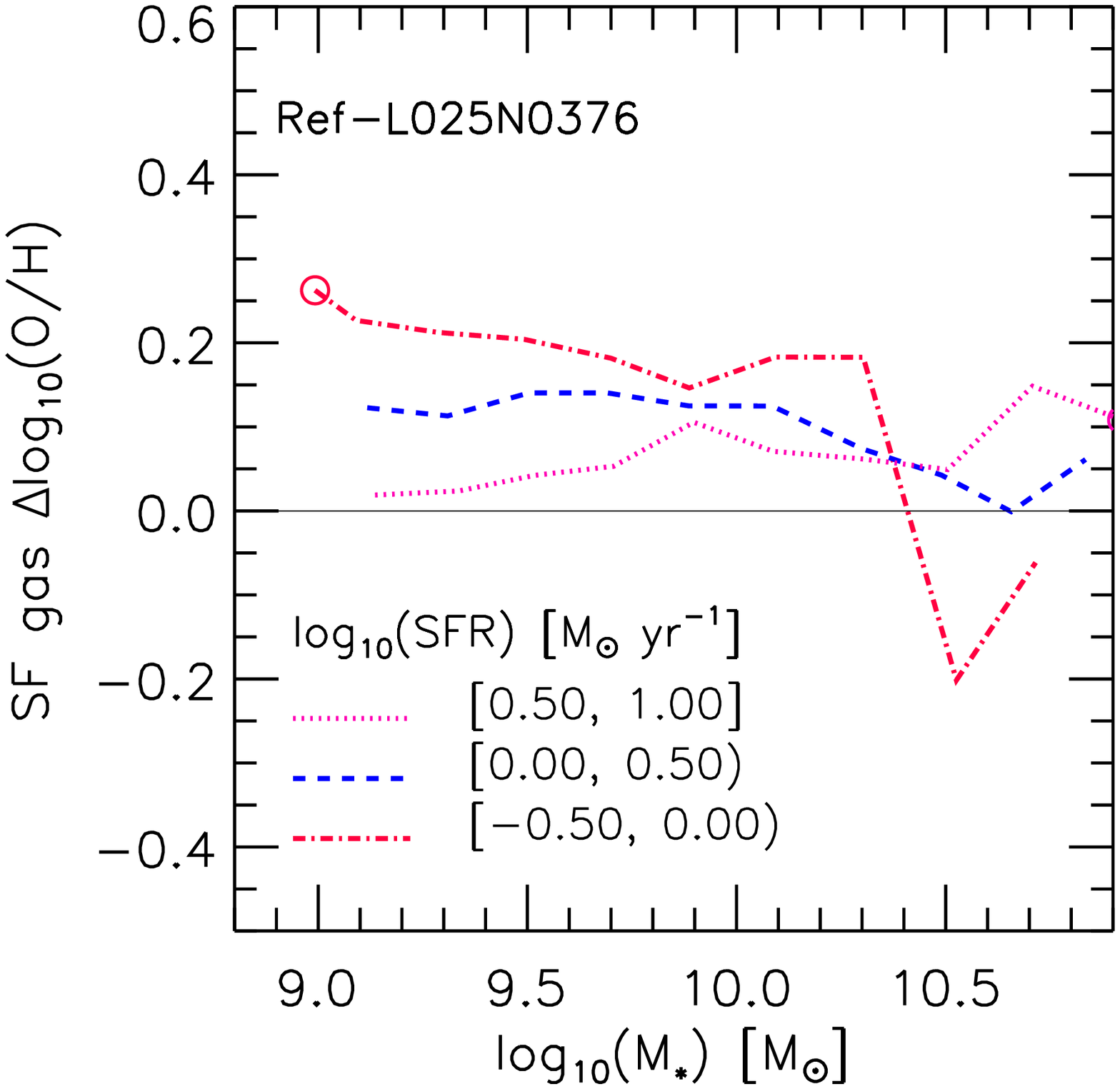}}\vspace{-0.5cm}
\resizebox{6.0cm}{!}{\includegraphics{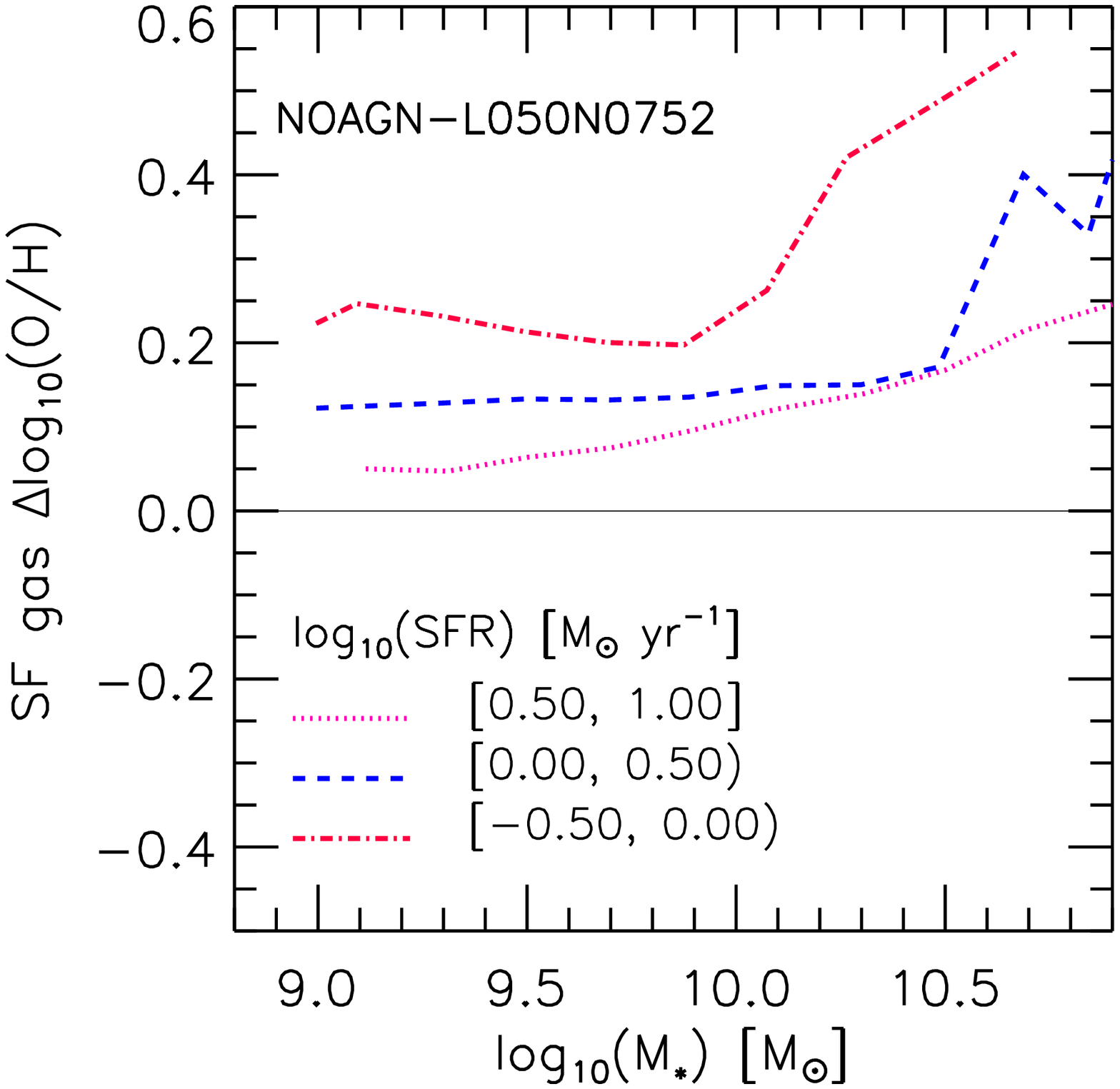}}
\hspace{-0.5cm}\resizebox{6.0cm}{!}{\includegraphics{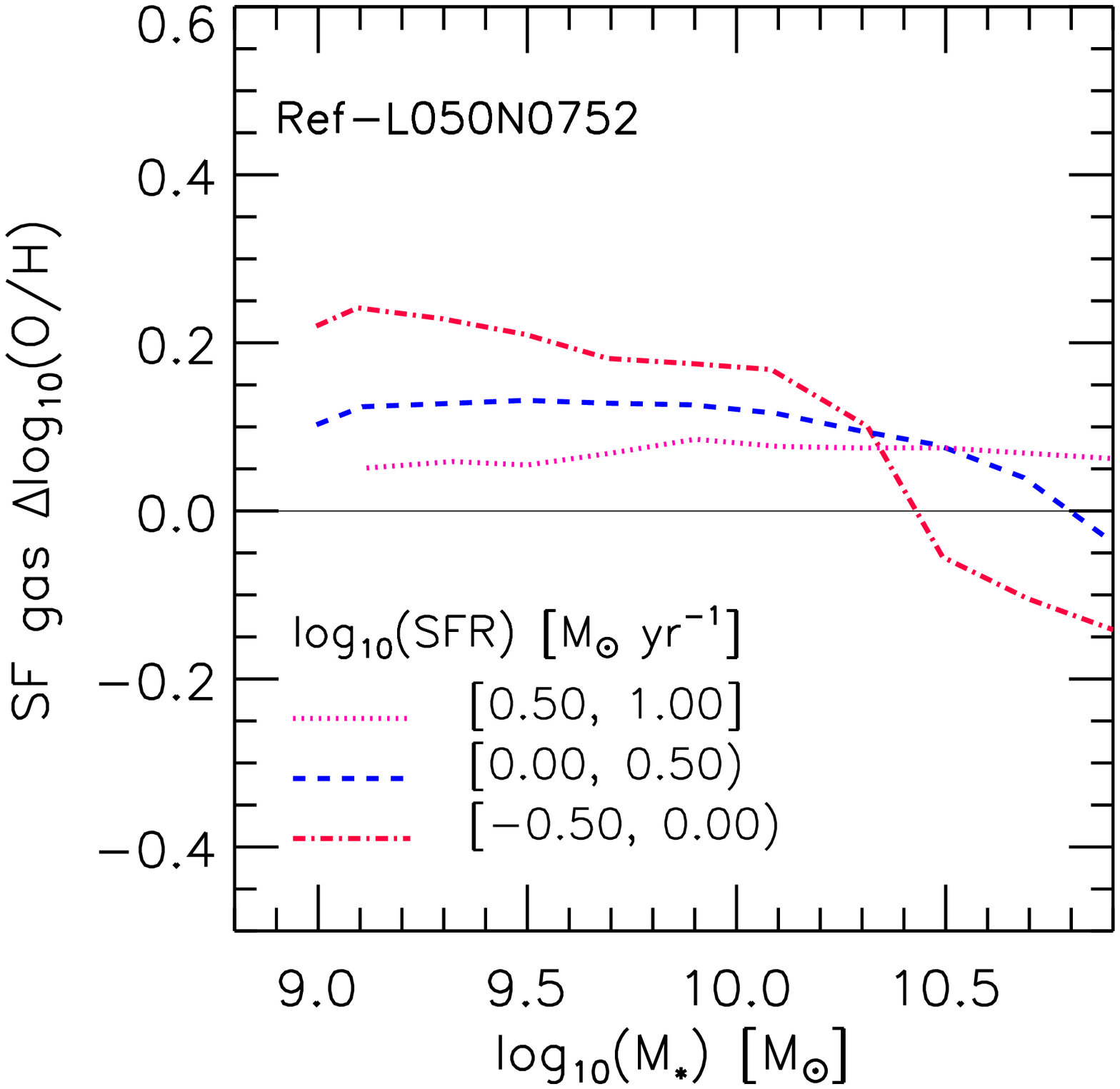}}
\hspace{-0.5cm}\resizebox{6.0cm}{!}{\includegraphics{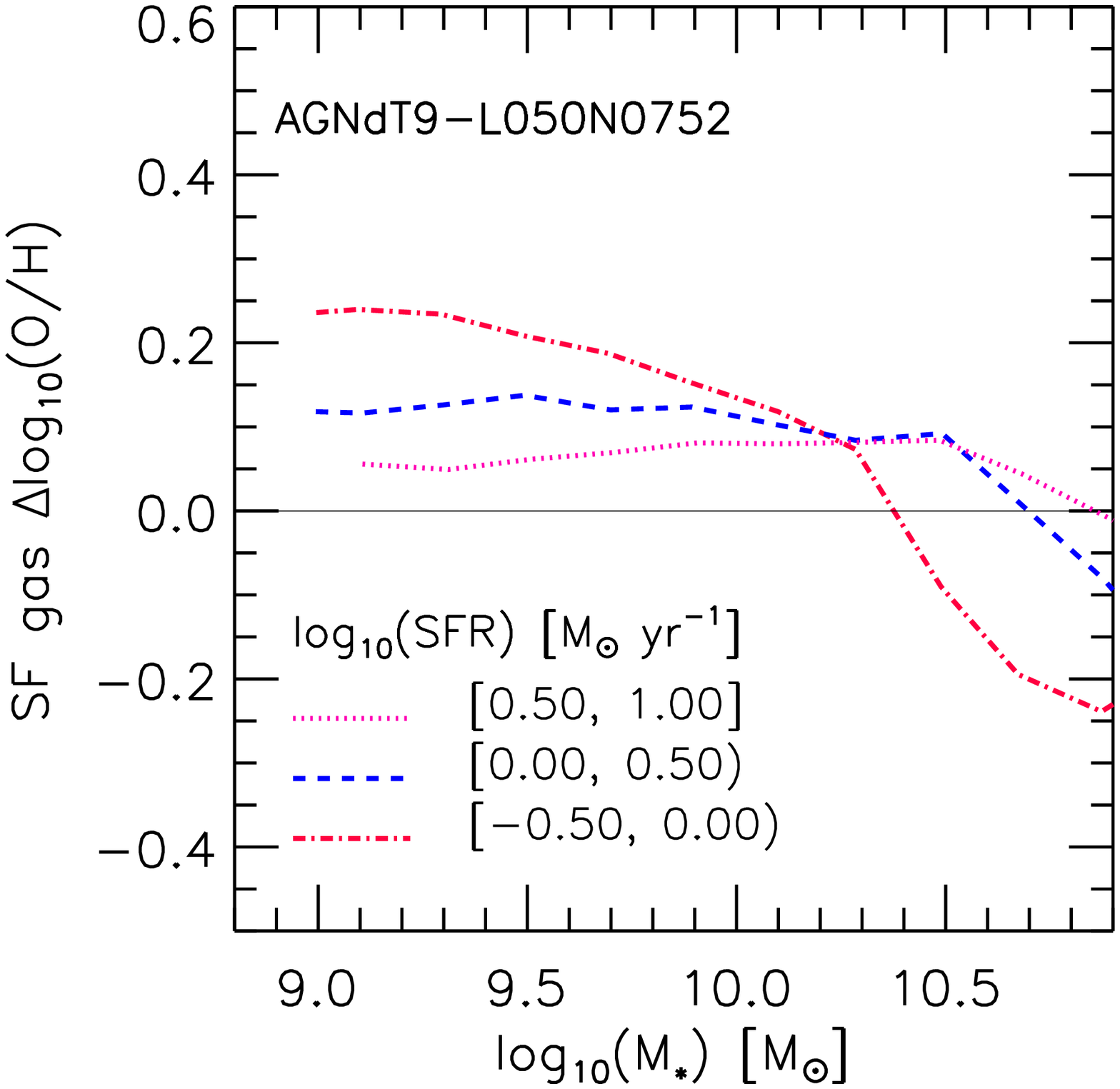}}\vspace{-0.5cm}
\resizebox{6.0cm}{!}{\includegraphics{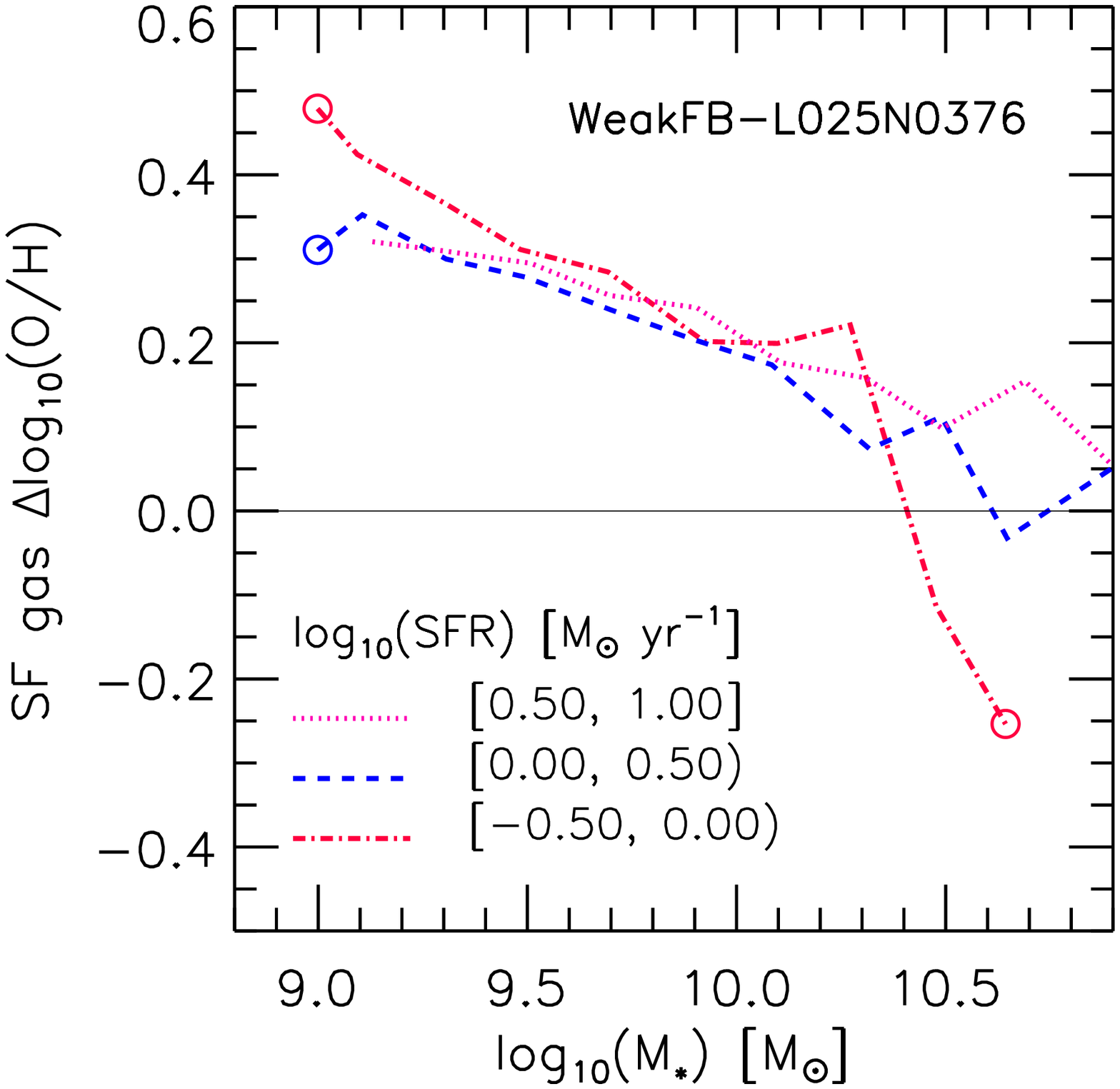}}
\hspace{-0.5cm}\resizebox{6.0cm}{!}{\includegraphics{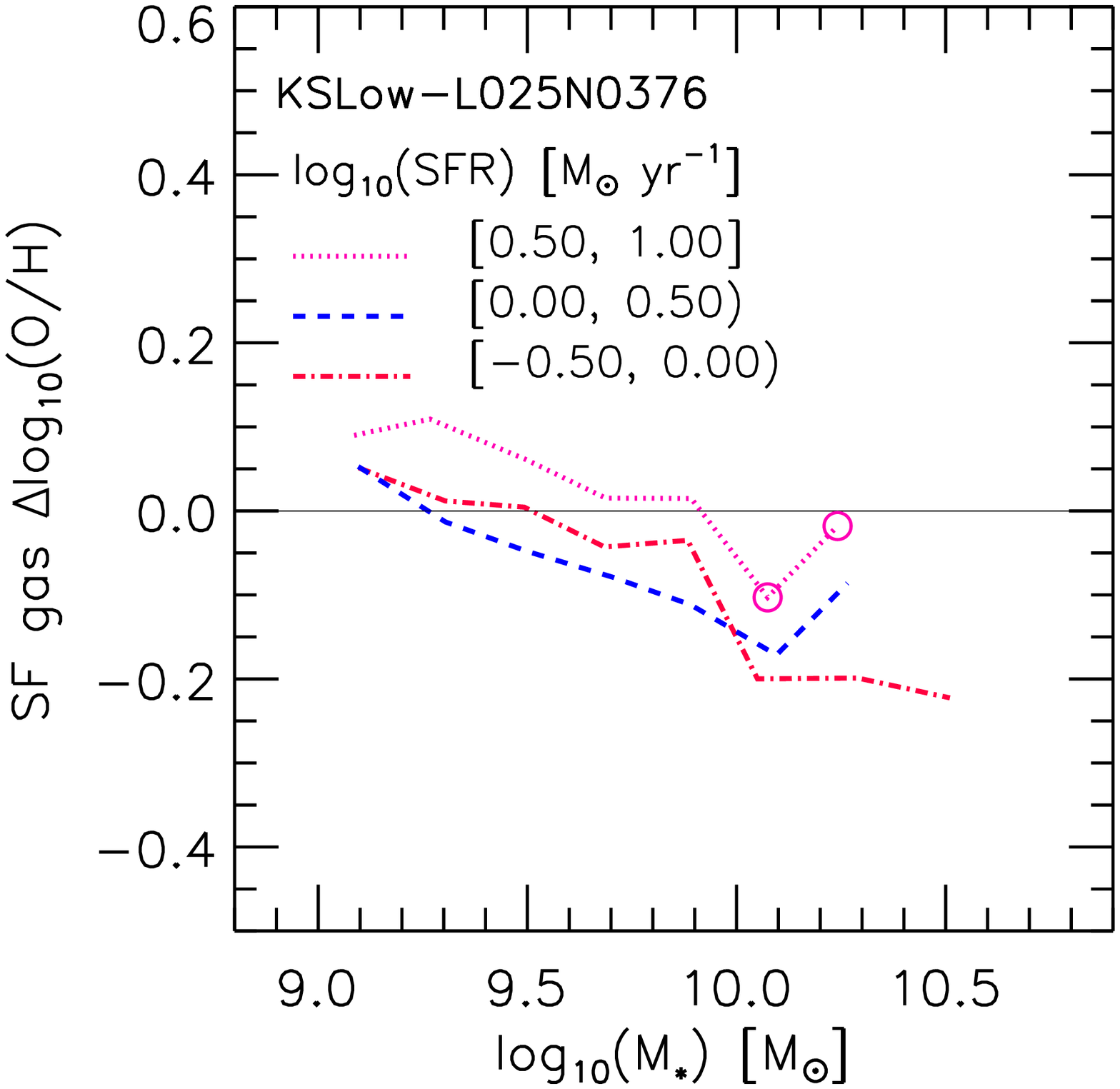}}
\hspace{-0.5cm}\resizebox{6.0cm}{!}{\includegraphics{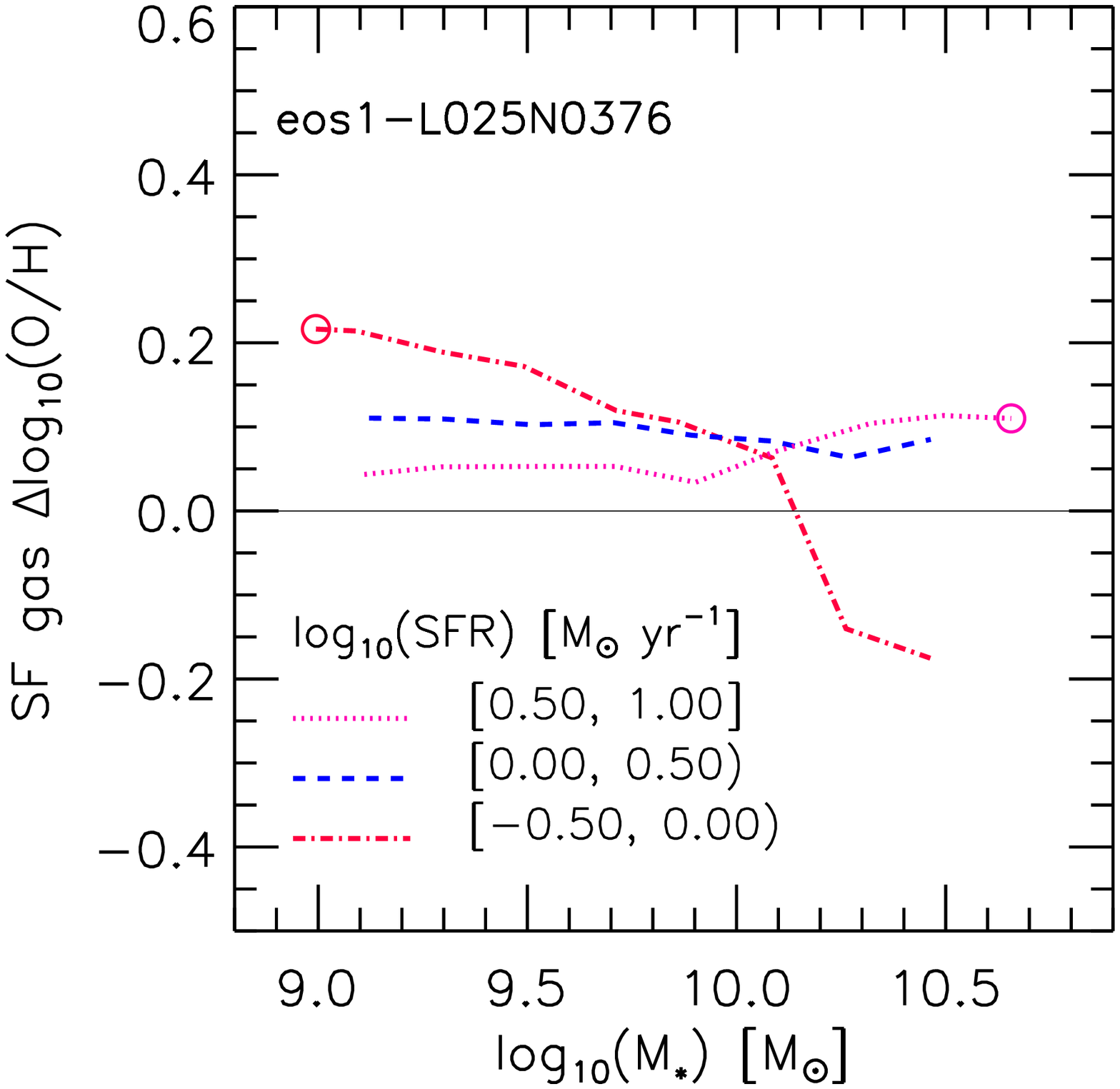}}\vspace{-0.5cm}
\resizebox{6.0cm}{!}{\includegraphics{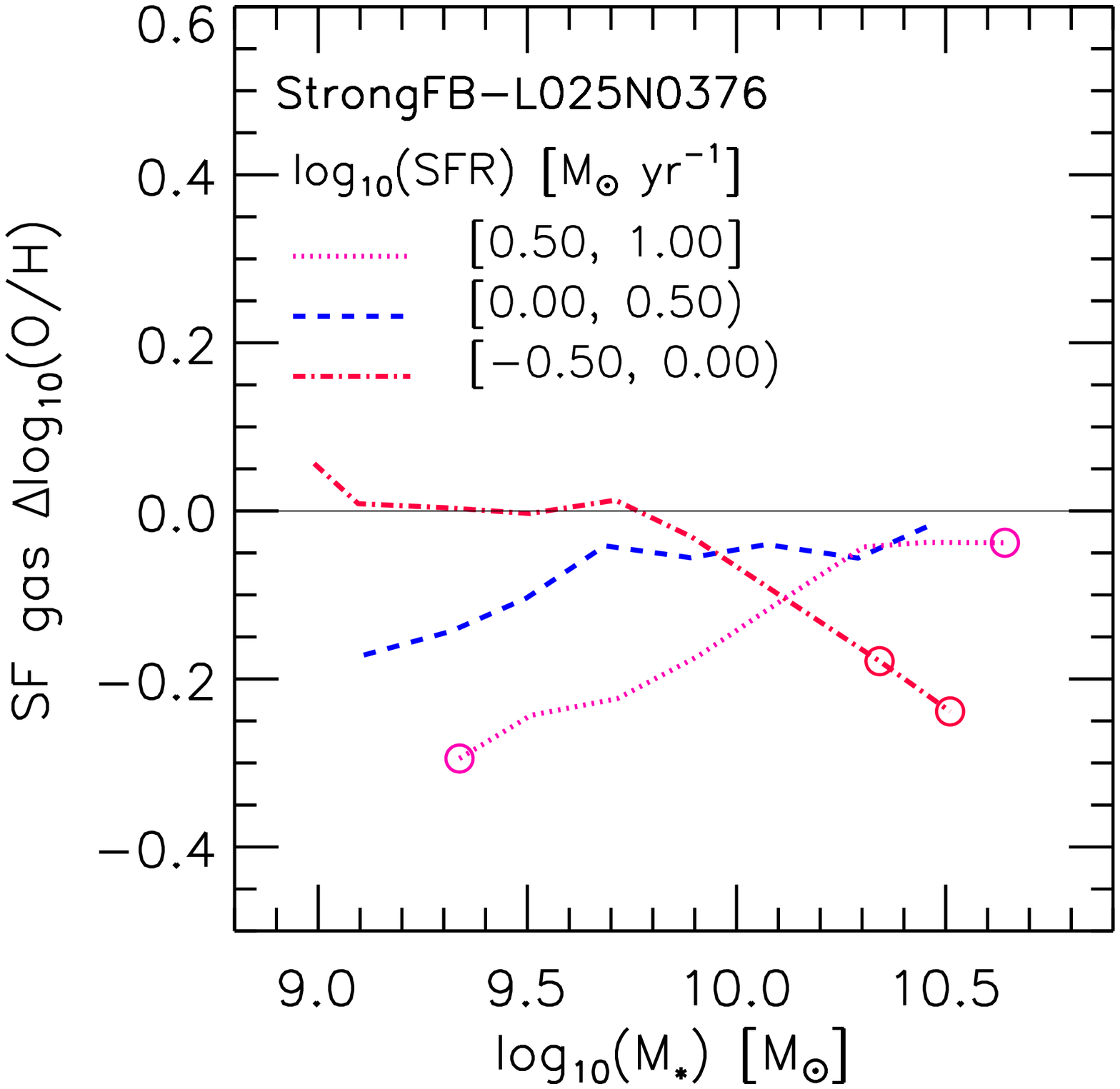}}
\hspace{-0.5cm}\resizebox{6.0cm}{!}{\includegraphics{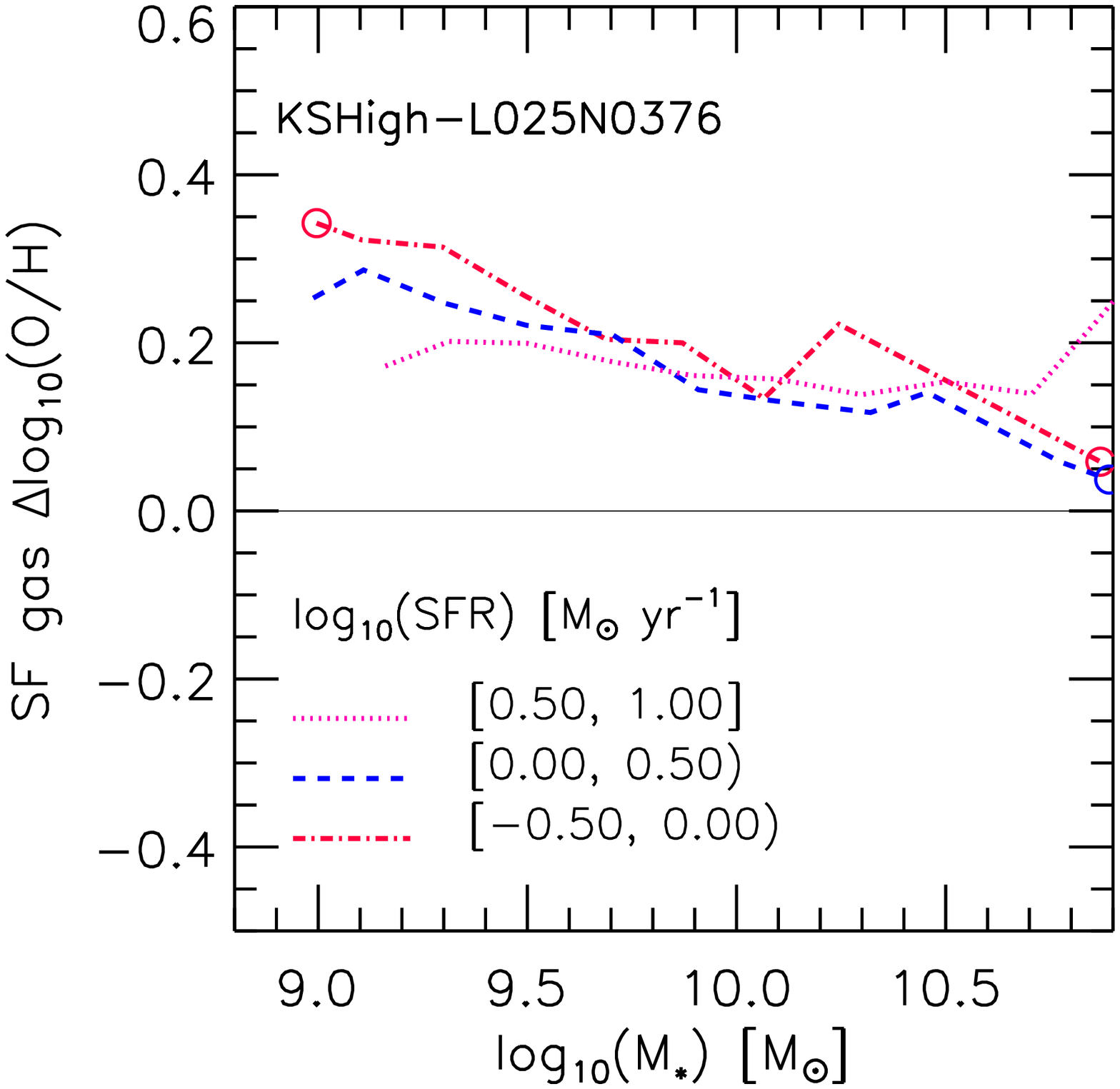}}
\hspace{-0.5cm}\resizebox{6.0cm}{!}{\includegraphics{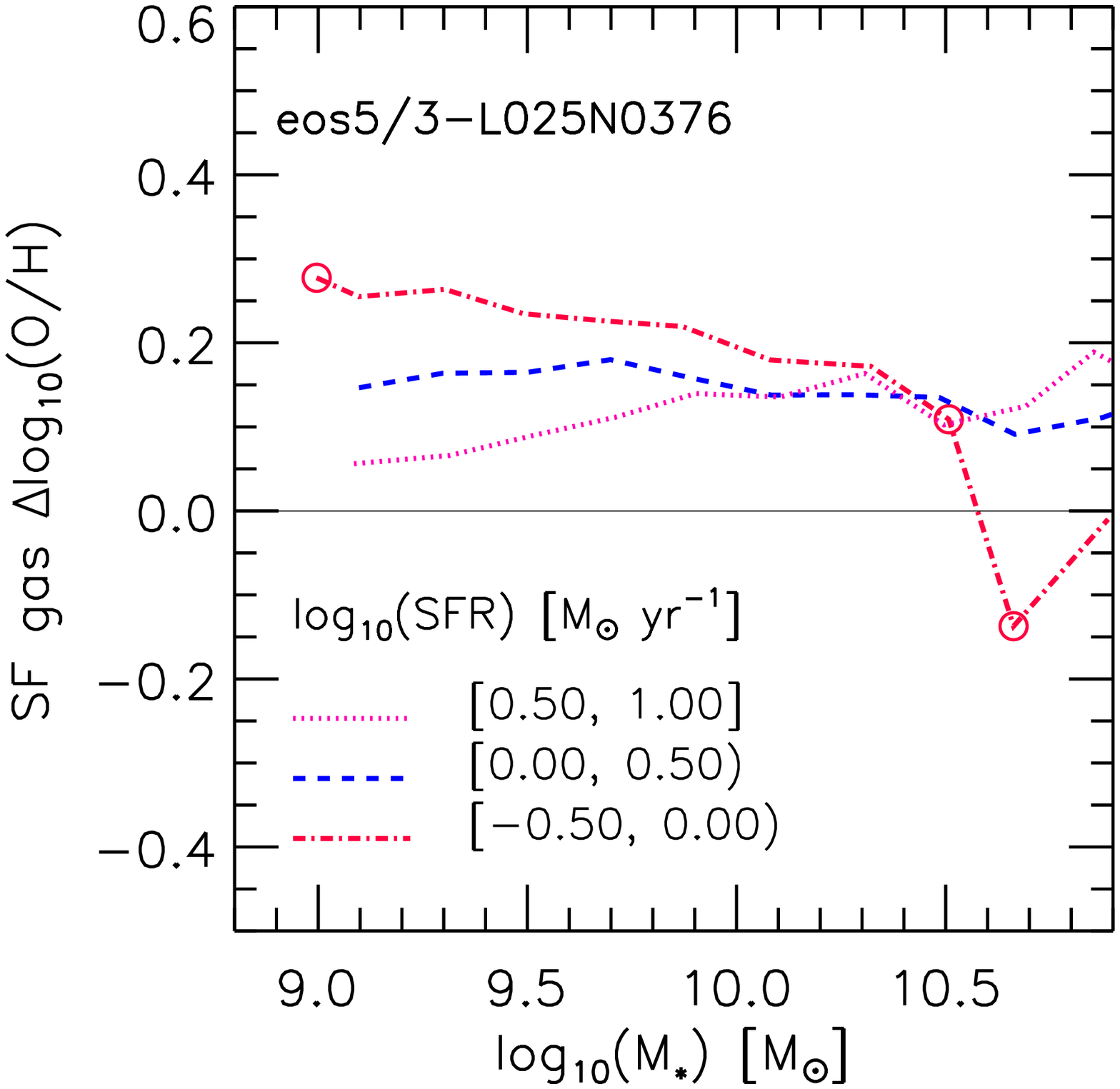}}
\end{center}
\caption[]
{
Residuals of the simulated median $M_* - {\rm O/H}|_{\rm SF,gas}$ relation, binned according to SFR, with respect to the fit for the observed FMR
given by \citet[][see shaded surface in Fig. \ref{fig:FMR}]{mannucci2010}.
All simulated galaxies at $z \la 5$ have been considered for the analysis.  
Less populated mass bins ($5 \le N_{\rm bin} <10$) are indicated with circles.
The convention used for lines
and colours is indicated in the figure.  Each panel corresponds to a different {\sc EAGLE} simulation
(see Table \ref{tab:simus}).
}
\label{fig:simus}
\end{figure*}

\begin{figure*}
\begin{center}
\resizebox{8.75cm}{!}{\includegraphics{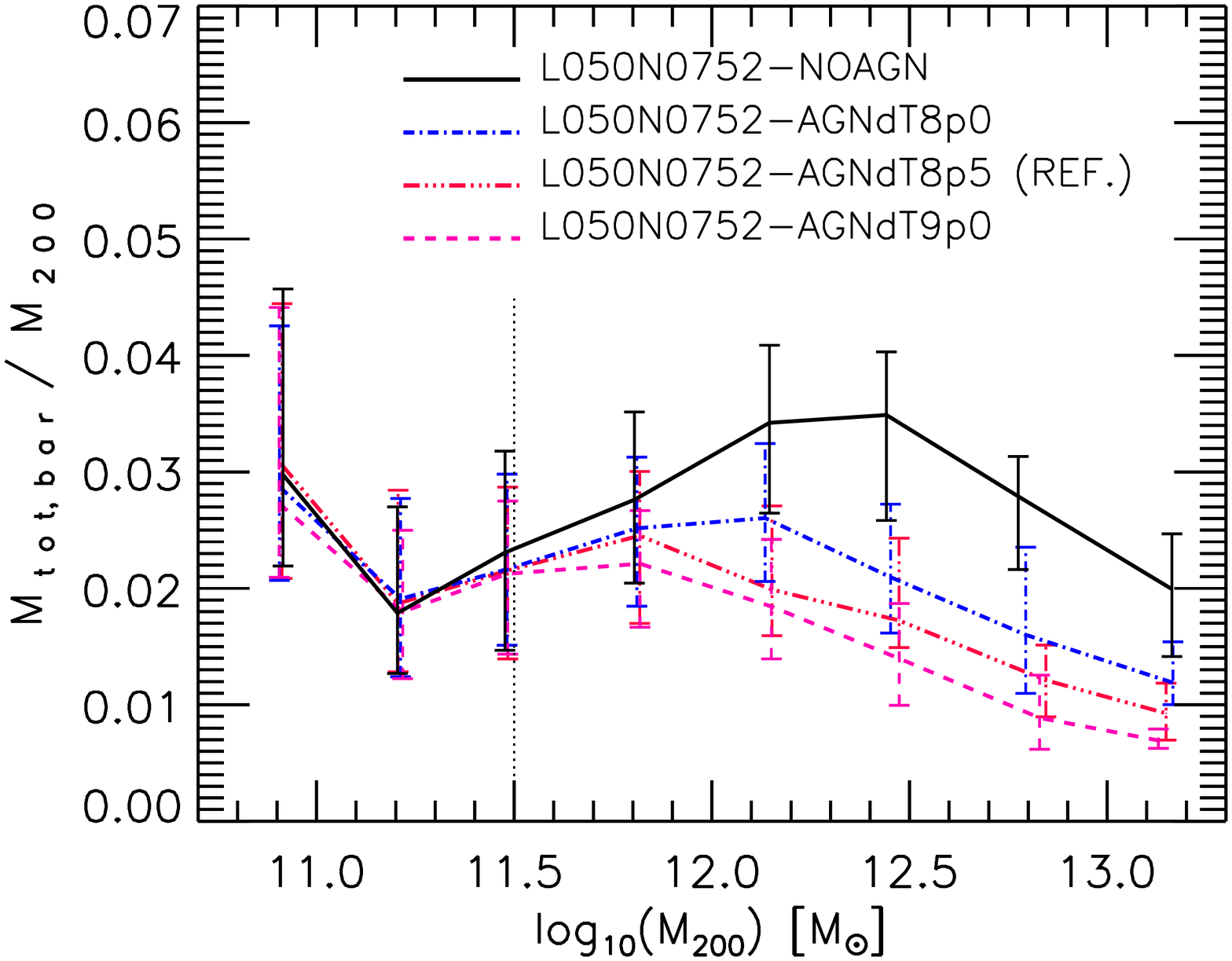}}
\resizebox{8.75cm}{!}{\includegraphics{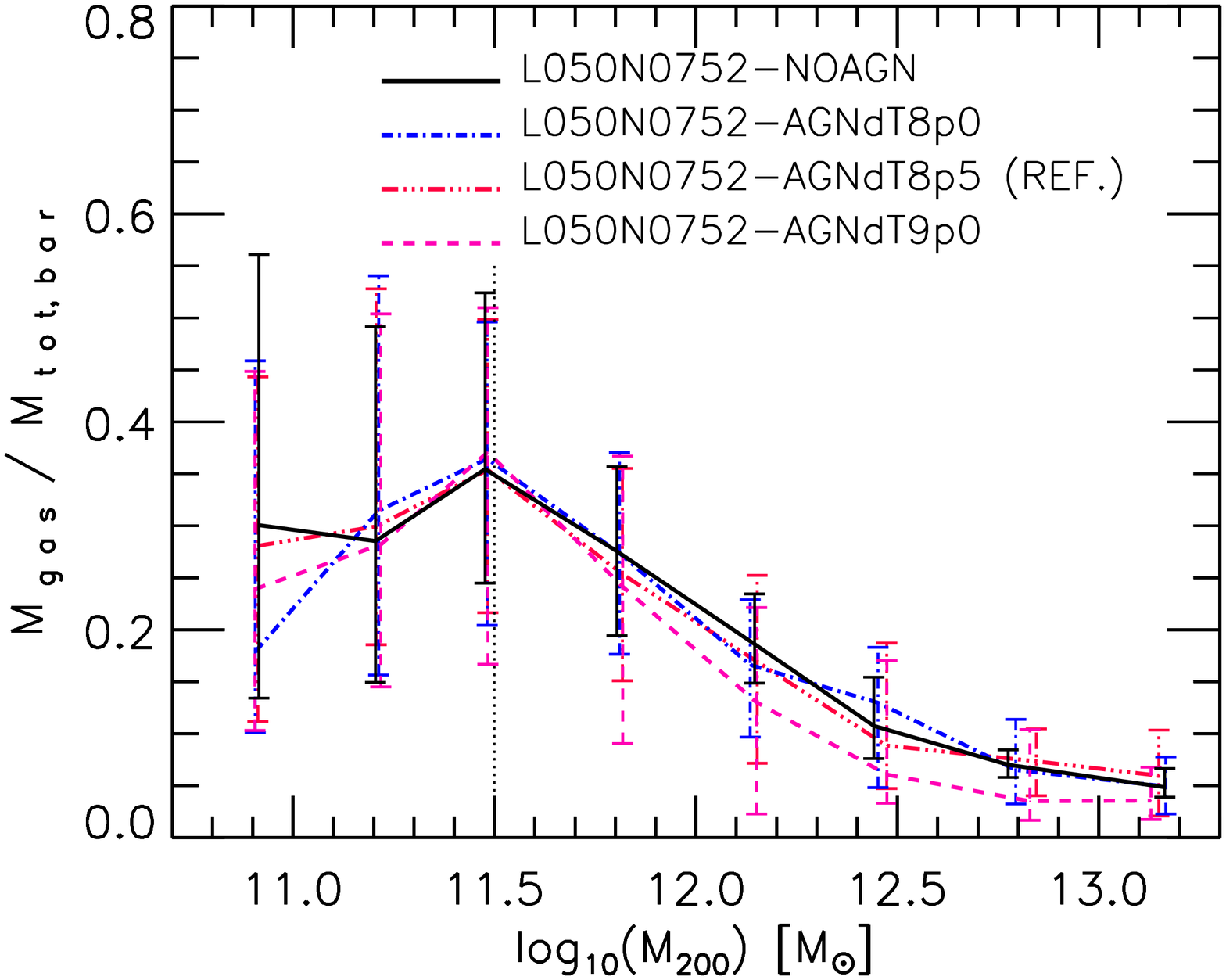}}\\
\vspace{-0.75cm}
\resizebox{8.75cm}{!}{\includegraphics{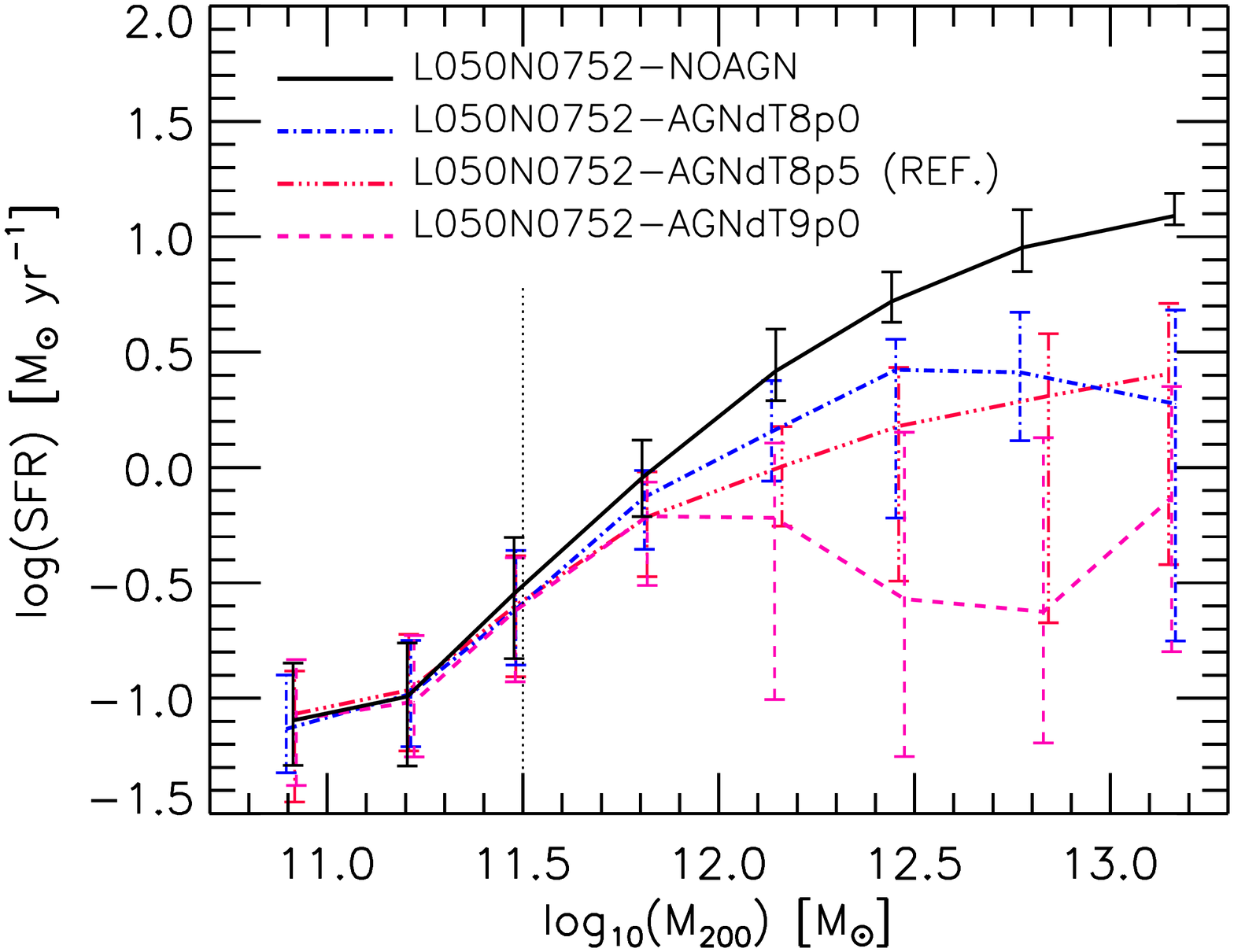}}
\resizebox{8.75cm}{!}{\includegraphics{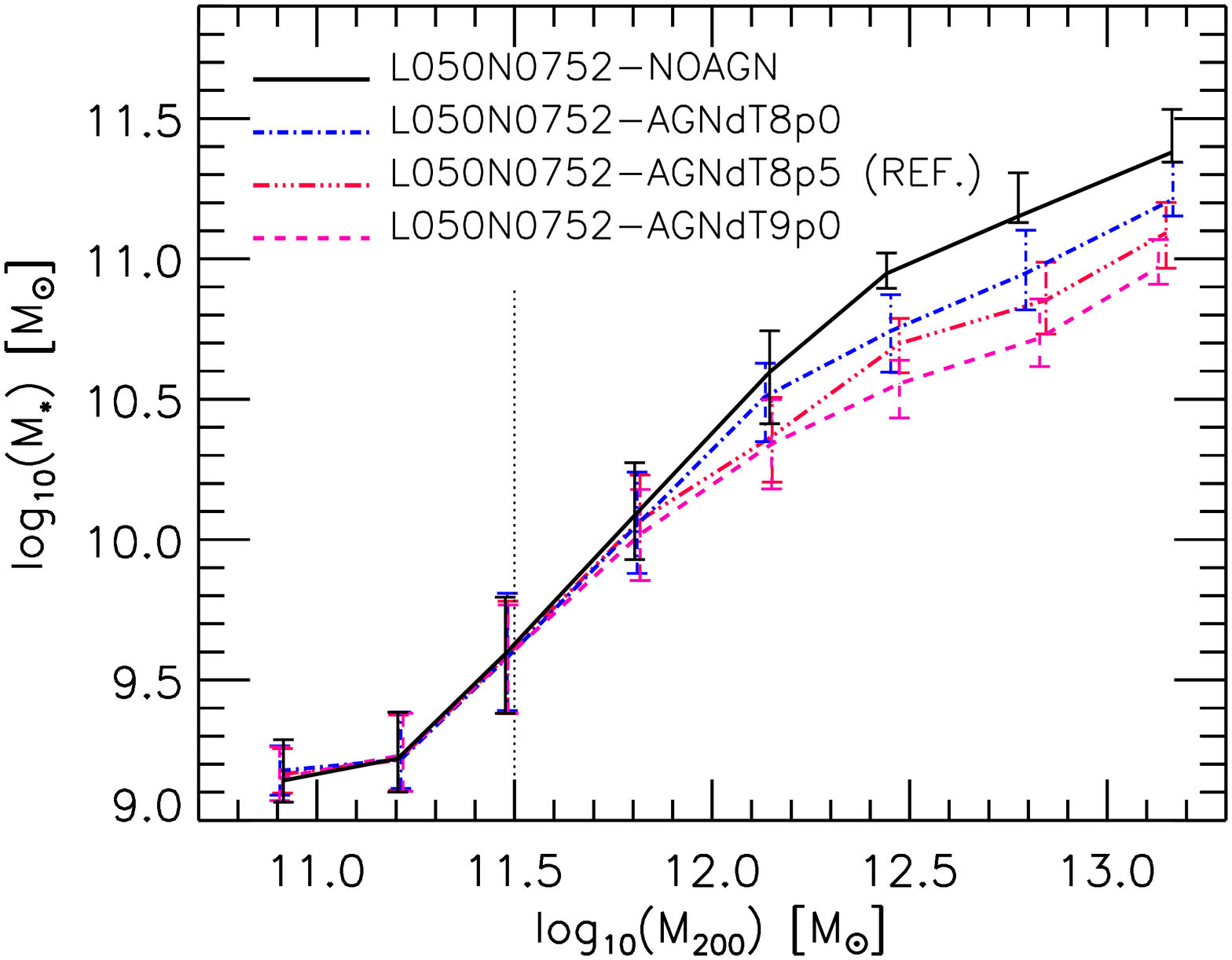}}\\
\vspace{-0.75cm}
\resizebox{8.75cm}{!}{\includegraphics{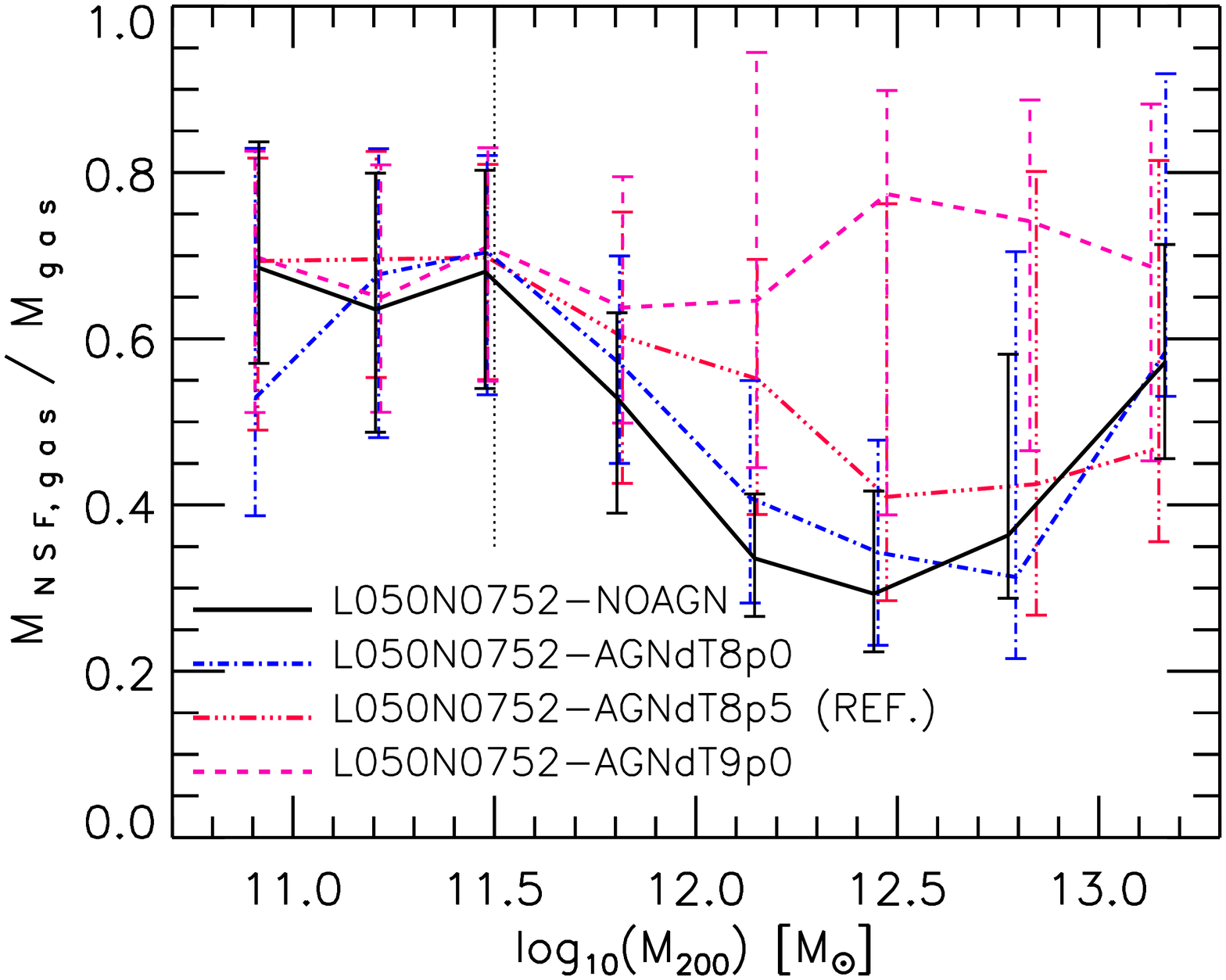}}
\resizebox{8.75cm}{!}{\includegraphics{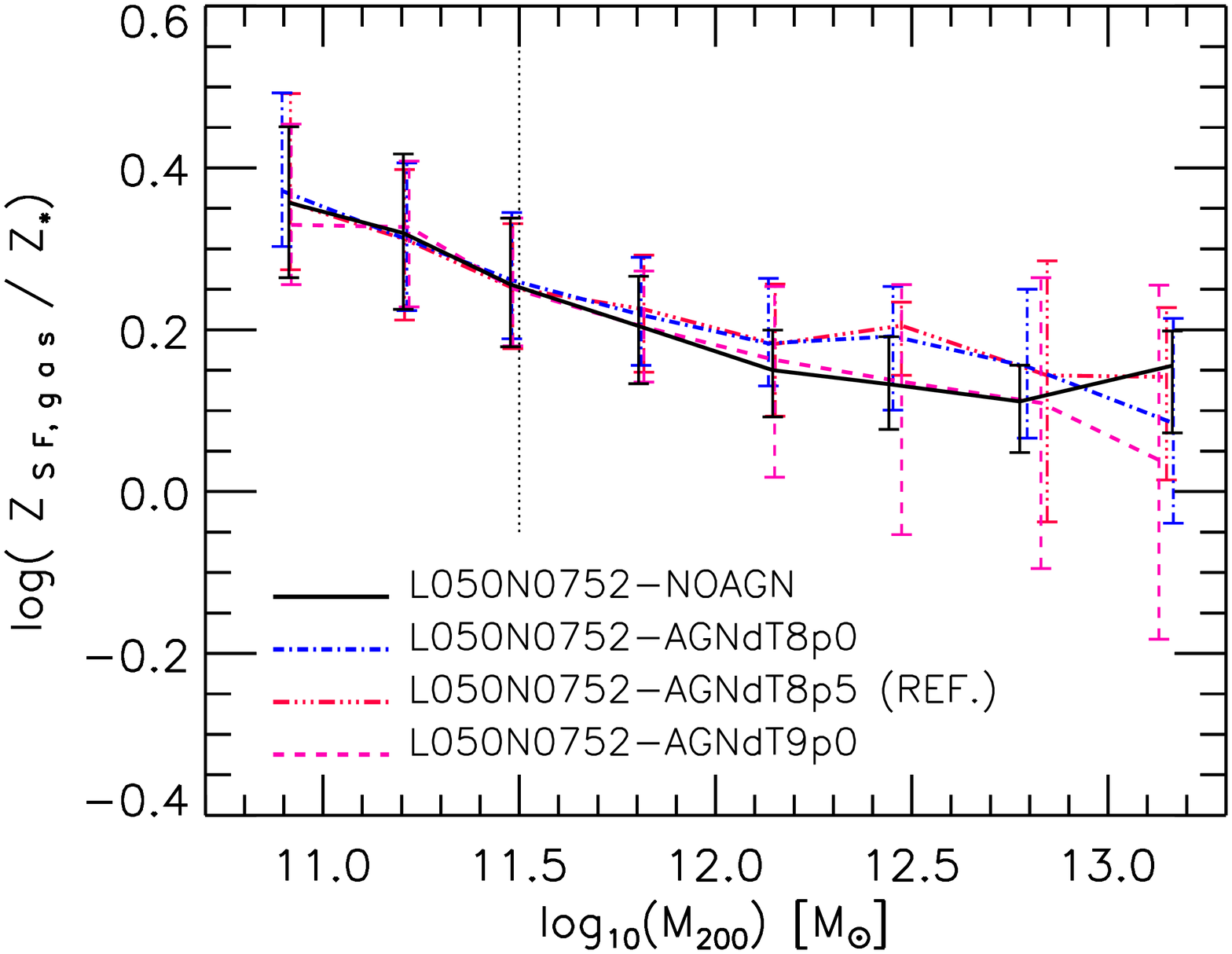}}\\
\vspace{-0.5cm}
\end{center}
\caption[]
{
Relations between different properties and halo mass ($M_{200}$) for {\em central}
simulated galaxies at $z=0$ for the same models shown in Fig. \ref{fig:mzr_agns2}.
All galaxies properties were estimated inside an aperture of 30 pkpc.
Top left panel: Ratio between total baryonic masses in galaxies and $M_{200}$ vs. $M_{200}$.
Top right panel: Ratio between total gas and total baryonic masses in galaxies vs. $M_{200}$.
Middle left panel: SFR in galaxies vs. $M_{200}$.
Middle right panel: $M_*$ in galaxies vs. $M_{200}$.
Bottom left panel: Ratio between NSF-gas and total gas masses in galaxies vs. $M_{200}$.
Bottom right panel: Ratio between SF gas and stellar metallicities in galaxies vs. $M_{200}$. 
Dotted vertical lines show the mass ($M_{200} \sim 10^{11.5}\ {\rm M}_{\sun}$, corresponding
to $M_* \sim 10^{10}\ {\rm M}_{\sun}$) above which AGN effects affect significantly
the $M_* - {\rm O/H}|_{\rm SF,gas}$ relation (Fig. \ref{fig:mzr_agns1} and \ref{fig:mzr_agns2}).
}
\label{fig:agn_effects}
\end{figure*}

\subsubsection{Residuals to the observed FMR}
\label{sec:residuals_models}

In Fig. \ref{fig:simus}, we compare the residuals of the simulated
$M_* - {\rm O/H|}_{\rm SF,gas}$ relation, in different SFR bins, with respect to the observed FMR given by 
\citet[][see shaded surface in Fig. \ref{fig:FMR}]{mannucci2010}.
In the three upper panels, we analyse the effects of changing the resolution (low resolution
"L025N0356" vs high-resolution "L025N0752" runs)
and model (reference vs recalibrated subgrid parameters, see Table \ref{tab:simus}) for a fixed simulation volume\footnote{
We have verified that similar trends are obtained if the simulation Ref-L100N1504 is used instead of
Ref-L025N0356.} 
$(25 \ {\rm cMpc})^3$.
In Sec. \ref{sec:gas_mzr}, we have already 
compare the MZRs predicted by these simulations.
As noted before, the best overall agreement with the observed FMR is obtained for 
Recal-L025N0752.  
The residuals of this simulation are below $\approx 0.1$ dex
for smaller galaxies ($M_* \la 10^{10.2} {\rm M}_{\sun}$) and below $\approx 0.2$ dex for more massive ones.

Comparing the results obtained from the Recal- and Ref-L025N0752 runs, 
it is clear that the larger differences are obtained at low masses and 
for systems with SFR $\la 1 \ {\rm M}_{\sun} \ {\rm yr}^{-1}$, where the residuals derived from
the reference model are $\approx 0.1$ dex higher than those resulting from the recalibrated model.  
Thus, the averaged low-mass $M_* - {\rm O/H}|_{\rm SF,gas}$ relation predicted by the Ref-L025N0752 simulation is flatter
than that associated with Recal-L025N0752, as already noted by \citet{schaye2015}.
The stronger outflows generate a decrease in the metallicity of the ISM in the recalibrated model,
producing a better match to the observed FMR.

To analyse the effects of resolution without changing subgrid parameters (the so-called strong convergence test, in the terminology introduced by \citealt{schaye2015}) and volume, we can compare
the simulations Ref-L025N0752 and Ref-L025N0376, shown in the middle and right upper panels 
of Fig. \ref{fig:simus}, respectively.  
The results seem to be reasonably robust against numerical
resolution as differences in the metallicity residuals 
are below $\approx 0.1$ dex considering the whole mass range.

In Fig. \ref{fig:simus}, second row, we explore how the parameter that regulates the
efficiency of AGN feedback events
affect the FMR.  
We make use of simulations NOAGN-L050N0752 (AGN feedback suppressed) and AGNdT9-L050N0752 ($\Delta T_{\rm AGN} = 10^9$K),
which yield higher AGN feedback heating temperature than the reference model ($\Delta T_{\rm AGN} = 10^{8.5}$K). 
We can see that variations in AGN feedback
have negligible effects at low stellar masses ($M_* \la 10^{10.2} {\rm M}_{\sun}$).  
At the
massive end ($M_* \ga 10^{10} {\rm M}_{\sun}$) of the FMR, AGN feedback produces a significant
decrease of the metallicity at a given $M_*$ with respect to the model without AGN, leading
to a better agreement with the data.
A higher heating temperature
drives lower oxygen abundances, especially for systems with lower SFR.  
This last issue might be related to the fact that lower
SFRs can also be related to a more efficient heating (see Section \ref{sec:mzr_agns}). 
The increase of $\Delta T_{\rm AGN}$
reduces the overall metallicities of galaxies leading to a better agreement with the 
data at ${\rm SFR} \ga 1 {\rm M}_{\sun} \ {\rm yr}^{-1}$. In the case of systems with low
SFRs (${\rm SFR} \la 1 {\rm M}_{\sun} \ {\rm yr}^{-1}$),  the decrease in metallicities obtained
for higher $\Delta T_{\rm AGN}$ generates a more significant departure from the observed FMR.  
Thus, increasing the AGN feedback heating temperature tends to strengthen the anti-correlation 
between metallicity and SFR at high stellar masses.

The panels in the last two rows of Fig. \ref{fig:simus} show the effects of 
halving and doubling the stellar feedback (first column), of changing the
power-law dependence index of the star formation law ($n$, second column) and 
the power law slope of the polytropic equation of state imposed on the ISM ($\gamma_{\rm eos}$, third column).  
See Table \ref{tab:simus} for details.
The efficiency of stellar feedback has a significant influence on the slope of the $M_* - {\rm O/H}|_{\rm SF,gas}$ relation 
(see also \citealt{crain2015}).  Weak stellar feedback yields very high metallicities towards lower stellar masses 
generating an important departure from the observed FMR ($\approx 0.4$ dex). 
Interestingly, that departure does not depend on SFR at fixed stellar mass. 
However, weaker stellar feedback does not significantly affect the metallicities of higher-mass galaxies 
(compare WeakFB-L025N0376 simulations with the Ref-L025N0376 set, in the upper right panel).  
As we have seen, the abundances in massive galaxies are mostly regulated by AGN feedback.   
On the other hand, strong stellar feedback alters the abundances of systems spanning our whole stellar mass range, 
which are reduced by up to $\approx 0.3$ dex with respect to those predicted by the reference model. 
This decrease leads to a better agreement with the data in the case of systems with lower SFR and low stellar masses.  
However, the MZR obtained for the intermediate and high-SFR bins moves below the observed relation.   
We can also deduce that an increase of the stellar feedback efficiency tends to strengthen the anti-correlation between SFR and metallicity at low stellar masses.  
In the case of the population of massive galaxies in the lowest SFR-bin, variations in the stellar feedback parameters do not 
generate important effects on the metallicities.  Because of their deeper potential-wells and lower SFRs,
they probably do not
experience significant outflows of metal-enriched material driven by stellar feedback.

Simulations KSLow-L025N0376 and KSHigh-L025N0376 were run using 
power-law indices in the star formation law (Eq. \ref{eq:sfr}, section \ref{sec:simulation_details})
of $n=1.0$ and $n=1.7$, respectively.
In the reference model, $n=1.4$. A higher value of $n$ corresponds to a higher star formation efficiency, which 
yields lower SF gas fractions if star formation is self-regulated \citep{schaye2010, haas2013}.
Because of the fundamental relation between SF gas fraction and metallicity, we thus expect metallicity to increase
with $n$ at fixed mass.
We can compare the effects of changing $n$ in the two lower panels in the middle column of Fig. \ref{fig:simus}.
Increasing $n$ generates similar effects,
at low stellar masses,
as those obtained by reducing the stellar feedback efficiency.  For higher $n$,
the metallicities of smaller systems increase with respect to the data and the slope of 
the $M_* - {\rm O/H}|_{\rm SF,gas}$ relation tends to flatten.  
In the case of a lower $n$, the main effects can be seen at the lower SFR-bins, which
describe $M_* - {\rm O/H}|_{\rm SF,gas}$ relations displaced 
towards lower abundances that those predicted by the reference model.  
Decreasing $n$ below the reference value also weakens the anti-correlation between metallicity
and SFR at low-masses but reinforces the correlation between both variables at the massive-end.

In the two lower right panels of Fig. \ref{fig:simus}, we compare simulations
eos1-L025N0376 and eos5/3-L025N0376, for which
the power law slope of the polytropic equation of state imposed on the ISM ($\gamma_{\rm eos}$) 
has been set at $\gamma_{\rm eos} = 1.0$ and $\gamma_{\rm eos} = 5 / 3$, respectively.
For the reference model, $\gamma_{\rm eos} = 4/3$.
It is clear that the effects of varying $\gamma_{\rm eos}$ 
do not generate significant variations in the metallicity relations.
This is consistent with the negligible effect on the gas fraction found by \citet{haas2013}.

Finally, from Fig. \ref{fig:simus}, it is clear that the reversal of the SFR-${\rm O/H}|_{\rm SF,gas}$
correlation at $M_* \sim 10^{10.3} \ {\rm M}_{\sun}$ is present in all {\sc EAGLE} models
considered here, with the only exception of the NOAGN run.  Varying the stellar feedback
efficiency, the equation of state imposed on the ISM or the SF law do not seem to alter this 
behaviour.  As discussed in section \ref{sec:mzr_scatter}, our findings suggest that AGN feedback 
is responsible for these trends.  Only when AGN feedback is turned off, the metallicity anticorrelates
with the SFR at the high-mass end, at least in these simulations.

\subsubsection{AGN feedback effects}
\label{sec:agn_effects}

As discussed above, AGN feedback plays an important role in the regulation of the chemical
enrichment of massive galaxies in {\sc EAGLE} simulations.  The flattening of the 
$M_* - {\rm O/H}|_{\rm SF,gas}$ relation at $M_* \ga 10^{10} \ M_{\sun}$, the decrease of effective yields as well as the
reversal of the SFR-${\rm O/H}|_{\rm SF,gas}$ relation at high $M_*$ are partly driven 
by the influence of AGN.  In particular, in section \ref{sec:mzr_agns}, we have shown that the lower
metallicities of more massive galaxies ($M_* \ga 10^{10} \ {\rm M}_{\sun}$,
corresponding to $M_{200} \ga 10^{11.5} \ {\rm M}_{\sun}$) can be associated to their lower
SFRs, caused by the heating of the gas by AGNs.
In this section, we will try to get more insight into how AGN affects the ISM and 
the metal content of simulated galaxies.  

In order to explore if AGN feedback may have led to a net metal depletion in simulated galaxies,
we studied how efficient AGNs are at ejecting gas from these systems.  In the top left panel
of Fig. \ref{fig:agn_effects}, we analyse the ratio of the total baryonic component 
(i.e. $M_{\rm tot,bar} = M_* + M_{\rm SF,gas} + M_{\rm NSF,gas}$, where the different masses were
calculated inside our standard aperture of 30 pkpc) 
of {\em central} galaxies relative to their host halo mass ($M_{\rm 200}$) 
as a function of halo mass. By analysing this quantity
for different AGN models (see Table \ref{tab:simus}), we can obtain a rough estimate of how
efficient AGNs were at ejecting mass from galaxies.  
From Fig. \ref{fig:agn_effects}, it is clear that when AGN feedback is turned off (NOAGN model), 
a higher percentage of baryons are present in central galaxies. 
In the presence of AGN feedback, the amount of baryons in {\em central} galaxies 
is reduced by $\approx 50$ percent for most massive systems. 
This percentage increases with $\Delta T_{\rm AGN}$,
as expected. As model BHs inject energy into neighbouring gas particles,
these particles are located in the inner regions of galaxies and, so, they are expected to be metal enriched.
Thus, our results suggest that AGN feedback is efficient at driving the ejection of metal-enriched material
from simulated galaxies.
Therefore, part of the decrease of the metallicity obtained at the high-mass
end of the simulated $M_* - {\rm O/H}|_{\rm SF,gas}$ relation when AGN feedback is turned on, might be caused 
by the ejection of metal-enriched material. A more detailed investigation regarding
this issue will be the subject of a future work.
   
In the following, we analyse how AGN feedback affects the state of baryons that remain inside simulated 
galaxies.
In the top right panel of Fig. \ref{fig:agn_effects},
we calculate the fraction of gas relative to the total amount of baryons inside these systems
(i.e. $M_{\rm gas} / M_{\rm tot,bar}$, with $M_{\rm gas} = M_{\rm SF,gas}+M_{\rm NSF,gas}$ and different masses 
calculated inside our standard aperture of 30 pkpc)
as a function of halo mass.
We can see that, at a given $M_{200}$, different AGN models yield similar gas fractions. 
However, at a given $M_{200}$, the SFR of galaxies decreases as $\Delta T_{\rm AGN}$ increases (middle left panel),
leading to lower $M_*$ at a fixed $M_{200}$ for higher $\Delta T_{\rm AGN}$ (middle right panel).  Thus, since $M_{\rm gas} / M_{\rm tot,bar}$
is not significantly affected by AGN feedback,
the decrease of $M_*$ caused by AGNs seems to have been compensated by the ejection of $M_{\rm gas}$
that they generate, as discussed above. 
In this context, the chemical enrichment of galaxies is quenched by AGN feedback by decreasing their SFR
and by generating the ejection of metal-enriched gas from these systems.
In the bottom left panel of Fig. \ref{fig:agn_effects}, we analyse the percentage of NSF gas
relative to the total gas component (inside an aperture of 30 pkpc) of galaxies for the different AGN models.  Despite the large dispersion,
there is evidence for an increase of the NSF-gas fraction at higher $\Delta T_{\rm AGN}$,
as expected.  This is consistent with the lower SFR obtained for higher $\Delta T_{\rm AGN}$.
Thus, AGNs also play a role in heating the gas remaining in the galaxies, shutting-down the SFR
activity and, thus, preventing further metallicity evolution from star formation.

Finally, the flattening of the
$M_* - {\rm O/H}|_{\rm SF,gas}$ relation at the massive end might be partly caused
by net metal dilution after AGNs suppress SF in these systems.
In order to explore this scenario, we followed \citet{yates2014} and
used the parameter $Z_{\rm SF,gas} / Z_*$ as a diagnostic for
dilution of the ISM after the last burst of star formation. 
$Z_{\rm SF,gas}$ and $Z_*$ were calculated inside our standard aperture of 30 pkpc.
When dilution takes place there
is a decrease in the gas-phase metallicity
without a corresponding decrease in the stellar metallicity, hence, lower values
of $Z_{\rm SF,gas} / Z_*$ should be obtained.
In the bottom right panel of Fig. \ref{fig:agn_effects}, we can see that the
median $Z_{\rm SF,gas} / Z_* - M_{200}$ relation is not significantly affected by changes in
the AGN feedback model.
However, at high masses, the scatter in the $Z_{\rm SF,gas} / Z_* - M_{200}$ relation 
spreads down to negative values of $Z_{\rm SF,gas} / Z_*$ as 
 $\Delta T_{\rm AGN}$ increases,  with a clearly asymmetric dispersion about the median for at least the AGNdT9p0 simulation.
These findings agree with those obtained by \citet{yates2014} (e.g. their Fig. 6), who showed that
dilution does not lead to a significant decrease of $Z_{\rm SF,gas} / Z_*$ for massive AGN host galaxies,
but rather to an increase of the scatter in the $Z_{\rm SF,gas} / Z_* - M_{200}$ down to low (even negative) 
values at old ages or high masses.
These results could suggest that dilution in AGN hosts also plays a (possibly
minor) role in the lower metallicities seen in massive systems, alongside shutting­down star formation and metal­rich
ejection.

\section{Conclusions}
\label{sec:conclusions}

We have studied "fundamental" metallicity relations in the 
{\sc EAGLE} suite of cosmological hydrodynamical simulations. 
These simulations were performed using a modified version of the {\sc GADGET}-3
SPH code and includes prescriptions for different baryonic processes such as
radiative cooling and heating, star formation, feedback from star formation and AGNs
and abundance evolution of 11 elements (see section \ref{sec:simulation}).
We have analysed the evolution
of metal abundances of galaxies as a function of mass and redshift and
compared our results with different observational works.
Given the 
discrepancies between observed metallicities obtained from different calibrators 
($\la 0.7$ dex), we focused on the shape and level of evolution of the relations
when comparing to simulations.
We have focused mainly on the high-resolution Recal-L025N0752 simulation, which
shows better agreement with the slope and normalization of certain 
observed metallicity scaling relations.
We have also explored other simulations to evaluate resolution effects as
well as variations in the subgrid physical parameters.

Our main findings can be summarized as follows:

\begin{itemize}
\item The high-resolution Recal-L025N0752 simulation predicts a correlation between
star forming gas metallicity and stellar mass in 
agreement with the observed trend (Fig. \ref{fig:gas_mzr_z0}).  
Metallicity tends to increase with
$M_*$, exhibiting a shallower slope towards high stellar masses ($M_* \ga 10^{10} {\rm M}_{\sun}$). 
At a given $M_*$, metallicity tends to decrease with $z$, consistent with the observed
trend, showing an overall variation of $\approx 0.5$ dex below $z\approx 3$ (Fig. \ref{fig:gas_mzr_z}).
The dependence of energy feedback on the local density and metallicity of the ISM seems to drive
the MZR evolution, at least in these simulations.
\item The flattening of the slope of the simulated $M_* - {\rm O/H}|_{\rm SF,gas}$ relation at the high-mass end 
is mainly the result of AGN feedback (see also \citealt{segers2016b}). 
To explore the role of AGN feedback, we used intermediate-resolution simulations
as we needed to focus on the trends at high masses. We note that, in the
overlap region at high masses, different resolution runs agree.
Higher values of the temperature increase ($\Delta T_{\rm AGN}$) decrease the slope of the 
$M_* - {\rm O/H}|_{\rm SF,gas}$ and $M_{200} - {\rm O/H}|_{\rm SF,gas}$ relations 
at  $M_* \ga 10^{10} {\rm M}_{\sun}$ and $M_{200} \ga 10^{11.5} {\rm M}_{\sun}$, respectively 
(Figs. \ref{fig:mzr_agns1} and \ref{fig:mzr_agns2}).
Because AGN feedback prevents further star formation, the rate of metallicity evolution decreases 
as the heating temperature increases. A higher $\Delta T_{\rm AGN}$ also generates a larger
scatter in the  $M_* - {\rm O/H}|_{\rm SF,gas}$ relation at high masses.
\item The simulated $M_* - Z_*$ relation is broadly consistent with
observations below $z \approx 3$ (Fig. \ref{fig:mzr_stars}).  In particular,  at a given $M_*$, $Z_*$ tends to be lower
than the ISM metallicity, especially for low-mass systems.  This is expected as
current star formation uses more metal-enriched material than was used to form the
older stellar components of galaxies.  For a study of the recycled stellar ejecta as fuel 
for star formation in {\sc EAGLE}, the reader is referred to \citet{segers2016b}.
\item At a given stellar mass, metallicities associated with the non-star forming (NSF) gas components 
are lower than stellar ones (Fig. \ref{fig:ZgZs_vs_M}).  
The ratio $Z_{\rm gas}/Z_*$ for the NSF gas-phase increases with decreasing $z$
and decreases with stellar mass, showing a stronger level of evolution at the high-mass
end. In particular, massive galaxies show the most homogeneous metallicity distribution
between the different baryonic phases.  
\item Simulated effective yields tend to increase with baryonic mass 
($M_{\rm bar} = M_* + M_{\rm SF,gas}$) at low masses ($M_{\rm bar} \la 10^{10} \ {\rm M}_{\sun}$),
consistent with the observed trend (Fig. \ref{fig:yeff_vs_M}). At higher $M_{\rm bar}$ and $z=0$, the predicted
effective yields decrease with $M_{\rm bar}$, while observations suggest a flattening of the relation. 
Stronger AGN feedback leads to a larger scatter and lower
average values of effective yields at higher masses.
In addition, the simulated relation between effective yields and $M_{\rm bar}$ does not evolve significantly with
$z$ at the low-mass end but, at $M_{\rm bar} \ga 10^{10} \ {\rm M}_{\sun}$, effective yields increase with $z$. 
\item At a given $M_*$, there are secondary dependencies of metallicity on the star-forming gas fraction,
specific SFR and mass-weighted stellar age (Fig. \ref{fig:MZR_bin}). Galaxies with higher star-forming gas fractions, higher
specific SFRs or lower ages are less metal enriched, with larger variations at lower stellar masses.
At the high-mass end, the dependence of metallicity on star-forming gas fraction and
specific SFR turns from an anti-correlation into
a correlation due to the effects of AGN feedback.
Simulations reproduce the observed fundamental metallicity relation between $M_*$, metallicity
and SFR reported by \citet{mannucci2010} quite well (Fig. \ref{fig:FMR}).
However, the simulated fundamental metallicity relation seems to be in place since $z\approx5$ while, according to some
observations, the observed one is established only at $z \le 3$. 
This discrepancy might be partly due to selection biases in observational studies.
\item In the case of {\sc EAGLE}, the most fundamental metallicity scaling relation
seems to be the correlation between oxygen abundance in star-forming gas and star-forming gas fraction (Fig. \ref{fig:OH_vs_fg})
as it presents low scatter and does not evolve significantly with $z$ within the scatter.  Besides, it involves only
two global properties of galaxies.  This is consistent with observational findings by \citet{bothwell2015}, who
reported a dependence of metallicity on the molecular gas fraction of galaxies. Note also that 
\citet{lagos2015} found a 
strong dependence of metallicity on neutral gas fractions from intermediate-resolution {\sc EAGLE} simulations.  
By means of empirical-constrained analytical models, 
\citet{zahid2014a} also reported a universal relation between metallicity and stellar-to-gas ratio.
\item  By comparing models with different subgrid physics parameters, 
we found that the slope
of the mass-metallicity relation is mainly modulated by stellar feedback at low-masses ($M_* \la 10^{10} {\rm M}_{\sun}$),
while AGN feedback regulates the slope at the high-mass end.
Increasing the efficiency of either type of feedback strengthens the dependence of metallicity
on SFR at a given stellar mass, increasing the scatter in the mass-metallicity relation.
Increasing the power-law index $n$ in the star formation law seems to generate similar effects,
at low stellar masses, to those obtained by reducing the stellar feedback efficiency.
On the other hand, variations in the power law slope of the polytropic equation of state
 (${\gamma}_{\rm eos}$) imposed on the ISM do not affect the features of the mass-metallicity relation.
\item Our findings suggest that AGN feedback 
generates a decrease in the global metallicity
of SF gas by quenching the star-formation activity of simulated
galaxies and also by leading to ejection of metal-enriched material. 
Net metal dilution in AGN hosts also seems to play a (possibly minor) role
in the lower metallicities obtained for massive systems.

\end{itemize}

We have seen that the {\sc EAGLE} simulation Recal-L025N0752 predicts metallicity scaling relations of galaxies 
as a function of stellar mass and redshift that are generally consistent with observations.
It also reproduces secondary dependencies of metal abundances on SFR, specific SFR,
star-forming gas fraction and stellar age at a fix stellar mass.
However, the reality of the detailed features of the observed metallicity
relations is still a matter of debate.  The use of different metallicity calibrators,
different methods for estimating stellar masses and gas fractions, selection biases,
redshift-dependent instrumental apertures, etc., prevent observational works from converging
to a consistent picture regarding the evolution of the mass-metallicity relation and the correlation between
mass, metallicity and SFR.
On the other hand, in the case of simulations, the uncertainties regarding the 
nucleosynthetic yields, resolution issues and the different subgrid prescriptions
implemented by different authors make the comparison between different models a difficult task.
In this context, continued comparison between observations and theoretical models
of galaxy formation provide a fruitful tool to  shed light on such a challenging topic.  
The sensitivity of the metallicity relations to the efficiencies of stellar and AGN feedback,
as well as to the star formation law, make such efforts worthwhile.

\section*{Acknowledgements}
We thank the referee for constructive remarks which improved the paper. 
We thank John Stott, Jabran Zahid and Sara Ellison for useful suggestions and comments.
M.E.D.R is grateful to Mar\'{\i}a Sanz and Guadalupe Lucia for their help and support during this project. 
We acknowledge support from PICT-2015-3125 of ANPCyT, PIP 112-201501-00447
of CONICET, UNLP G151 of UNLP (Argentina) and 
STFC consolidated and rolling grants ST/L00075X/1 (Durham, UK).
This work was supported by the Netherlands Organisation for Scientific Research (NWO), through VICI grant
639.043.409, and the European Research Council under the European Union's Seventh Framework Programme
(FP7/20072013)/ERC Grant agreement 278594-GasAroundGalaxies.
We acknowledge support from the European Commission's Framework Programme 7, 
through the Marie Curie International Research Staff Exchange Scheme LACEGAL (PIRSES-GA-2010-269264).
We acknowledge the Virgo Consortium for making their simulation data available. 
The EAGLE simulations were performed using the DiRAC-2 facility at Durham, managed by the ICC, 
and the PRACE facility Curie based in France at TGCC, CEA, Bruy\`res-le-Ch\^atel.
This work used the DiRAC Data Centric system at Durham University, operated by the Institute for Computational Cosmology on behalf of the STFC DiRAC HPC Facility (www.dirac.ac.uk). This equipment was funded by BIS National E-infrastructure capital grant ST/K00042X/1, STFC capital grants ST/H008519/1 and ST/K00087X/1, STFC DiRAC Operations grant ST/K003267/1 and Durham University. DiRAC is part of the National E-Infrastructure.

\bibliographystyle{mn2efix.bst}

\bibliography{references}

\begin{figure*}
\begin{center}
\resizebox{17.0cm}{!}{\includegraphics{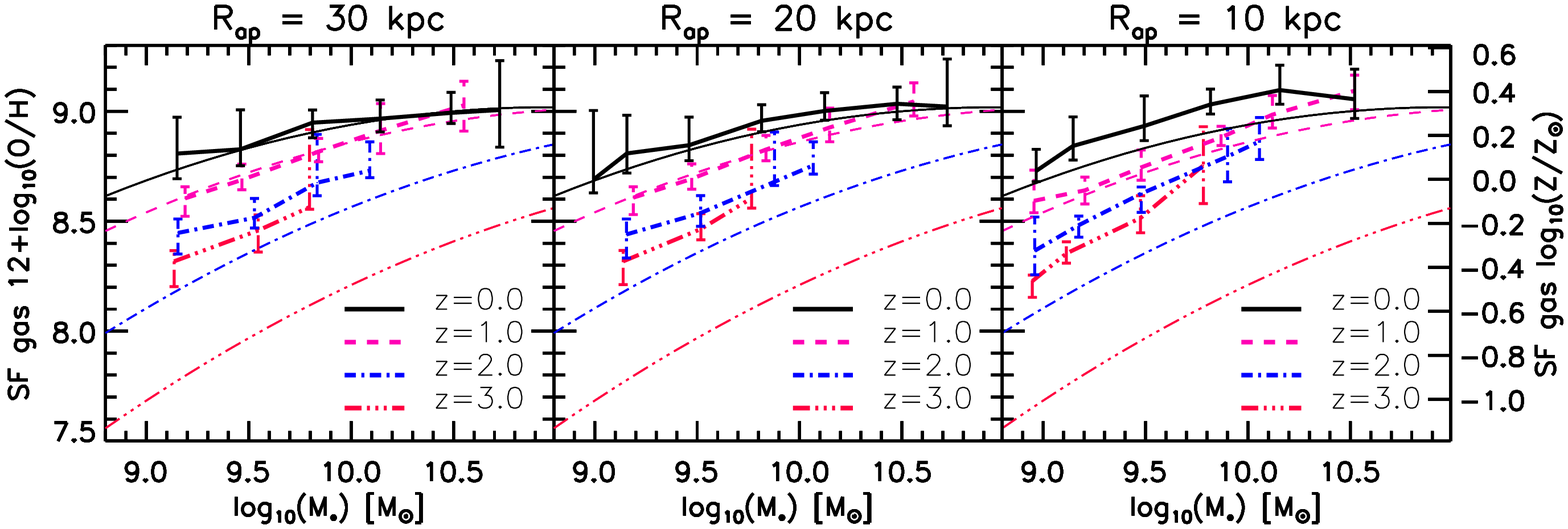}}
\end{center}
\caption[]
{
$M_* - {\rm O/H}|_{\rm SF,gas}$ relation at different $z$ for Recal-L025N0752 simulations.  
We compare results obtained by using different aperture radii, $R_{\rm ap}$: 30 kpc (left panel),
20 kpc (middle panel) and 10 kpc (right panel).
The simulated relations are shown as curves with error bars, depicting the median relation with the 25th and 75th
percentiles.  For reference, polynomial fits to observations reported by \citet{maiolino2008}
at different $z$ are shown with different lines without error bars.
The conversion between oxygen abundances along the left $y$ axes to total metallicities
shown along the right $y$ axes has been carried out assuming
$12+ \log_{10} ({\rm O/H})\sun = 8.69$ \citep{allende2001}.
}
\label{fig:mzr_ap}
\end{figure*}

\begin{figure*}
\begin{center}
\resizebox{17.cm}{!}
{\includegraphics{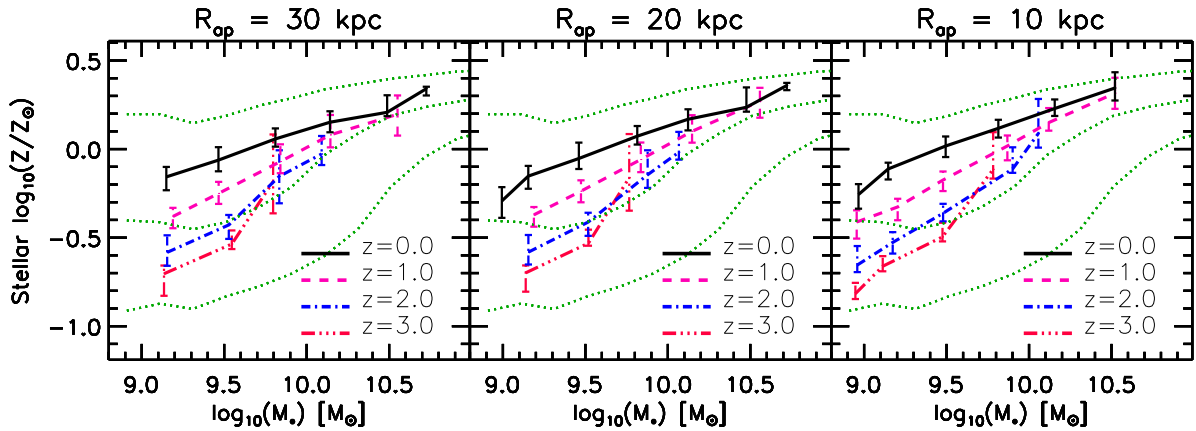}}
\end{center}
\caption[]
{
$M_* - Z_*$ relation at different $z$ for Recal-L025N0752 simulations. 
We compare results obtained by using different aperture radii, $R_{\rm ap}$: 30 kpc (left panel),
20 kpc (middle panel) and 10 kpc (right panel).
The simulated relations are shown as curves with error bars, depicting the median relation with the 25th and 75th
percentiles.  
For reference, observational data at $z\approx 0$ from \citet{gallazzi2005} are shown as
green dotted lines and indicate the median relation and 16th and 84th percentiles.
We assumed $Z_{\sun} = 0.0127$ \citep{wiersma2009b} and re-scaled observational data accordingly.
}
\label{fig:stellar_mzr_ap}
\end{figure*}

\begin{figure*}
\begin{center}
\resizebox{8.0cm}{!}{\includegraphics{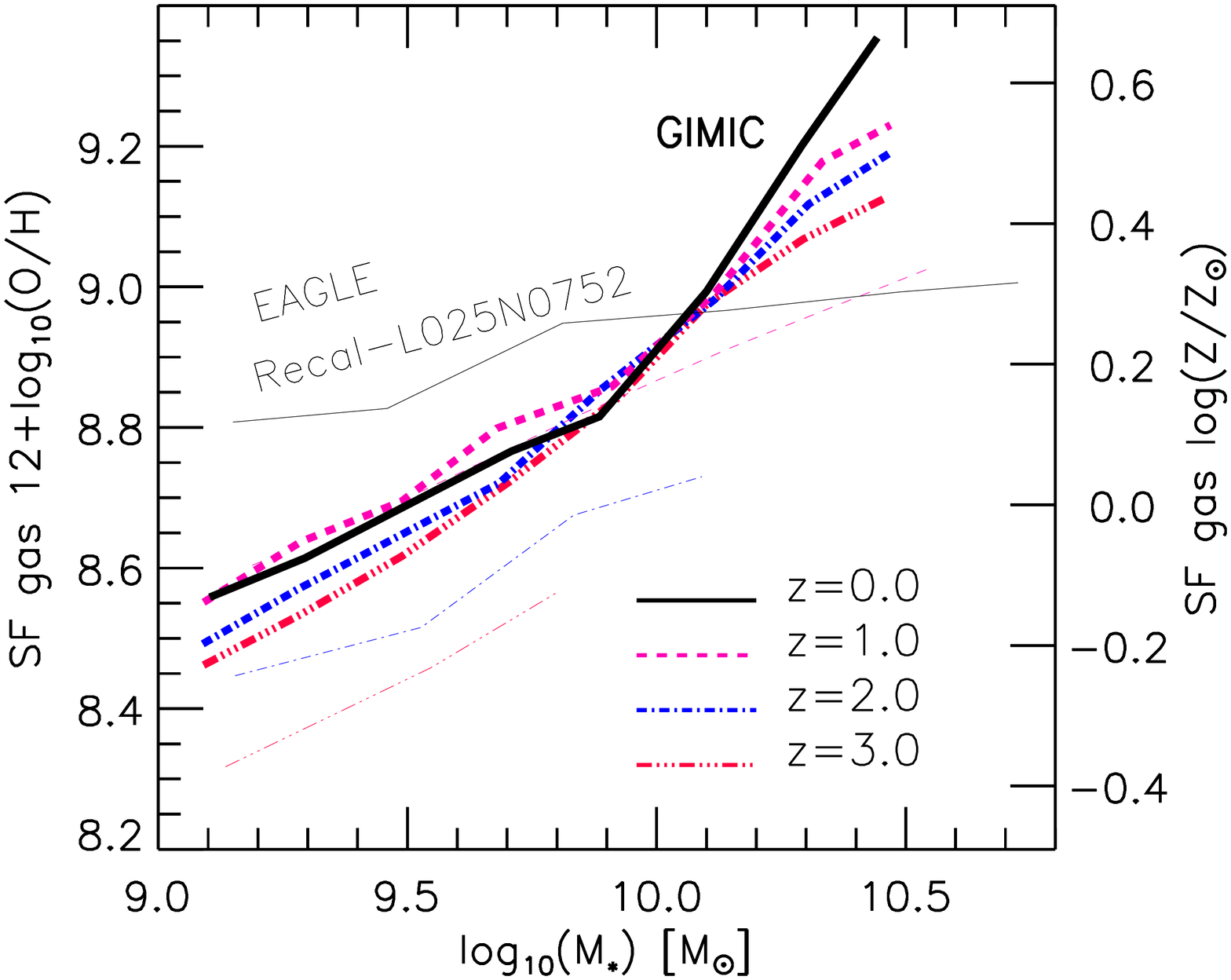}}
\resizebox{8.0cm}{!}{\includegraphics{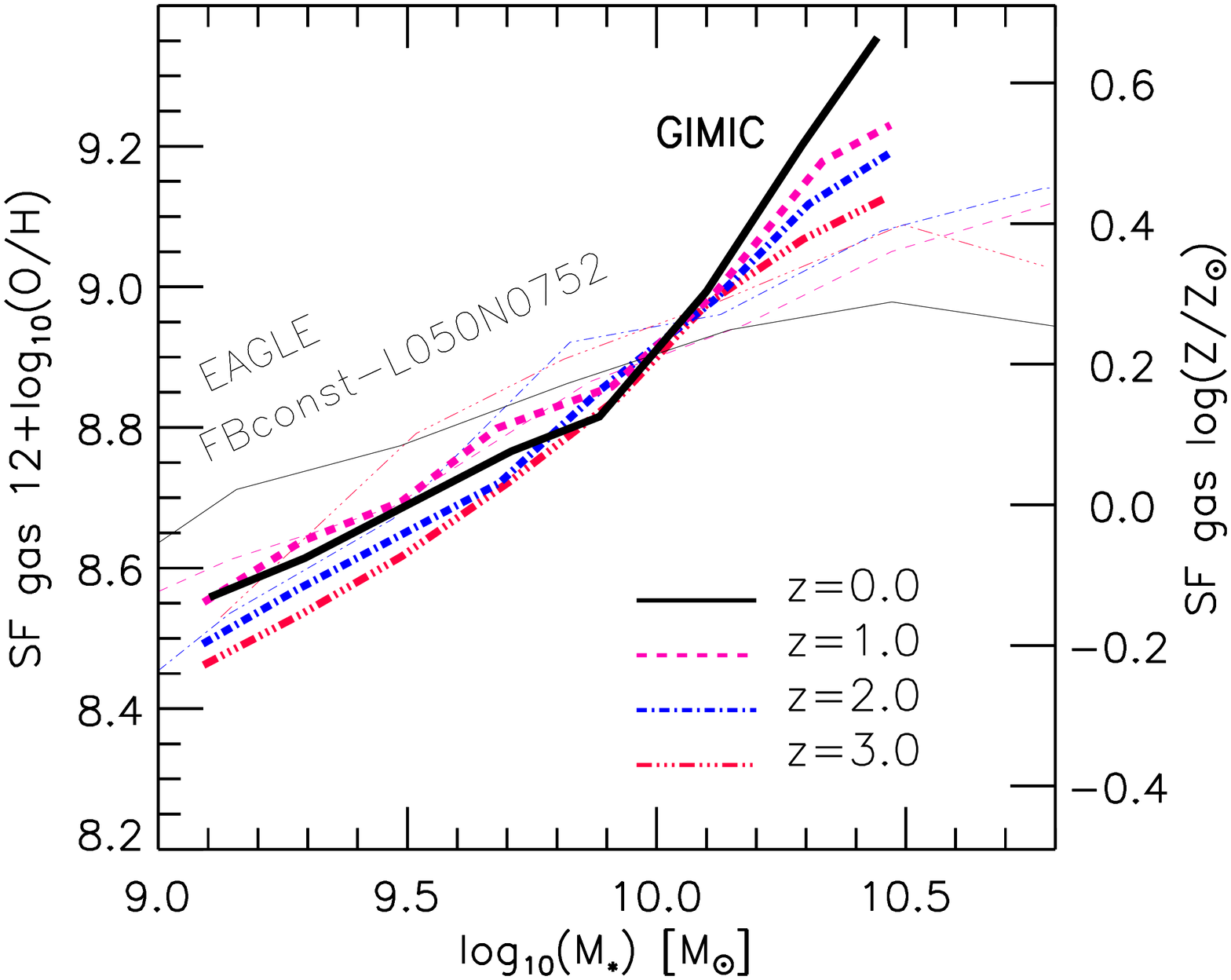}}
\end{center}
\caption[]
{
Median $M_* - {\rm O/H}|_{\rm SF,gas}$ relations at different $z$ obtained
from {\sc EAGLE} (thin lines) and GIMIC (thick lines) simulations. 
We compare results from the Recal-L025N0752 (left panel) and FBconst-L025N0752 (right panel)
{\sc EAGLE} simulations.
The conversion between oxygen abundances along the left $y$ axis and total metallicities
shown along the right $y$ axis has been carried out assuming
$12+ \log_{10} ({\rm O/H})\sun = 8.69$ \citep{allende2001}.
}
\label{fig:gimic_mzr}
\end{figure*}

\section*{Appendix A: Aperture effects}

Throughout this work we have determined global properties of galaxies
within 3-D aperture radius $R_{\rm ap} = 30$ proper kpc.  We have checked that changing
$R_{\rm ap}$ does not affect the main trends found for metallicity scaling relations.
However, changes in $R_{\rm ap}$ can generate moderate variations of the slope and
normalization of those relations.

In Figs. \ref{fig:mzr_ap} and \ref{fig:stellar_mzr_ap} we study the impact of 
changing $R_{\rm ap}$ on the $M_* - {\rm O/H}|_{\rm SF,gas}$ and $M_* - Z_*$ relations,
respectively.
Results for Recal-L025N0752 are analysed at different $z$. It is clear that varying $R_{\rm ap}$ between 20
and 30 proper kpc yields negligible modifications in the MZR for all considered $z$.  This suggests
that most of the mass in these galaxies is located at radii $< 20$ kpc.
On the other hand, when decreasing $R_{\rm ap}$ towards 10 kpc, 
there is a slight overall increase of gas and stellar metallicity at all studied $z$ and, especially, at high stellar masses.

Thus, according to our results, simulated metallicity scaling relations 
tend to be convergent at $R_{\rm ap} \ga 20$ kpc while moderate changes in slope
and normalization are obtained when $R_{\rm ap}$ is reduced to $\approx 10$ kpc,
especially at high masses.

We have also analysed the evolution of the MZR 
by measuring metallicities inside the apertures reported by \citet{maiolino2008} for observed galaxies, 
which are redshift-dependent.
Such changes tend to slightly increase the slope of the simulated MZR, especially for massive galaxies
at $z\approx 2$.  Nevertheless, the flattening of the MZR at $z\sim 0$ is still present when applying 
the aforementioned apertures.
In addition, when using those redshift-dependent apertures,
the median simulated MZR obtained at a given $z$ is displaced by $\approx 0.1$ dex towards higher metallicities.  
However, the level of evolution of the MZR between $z=0$ and $z=3$ remains at $\approx 0.5$ dex.

\section*{Appendix B: Comparison with {\sc GIMIC} simulations}
\label{sec:gimic}

The subgrid physics implemented in {\sc EAGLE} is based on that developed for OWLS
\citep{schaye2010}, and applied also in {\sc GIMIC} \citep{crain2009} and 
cosmo-OWLS \citep{lebrun2014}. 
As commented before,
here we extended the work of \citet{derossi2015}, who studied galaxy metallicity scaling relations
by analysing {\sc GIMIC} simulations.
The most important differences between the {\sc GIMIC} and {\sc EAGLE} subgrid models concern the 
implementation of energy feedback from star formation (which is now thermal rather than kinetic and constant
and depends on the local density and metallicity), 
the star formation law (which now depends on metallicity) and the inclusion of an AGN feedback model.

As the subgrid physics implemented in {\sc EAGLE} is based on that applied
in {\sc GIMIC} simulations, it is worth a comparison between the MZR derived from both models.
For more details about metallicity scaling relations in {\sc GIMIC} simulations,
the reader is referred to \citet{derossi2015, derossi2015b, derossi2016}.
In the following, we will compare results from high-resolution {\sc GIMIC} simulations, 
high-resolution {\sc EAGLE} simulation Recal-L025N0752 and intermediate-resolution {\sc EAGLE} simulation
FBconst-L050N0752 (see Table \ref{tab:simus}).
We note that high-resolution {\sc GIMIC} simulations have a resolution similar to that corresponding
to {\sc EAGLE} intermediate-resolution simulations.
We have checked that similar conclusions to the ones presented below are obtained if 
reference intermediate-resolution {\sc EAGLE} simulations are used instead of the Recal-L025N0752 run.

In the left panel of Fig. \ref{fig:gimic_mzr}, we show {\sc GIMIC} (thick lines) and 
{\sc EAGLE} Recal-L025N0752 (thin lines) $M_* - {\rm O/H}|_{\rm SF,gas}$ relations at different $z$, as indicated in the figure.
We will not concern about the normalization of the relations as it is not well constrained
observationally (see Fig. \ref{fig:gas_mzr_z0}).
We can see that, at low masses, the slope predicted by {\sc GIMIC} is slightly steeper than that
obtained from {\sc EAGLE} Recal-L025N0752.
As can be inferred from Fig. \ref{fig:gas_mzr_z0}, 
intermediate-resolution {\sc EAGLE} simulations predict even flatter slopes than {\sc GIMIC} 
for less massive galaxies.
The {\sc EAGLE}-{\sc GIMIC} differences at
the low-mass end of the MZR are consistent with a weaker net effect of SN feedback in {\sc EAGLE}.

Towards higher masses and lower $z$, the slope of the {\sc EAGLE} Recal-L025N0752 MZR flattens,
in agreement with observations.
There is evidence of a saturation of 
the {\sc EAGLE} galaxies metallicities 
at an asymptotic value of $12+ \log_{10} ({\rm O/H}) \approx 8.9-9.0$.
However, there is no signature of this flattening in the case of {\sc GIMIC}. \citet{derossi2015}
claimed that this last issue
is related to the lack of an AGN feedback model in GIMIC. 
As discussed previously in section \ref{sec:mzr_agns}, our findings validate that conclusion as
the AGN feedback model implemented in {\sc EAGLE} seems to be responsible for the 
shallow slope of the MZR at the high-mass end \citep[see also][]{crain2015,segers2016b}.  
AGN feedback prevents further
star formation in massive galaxies and, hence, decelerates the metal enrichment process of
massive systems.  AGN feedback also can drive the ejection of metal-enriched gas.
In addition, 
\citet{derossi2015} obtained a very steep slope for the FMR at high stellar masses
because of the lack of AGN feedback in the {\sc GIMIC} simulations.
We have seen that the {\sc EAGLE} simulations, 
which include AGN, seem to have resolved
that issue too (Section \ref{sec:mzr_scatter}).

As reported by \citet{derossi2015}, Fig. \ref{fig:gimic_mzr} shows a negligible evolution  
of the {\sc GIMIC} MZR between $z\approx 0$ and $z\approx 3$.  
The simulated evolution of the {\sc EAGLE} Recal-L025N0752 MZR between $z\approx 0$ and 
$z\approx 3$ reaches $\approx 0.5$ dex, 
in better agreement with observed data (see Fig. \ref{fig:gas_mzr_z}).
However, the weaker MZR evolution derived from {\sc EAGLE} at $z \ga 2$ compared to some observations
\citep[e.g.][]{maiolino2008,onodera2016}
suggests that stellar feedback might still be inefficient to model metal-enriched outflows in 
low-mass galaxies at $z > 2$. However, as there is not consensus about the observed level of evolution of the MZR yet, 
we are not able to validate or refute findings from simulations.

We found that the evolution of the MZR in EAGLE is driven by the dependence of
the energy injection from core-collapse supernovae on the local density and metallicity of the ISM.
We verified this by analysing the simulation FBconst-L050N0752, which implements a model that
injects into the ISM a fixed quantity of energy per unit stellar mass formed, 
similarly to the model implemented in {\sc GIMIC} simulations.  Results can be seen in 
the right panel of Fig. \ref{fig:gimic_mzr}.  Clearly, when the injected energy is independent of
local conditions, we obtain a negligible evolution for the {\sc EAGLE} MZR.
At low masses, the MZR slope predicted by the FBconst-L050N0752 simulation is also consistent with {\sc GIMIC} simulations.
As mentioned, at high masses, the discrepancies between {\sc EAGLE} and {\sc GIMIC} are due to the
implementation of an AGN feedback in the former simulations.

\end{document}